\documentclass[a4paper,11pt,final]{article}

\usepackage[colorlinks=true]{hyperref} 
\usepackage{url}
\usepackage[english]{babel}
\usepackage[utf8]{inputenc}
\usepackage[T1]{fontenc}
\usepackage{lmodern}
\usepackage[pdftex]{graphicx}
\usepackage[top=2cm, bottom=2cm, left=2cm, right=2cm]{geometry}
\usepackage{setspace}
\usepackage[french]{varioref}
\usepackage{lastpage}
\usepackage{fancyhdr}
\usepackage[table]{xcolor}
\usepackage{tikz}
\usepackage{titling}
\usepackage{graphicx}
\usepackage{amsmath}
\usepackage{amsthm}
\usepackage{amssymb}
\usepackage{amsfonts}
\usepackage{bm}
\usepackage{dsfont}
\usepackage{bbold}

\usepackage{multirow}
\usepackage{rotating}
\usepackage{subcaption} 
\usepackage[gen]{eurosym}
\usepackage{dirtree}

\usepackage{rotating}
\usepackage{colortbl}

\usepackage{footmisc}

\usepackage{diagbox} 

\usepackage{pdflscape}

\usepackage[title]{appendix}

\usepackage[T1]{fontenc}
\DeclareFontFamily{T1}{calligra}{}
\DeclareFontShape{T1}{calligra}{m}{n}{<->s*[1.44]callig15}{}
\DeclareMathAlphabet\mathcalligra   {T1}{calligra} {m} {n}
\DeclareMathAlphabet\mathzapf       {T1}{pzc} {mb} {it}
\DeclareMathAlphabet\mathchorus     {T1}{qzc} {m} {n}
\DeclareMathAlphabet\mathrsfso      {U}{rsfso}{m}{n}

\usepackage{enumitem}
\newcommand{\fuda}[1]{%
    \begin{minipage}[t]{\linewidth}
    \begin{itemize}[nosep]
    #1%
    \end{itemize}
    \vspace{0.05cm}%
    \end{minipage}
}

\definecolor{greycolor}{cmyk}{0,0,0,1}

\usepackage{relsize} 

\newcommand\ssigma{{\mathlarger{\sigma}}} 
\newcommand\sssigma{{\mathlarger{\ssigma}}}

\DeclareMathOperator*{\argmax}{argmax}

\usepackage[linesnumbered,ruled,vlined,boxed]{algorithm2e}
\DontPrintSemicolon 
\SetKw{Continue}{continue}

\usepackage{tabularx}

\newcolumntype{C}[1]{>{\centering\let\newline\\\arraybackslash\hspace{0pt}}m{#1}}
\newcolumntype{L}[1]{>{\raggedright\let\newline\\\arraybackslash\hspace{0pt}}m{#1}}
\newcolumntype{R}[1]{>{\raggedleft\let\newline\\\arraybackslash\hspace{0pt}}m{#1}}

\usepackage{doi}
\usepackage[autostyle]{csquotes}
\usepackage[style=numeric, citestyle=numeric-comp, 
	maxcitenames=2, maxbibnames=99, 
    backend=bibtex
]{biblatex}
\title{An evaluation framework for comparing epidemic intelligence systems}
\author{Nejat Ar\i n\i k, Roberto Interdonato, Mathieu Roche \& Maguelonne Teisseire}

\hypersetup{
    pdftitle={\thetitle},
    pdfauthor={\theauthor},
    bookmarksnumbered=true,bookmarksopen=true,
	unicode=true,colorlinks=true,linktoc=all,
	linkcolor=blue,citecolor=blue,filecolor=blue,urlcolor=blue,
	pdfstartview=FitH
}

\usepackage{enumitem}
\setlist{nolistsep}

\begin{document}
\maketitle
\sloppy

\abstract{In the context of Epidemic Intelligence, many Event-Based Surveillance (EBS) systems have been proposed in the literature to promote the early identification and characterization of potential health threats from online sources of any nature. Each EBS system has its own surveillance definitions and priorities, therefore this makes the task of selecting the most appropriate EBS system for a given situation a challenge for end-users. In this work, we propose a new evaluation framework to address this issue. It first transforms the raw input epidemiological event data into a set of normalized events with multi-granularity, then conducts a descriptive retrospective analysis based on four evaluation objectives: spatial, temporal, thematic and source analysis. We illustrate its relevance by applying it to an Avian Influenza dataset collected by a selection of EBS systems, and show how our framework allows identifying their strengths and drawbacks in terms of epidemic surveillance.}

\textbf{Keywords:} Epidemic Intelligence, Event-Based Surveillance, Retrospective Analysis.

\textcolor{red}{\textbf{Cite as:} N. Arinik, R. Interdonato, M. Roche \& M. Teisseire. \href{www.doi.org/10.1109/ACCESS.2023.3262462}{An evaluation framework for comparing epidemic intelligence systems}. IEEE Access, 2023. DOI: \href{www.doi.org/10.1109/ACCESS.2023.3262462}{10.1109/ACCESS.2023.3262462}}

\section{Introduction}
\label{sec:introduction}
At least 60\% of infectious human diseases originated in animals\footnote{\url{www.cdc.gov/onehealth/basics/zoonotic-diseases.html}}. The emergence and spread of any animal infectious disease, such as Avian Influenza, has serious consequences for animal health and a substantial socio-economic impact for agriculture. For instance, the 2021–2022 season have experienced the largest observed highly pathogenic avian influenza (HPAI) cases in Europe so far, with a total of 2,467 outbreaks in poultry, 3,573 HPAI events in wild birds, and 48 million birds culled in the affected establishments\footnote{\url{www.ecdc.europa.eu/en/news-events/2021-2022-data-show-largest-avian-flu-epidemic-europe-ever}}. Due to this highly contagious nature, it is critical to monitor new and emergent infectious animal diseases. To this aim, epidemic intelligence has been used to remedy this public health issue.

Traditionally, a public health surveillance system has long used Indicator-Based Surveillance (IBS) for a global epidemic monitoring approach, the well-known ones being the World Organisation for Animal Health (WOAH)\footnote{\url{www.woah.org}} and the Food and Agriculture Organization of the United Nations (FAO)\footnote{\url{www.fao.org}}. This type of surveillance consists in collecting structured and verified official health threats, hereafter referred to as epidemiological events (or events for short), through routine national surveillance systems and public health authorities. However, IBS typically undergoes some reporting delay in the detection of these data, as it relies only on laboratory confirmed animal cases. To improve this timeliness issue, several Event-Based Surveillance (EBS) systems have been proposed with the aim of promoting the early identification and characterization of potential epidemiological events from online sources of any nature, including online news outlets and social media, thanks to the recent developments in internet and digital technologies~\cite{Zeng2021}. Recently, several EBS platforms have shown their effectiveness by detecting the first signals of emerging infectious disease outbreaks in a timely manner and providing alerts within previously unaffected areas (e.g.~\cite{Valentin2021a}).

In the literature, there exist two categories of EBS systems by their functioning nature: 1) moderated (i.e. human-curated) and 2) automated. The first type of systems are human-curated ones that rely on pure manual data collection and analysis. The data can be provided by official or unofficial data sources, but in any case their accuracy is manually assessed by moderators. The Program for Monitoring Emerging Diseases (ProMED) is such an example of a moderated system~\cite{Carrion2017}. The second type of systems differs from the first one in that it includes in some or all of their pipelines automated text-mining based steps for data collection and processing. Furthermore, automated systems are also categorized into semi- and fully-automated systems. The main difference between them is that the former includes a dedicated team of curators to assess and verify the outputs, whereas the latter does not. An example of semi-automated system is the Canadian Public Health Agency Global Public Health Intelligence Network (GPHIN)~\cite{Mikanatha2013}. Likewise, fully-automated systems include BioCaster~\cite{Collier2008, Meng2022}, HealthMap~\cite{Freifeld2008}, MediSys~\cite{Linge2010}, PADI-web~\cite{Valentin2021}, DANIEL~\cite{Lejeune2015}, Sentinel~\cite{Serban2019} and Epitweetr~\cite{Espinosa2022}.


Each EBS system has its own priorities (e.g. geography, disease) and surveillance definitions (e.g. collected epidemiological information), so there is no such candidate as a \textit{best} EBS system, that would fit all situations. However, due to the profusion of available EBS systems, selecting the most appropriate one(s) for an effective surveillance system of a given situation is a challenge for end-users. Some existing works try to compare them according to the guideline of the Centers for Disease Control and Prevention (CDC)~\cite{Schwind2014, Barboza2013}, but they either focus only on few evaluation aspects or require human resources for manual assessment, which brings some cost to practitioners. Furthermore, there exist many other studies conducting a retrospective analysis using surveillance dataset, without any objective of comparison. These works deal with additional evaluation points that are not considered in the CDC's guideline, which would bring valuable additional information for evaluation purposes.

In this work, we propose a new automatic evaluation framework to solve all these issues. It is based on four evaluation objectives: 1) spatial analysis (how the events are geographically distributed), 2) temporal analysis (how the events evolve over time and what temporal aspects characterize it), 3) thematic entity analysis (what thematic entities are extracted from the events and how they are related to spatio-temporal analysis) and 4) news outlet analysis (what news sources play key role in epidemiological information dissemination). For each aspect, we compare the obtained results with a reference gold standard database, along with an appropriate visualization for end-users. All these analyses aim to highlight the strengths and drawbacks of the considered EBS systems in terms of epidemic surveillance.  We illustrate its relevance by applying it to a selection of EBS systems. Our main contribution is essentially threefold. First, we propose a generic evaluation framework, which is not tied to any specific disease, geographical region, or surveillance definition, so it can be applied to any situation, as long as we have access to a gold standard database. Second, we model the studied epidemiological events with multi-granularity in order to better understand the spatial and temporal evolution of disease events, as well as their thematic characterization. Third, we take into consideration in our framework the fact that there exist some gaps between EBS systems in disease detection and collection, an issue so-called \textit{reporting bias}~\cite{Jones2008, Huang2016}. 

The rest of the article is organized as follows. First, in Section~\ref{sec:RelatedWork}, we review the literature on EBS systems, focusing on different evaluation strategies. Next, in Section~\ref{sec:EvaluationFramework}, we introduce our evaluation framework designed to study and compare EBS systems and their outputs. We put it into practice on a selection of EBS systems in Section~\ref{sec:ExperimentalSetup} and discuss these results in Section~\ref{sec:Results}. Finally, we review our main findings in Section~\ref{sec:Conclusion}, and identify some perspectives for our work.

\section{Related Work}
\label{sec:RelatedWork}

In this section, we review the existing evaluation strategies for EBS systems. The performance assessment of these systems are traditionally performed according to the CDC's guideline, which aims at understanding the internal and external performances of EBS systems~\cite{Schwind2014, Barboza2013}. Nevertheless, most of these evaluation metrics are more in line with an end-user perspective, which require human resource for manual assessment. 


On the other hand, there exist many studies which conduct a retrospective analysis using surveillance dataset, i.e. the output of an IBS/EBS system, 
without performing any comparative study. These works deal with additional evaluation points that are not considered in the traditional evaluation methods~\cite{Barboza2013, Barboza2014a}. In particular, we are interested in those works performing a descriptive analysis, rather than predictive analysis, which is in line with our work. For this reason, we widen the scope of our review with these works. 

In the following, we overview the existing works in four parts: 1) Spatial (Section~\ref{subsubsec:RelatedWork_SpatialDimension}), 2) temporal (Section~\ref{subsubsec:RelatedWork_TemporalDimension}), 3) thematic (Section~\ref{subsubsec:RelatedWork_ThematicDimension}) and 4) source (Section~\ref{subsubsec:RelatedWork_SourceDimension}) dimensions. Note that although the surveillance data is naturally spatio-temporal, we review each dimension separately for the sake of clarity.

\subsection{Spatial dimension}
\label{subsubsec:RelatedWork_SpatialDimension}
The spatial dimension is the most studied dimension in the existing works. We summarize these works in two aspects: 1) geographic coverage and 2) hotspot analysis. The most widespread evaluation analysis is the assessment of geographic coverage of the surveillance data, and it is often time-invariant. This geographic coverage is calculated for either the whole world~\cite{Lyon2011, Jones2008, Goel2020, Boulos2020} or some particular regions/countries~\cite{Mukhi2016}.
Hence, this analysis allows showing to what degree the locations (e.g. countries) are covered by the data at hand. Moreover, it can be used to manually identify the events appearing in an unusual geographic zone in the context of early warning detection~\cite{Goel2020}.

Another analysis for the spatial dimension is hotspot analysis. The hotspots are the areas, where a substantial number of events are concentrated over time. The task of identifying the hotspots is also referred to as \textit{outbreak detection} in the literature of Epidemic Intelligence. Such hotspots are usually found through three different approaches. The first one is the exponentially weighted average method~\cite{Freifeld2008} by assigning large values to more recent alerts coming from multiple sources through the decay parameter of the exponential weight. The second one is the spatial auto-correlation analysis, which statistically identify the hotspots~\cite{Wang2018, Gehlen2019, Pakzad2018, Mao2019}. The most used technics are the Moran's I~\cite{Moran1950} and the Getis-Ord Gi$^\star$\cite{Getis2010}. The last approach is spatio-temporal clustering analysis, which aims at determining regions where the number of events is significantly higher than expected. Space-time scan statistics~\cite{Xue2022} and ST-DBSCAN~\cite{Birant2007} are two such well-known clustering methods. 

In our work, we only include the geographic coverage-based assessment. 
This is because we want to evaluate the epidemiological information collected by EBS systems at fine-grained level. This is only possible at event level, rather than at outbreak level. Nonetheless, as in hotspot analysis, we take the temporal aspect into account by adapting the traditional calculation of geographic coverage.

\subsection{Temporal dimension} 
\label{subsubsec:RelatedWork_TemporalDimension}
The temporal dimension is another important aspect in a retrospective analysis of surveillance data. Since our goal is to perform a descriptive analysis, in the following we focus only on it with two aspects: qualitative vs. quantitative assessments. 


There are two main approaches for the qualitative assessments to describe the temporal evolution of the events. The most widespread approach is trend analysis to capture underlying temporal features in time-series event data. This includes methods that can identify discriminatory information about a particular time-series data (e.g., shapelets~\cite{Hills2013}), those that look for temporally frequent sub-sequences that occur in a majority of time-series (e.g., temporal patterns~\cite{Mueen2014})), and those that investigate on seasonal~\cite{Matsubara2014} and periodic~\cite{Kiran2019} effects. The second approach is to identify anomalous cases in a time-series~\cite{Scales2013}. These anomalous cases can represent either the locations having a significantly high infection cases (e.g. outbreaks)~\cite{Serban2019} or the locations that have remarkably different infection history than neighbor locations (e.g. potential early signals)~\cite{Goel2020}.

Another evaluation analysis is through the quantitative assessments. The most widespread analysis is using the concept \textit{timeliness} in order to evaluate how timely the events are detected by an EBS system~\cite{Bahk2015, Arsevska2016, Valentin2020b, Oliveira2018, Schwind2014, Barboza2013}. If an EBS system reports the events in a timely manner, this would allow public authorities to mitigate potentially dangerous situations as soon as possible. Another approach for quantitative assessments aims at evaluating how two time series data are in a similar trend. This evaluation is usually done with the correlation analysis between the daily or weekly event time series derived from IBS/EBS systems using Pearson's correlation coefficients~\cite{Bhatia2021}. The final approach relies on the concept of \textit{transmissibility}. It is used to quantify how easily a disease can spread through a population, i.e. how rapidly an outbreak is growing or declining. It can be measured by estimating the basic~\cite{Dharmaratne2020}, effective or time-varying reproduction numbers~\cite{Cori2013, Bhatia2021a, Nash2022}. 

Regarding the connection with our work, we include a qualitative assessment based on frequent temporal events in time-series data, and we perform it with an appropriate frequent spatio-temporal pattern mining method. Moreover, we include a quantitative assessment based on timeliness. 
However, we do not include the other works presented above for the following reasons. First, EBS systems rely on unofficial data sources, therefore false alerts might be introduced in the data. This requires to handle it with a specific method, which is out of scope in this work. Second, each EBS system collects epidemiological data of different size, and their differences can be substantial. In which case, calculating the correlation coefficient of two time series data, each associated with a different source, can be biased towards the most populated source. 
Finally, estimating the transmissibility and the reproduction number are disease-dependent. This requires to develop a different model for each disease, which is also out of scope in this work.

\subsection{Thematic dimension}
\label{subsubsec:RelatedWork_ThematicDimension}
The thematic dimension is not always well elaborated in the existing surveillance systems. This is probably because the collected events are characterized by spatio-temporal attributes in practice, rather than their thematic attributes (e.g. disease and host)~\cite{Valentin2022}. Therefore, to the best of our knowledge, there exist only a handful of works for evaluating the thematic dimension of the existing EBS systems.

All the existing works in the literature are interested only in the ranking of thematic entities. 
This ranking can be obtained with the frequency~\cite{Goel2020}, a statistical measure (e.g. F-measure~\cite{Roche2004, Valentin2021}, chi-square~\cite{YomTov2014}) and a constraint based objective (e.g. temporal periodicity~\cite{Li2018}). In all these works, there are two factors, which directly affect the ranking results. The first one is related to the multidimensionality nature of the elements, for which the ranking is computed. 
In case of two or more dimensions, this corresponds to the identification of co-occurences in the same events. The second factor is related to the normalization of thematic entities, i.e. how they are individually expressed for comparison purposes. This normalization step consists in transforming a raw text into one of well-defined taxonomy classes, which results in hierarchical information. In the literature, most of the works focus only on onea fixed~\cite{Meng2022} or a few~\cite{Valentin2021, Valentin2022} hierarchical levels.

In this work, we also include the assessment based on the ranking of thematic entities. We use the combination of all mentioned approaches: frequent pattern mining, F-measure~\cite{Roche2004, Valentin2021} and temporal periodicity~\cite{Li2018}. Moreover, the thematic elements are normalized with  multi-granularity. Finally, we take temporal periodicity into account in the ranking results.

\subsection{Source dimension}
\label{subsubsec:RelatedWork_SourceDimension}
All EBS systems partially or completely rely on various online news and press agencies, news outlets for short, for ensuring their monitoring of emerging infectious diseases across the world. Nevertheless, there are not enough studies that characterize and assess the news sources involved in EBS systems. The existing works study these sources at two different levels: news aggregator and news outlet levels. 

On the one hand, the first level aims to assess the degree to which news aggregators contribute to the news collected by EBS systems. 
Lyon \textit{et al.}~\cite{Lyon2011} show based on the main EBS systems that the most contribution is provided by Google News, then to a lesser extent ProMED, MeltWater and Baidu.
On the other hand, the second level focuses on how countries are covered by the news outlets at hand. \cite{Schwind2014, Ao2016} show that international news outlets do not capture well news infection events occurring in some less-developed regions, which results in a reporting bias. In which case, local news outlets performs better, because these events are mostly reported in local television or recorded in local print media in local or regional languages. 
Finally, the news aggregators and news outlets are inherently related and dependent to each other, if an EBS system collects its news data from news aggregators. The authors of \cite{Valentin2022a} analyze this aspect with a network analysis approach by describing how outbreak-related information disseminates from a news outlet to a news aggregator.

In this work, we only analyze the publishing sources at news outlet level. This is because not all EBS systems rely on multiple news aggregators (e.g. PADI-web). Unlike the existing works, we rely on a ranking based assessment of news outlets with two different objectives: importance and timely detection. 

\section{Evaluation Framework}
\label{sec:EvaluationFramework}
In this section, we describe the framework that we propose to evaluate and compare a number of EBS systems based on the epidemiological data that they collect. Our goal here is to highlight the strengths and drawbacks of the considered EBS systems in terms of epidemic surveillance. Put differently, we want to know what we lose when we monitor a number of high-threat diseases with a single EBS platform, while there might be some different epidemiological information captured by other EBS systems.

\begin{figure*}
    \centering
    \includegraphics[width=0.85\textwidth]{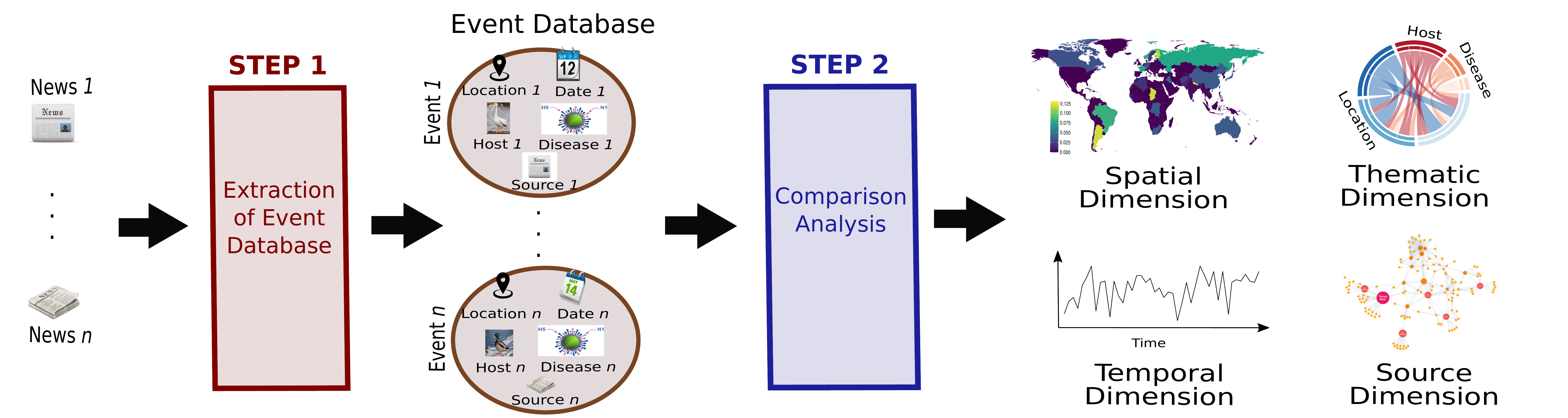}
    \caption{Workflow of our evaluation framework.}
	\label{fig:worfklow}
\end{figure*}

To this aim, we propose a two-step pipeline approach, which is illustrated in Figure~\ref{fig:worfklow}. The input of the pipeline is a set of unnormalized events, accompanied by the associated news documents. Since each EBS system can collect and extract epidemiological information from online sources in a different way, the first step is to extract the normalized events from the input. We detail this step in the Appendix (Section~\ref{secapx:CorpusEventExtraction}), for space matters. Then, the second step consists in performing a retrospective analysis of these events with four objectives: 1) spatial, 2) temporal, 3) thematic and 4) source dimensions. Each dimension allows answering a question that naturally arises in our analysis, and it is implemented through a well-known existing tool deemed appropriate for this purpose. Our methodological contribution is found in the combination of these tools. In the rest of this section, we describe the different steps of our framework in detail.

In the following, we first describe in Section~\ref{subsec:EventRelatedDefinitionsNotations}. how we define an event in our context, as well as event related definitions and notations. Second, we explain how to identify common events across IBS/EBS platforms in an automatic manner, a task that we call \textit{event matching} (Section~\ref{subsec:EventMatching}). Then, we evaluate an EBS system based on the spatial (Section~\ref{subsec:Method_SpatialDimension}), temporal (Section~\ref{subsec:Method_TemporalDimension}), thematic (Section~\ref{subsec:Method_ThematicDimension}) and source (Section~\ref{subsec:Method_SourceDimension}) dimensions. Except the last dimension, these evaluations are always performed with respect to a gold standard database, which is supposed to contain all events confirmed and notified by national and supranational authorities. Moreover, we take into consideration the fact that there exist some gaps between EBS systems in disease detection and collection, an issue so-called \textit{reporting bias}~\cite{Jones2008, Huang2016}. To do so, our evaluation relies on either ranking results (see Section~\ref{subsecapx:RankingEvaluation} in the Appendix) or the concept of representativeness (e.g. Section~\ref{subsec:Method_SpatialDimension}).

\subsection{Definitions and Notations Related to Events}
\label{subsec:EventRelatedDefinitionsNotations}
An event definition depends on the application at hand, and there is no unified standard. In the literature, an event is minimally defined as a disease-location pair, and associated with an infection time (or time period)~\cite{Lejeune2015, Valentin2021}. Although this minimal definition shows to what degree the locations are covered by the data at hand over time for a particular disease, the other works expand this definition with 1) the news outlets by which news documents are found~\cite{Lyon2011, Valentin2021} and 2) thematic information (e.g. disease serotype, hosts, symptoms)~\cite{Valentin2021, Mikanatha2013, Rortais2010, Collier2008, Meng2022, Serban2019}. Although the extracted thematic information can be very rich, depending on a system, relating thematic entities to the events can be challenging. This is because there can be multiple events in the same news document, even in the same sentence. 
Based on these previous event definitions, we define an event throughout this work as the detection of the virus for a specific host at a specific date and in a specific location. Moreover, we also consider the fact that an event is reported by a news outlet. We illustrate in Table~\ref{tab:event-example} how we define an event from the following text: "\textit{A highly pathogenic strain of bird flu (H7N9) has been detected in two captive birds of prey at a private property in Skelmersdale on March 31.}". 

\begin{table}
    \centering
	\scriptsize
	\renewcommand{\arraystretch}{1.2}
	\begin{tabular}{| p{1.5cm} p{1.25cm} p{1.4cm} p{1.5cm} p{1.25cm} |}
    	\hline
           Location & Date & Disease & Host & Source\\\hline
           Skelmersdale & 31-03-2021 & H7N9 serotype & captive bird & Liverpool Echo\\
        \hline
	\end{tabular}
 \captionsetup{width=.9\linewidth}
    \caption{The corresponding event for the text reported from Liverpool Echo\protect\footnotemark: "\textit{A highly pathogenic strain of bird flu (H7N9) has been detected in two captive birds of prey at a private property in Skelmersdale on March 31.}".}
	\label{tab:event-example}
\end{table}
\footnotetext{\url{www.liverpoolecho.co.uk/news/liverpool-news/highly-pathogenic-bird-flu-detected-20309813}}

Next, we introduce the definitions and notations related to events. An event database $\mathcal{E}$ is a finite set of events collected by an EBS system. Let $\mathcal{D} = \{D_1, \dots D_n\}$ be a set of dimensions to define the events in $\mathcal{E}$. Every event is expressed as a tuple $e = (d_1, \dots, d_n)$, where we call $d_i$ event attribute for every $i = 1, \dots, n$. Concretely, in this work, the set $\mathcal{D}$ contains five dimensions, and it is defined as $\{D_Z, D_T, D_D, D_H, D_S\}$. The dimension $D_Z$ is the location, where disease events have been occurred, and they are expressed as polygons (e.g. country or city polygons). The dimension $D_T$ is the notification date of events and it is a totally ordered domain. 
Moreover, the dimension $D_D$ is the disease which infects a number of hosts. The dimension $D_H$ is the host who have caught the viruses of a specific disease. Finally, the dimension $D_S$ is the news outlet publishing a given epidemiological event. 

Each dimension $D_i$ is associated with a domain of (discrete) values, denoted by $dom(D_i)$. Given an event database $\mathcal{E}$ over $\mathcal{D}$, for every $i = 1, \dots, n$, we denote by $Dom_\mathcal{E}(D_i)$ (or simply $Dom(D_i)$ if $\mathcal{E}$ is clear from the context), the active domain of $D_i$ in $\mathcal{E}$, which corresponds to the set of all values of $Dom(D_i)$ occurring in $\mathcal{E}$. In this work, we consider only values in active domains. Moreover, we assume that each dimension $D_i \in \mathcal{D}$ is associated with a hierarchy, denoted by $H_i$, in order to consider different granularity levels of domain values. 
Every hierarchy $H_i$ is a tree whose nodes are elements of $Dom(D_i)$ and whose root is $ALL_i$. For instance, for the spatial dimension $D_Z$, $ALL_Z$ corresponds to the whole world containing all existing locations. We illustrate in Table~\ref{tab:hierarchical-event-example} the hierarchical representation of the event from Table~\ref{tab:event-example}.

\begin{table}
    \centering
	\scriptsize
	\renewcommand{\arraystretch}{1.2}
	\begin{tabular}{| p{0.5cm} | p{2.5cm} p{1.5cm} p{2cm} p{1.15cm} p{0.8cm} |}
    	\hline
        Level & Location ($D_Z$) & Date ($D_T$) & Disease ($D_D$) & Host ($D_H$) & Source ($D_S$)\\\hline
         0 & $ALL_Z$ & $ALL_T$ & $ALL_D$ & $ALL_H$ & $ALL_S$\\
         1 & Europe & 2021 & avian flu & bird & Liverpool Echo\\
         2 & United Kingdom & 03-2021 & highly patho. & captive bird & \\
         3 & England & week 13 & H7N9 &  & \\
         4 & Lancashire & 31-03-2021 & & & \\
         5 & West Lancashire & & & & \\
         6 & Skelmersdale & & & & \\
        \hline
	\end{tabular}
    \caption{Hierarchical event representation for the event illustrated in Table~\ref{tab:event-example}.}
	\label{tab:hierarchical-event-example}
\end{table}

Moreover, we sometimes need to fix the spatial and temporal scales (i.e. hierarchical levels) of an event database $\mathcal{E}$. This operation amounts to discretize the dimensions $D_Z$ and $D_T$ over a set $Z$ of geographic zones and a set $T$ of time intervals, depending on the spatial and temporal scales. We denote this fixed scaled event database by $\mathcal{E}^{H_Z \sim l_Z}_{H_T \sim l_T}$ ($\mathcal{E}^{l_Z}_{l_T}$ for short), where $l_Z$ (resp. $l_Y$) represents a spatial (resp. temporal) scale in $H_Z$ (resp. $H_T$). When this fixed scaled database is ordered by time, then it is defined as $\mathcal{E}^{l_Z}_{l_T} = \{(t_{1}, X_1), (t_{2}, X_2), \dots , (t_{|T|}, X_{|T|})\}$, where $|T|$ represents the size of database, $X_j \subseteq Z$ is a set of spatial entities and $t_{j}$ represents a time interval for which $X_j \in Z$ occurs in $\mathcal{E}^{l_Z}_{l_T}$.
Note that if an event does not have precise information with respect to $l_Z$ and $l_T$ (e.g. an event occurring in France, while $l_Z = city$), we do not include it in $\mathcal{E}^{l_Z}_{l_T}$. We illustrate a fixed scale event database $\mathcal{E}^{l_Z}_{l_T}$ with an example in Table~\ref{tab:STpatterns_example}. This example relies on a toy fictional event database, in which we fix the spatial and temporal scales to \textit{country} and \textit{week}, respectively. Each row in Table~\ref{tab:STpatterns_example} includes the countries reporting at least one epidemiological event for a given weekly time interval. For instance, we observe the first disease cases in France, Italy, China and India during the first week. Then, in the second week the viruses spread over neighbor countries, which are Spain, India and Nepal. Finally, it is also possible to restrain all literal values of a dimension $D_i$, i.e. $Dom_\mathcal{E}(D_i)$, with a fixed spatial or temporal scale. For a given spatial (resp. temporal) scale, we denote it as $Dom_\mathcal{E}(D_i, l_Z)$ (resp. $Dom_\mathcal{E}(D_i, l_T)$).

\begin{table}
    \scriptsize
    \centering
    \begin{tabular}{|l|l|}
        \hline
        \textbf{Time Interval} & \textbf{Country} \\\hline
        week 1 & France, Italy, China, India \\\hline
        week 2 & France, Italy, Spain, China, India, Nepal \\\hline
        week 4 & France, Spain, Portugal, India, Nepal \\\hline
        week 6 & Spain, Portugal, India \\\hline
        week 7 & Spain, Portugal, India \\\hline
        week 8 & Portugal, India, Pakistan\\ \hline
        week 10 & India, Pakistan \\\hline
        week 11 & Italy, India, Pakistan \\\hline
    \end{tabular}
    \captionsetup{width=.9\linewidth}
     \caption{Illustrative example of $\mathcal{E}^{l_Z}_{l_T}$, where $l_Z = country$ and $l_T = week$. The first column describe the time intervals, for which epidemiological events occur, and the second column indicates the countries, in which epidemiological events occur.  Note that $\mathcal{E}^{l_Z}_{l_T}$ contains only the country information, even though more specific spatial information can be present in the data. 
     }
     \label{tab:STpatterns_example}
\end{table}

\subsection{Event matching}
\label{subsec:EventMatching}
In this section, our goal is to identify common events between two event databases in an automatic manner, which is not a trivial task. We propose here an approximation scheme by modeling this task as an assignment problem, also known as maximum weighted bipartite matching problem, as already done in the literature (e.g.~\cite{Ramshaw2012}). In the end, we obtain a set of "\textit{putatively}" associated events between two event databases.

Let $\mathcal{E}_1$ (resp. $\mathcal{E}_2$) be two event databases associated with IBS or EBS systems, containing $N_{\mathcal{E}_1}$ and $N_{\mathcal{E}_2}$ events, respectively. Also, we assume $N_{\mathcal{E}_1} \leq N_{\mathcal{E}_2}$ without loss of generality. Moreover, let $S$ be the $N_{\mathcal{E}_1}\ \times\ N_{\mathcal{E}_2}$ similarity matrix of $\mathcal{E}_1$ and $\mathcal{E}_2$. The term $S_{ij}$, with $1\leq i \leq N_{\mathcal{E}_1}$ and $1\leq j \leq N_{\mathcal{E}_2}$, represents the similarity score between events $e_i$ and $e_j$ and it is calculated with Equation~\ref{eq:FinalSim} (Section~\ref{secapx:EventSimilarity} in the Appendix). Then, we look for a bijection $f : \{ 1, 2, ..., N_{\mathcal{E}_1} \} \rightarrow \{ 1, 2, ..., N_{\mathcal{E}_2} \}$ such that the objective is to maximize the similarity between $\mathcal{E}_1$ and $\mathcal{E}_2$, as defined in Equation~\ref{eq:EventMatching_AssignmentProblem}.

\begin{equation}
	Max  \sum_{i=1}^{N_{\mathcal{E}_1}} S_{if(i)}.
	\label{eq:EventMatching_AssignmentProblem}
\end{equation}

Since this problem can be modelled as an assignment or a maximum weighted bipartite matching problem, it can be solved in various ways. One of them is through the well-known Hungarian algorithm, whose complexity is $O(n^3)$~\cite{Kuhn1955}.

Finally, in the solution of the assignment problem, some events might be assigned to other events with negative or weak positive similarity scores. Therefore, we perform a post-processing by removing the assignment results, whose similarity scores are lower than some threshold value. 

\subsection{Spatial Dimension}
\label{subsec:Method_SpatialDimension}
Our evaluation strategy for the spatial dimension relies on the concept \textit{representativeness}. Barboza \textit{et al.}~\cite{Barboza2013} define this concept as the ability of describing accurately the distribution of events in terms of place, time and host. Particularly, geographic representativeness constitutes an important aspect in Epidemic Intelligence. For this reason, we propose to compare the spatial dimension of the events collected by an EBS system through geographic representativeness by taking the temporal aspect into account. We call it \textit{spatio-temporal representativeness}, and it allows measuring how well the event database $\mathcal{E}$ of an EBS system represents geographic zones (e.g. country, regions) in terms of the events found in a gold standard database $\mathcal{E_R}$, for a given time period. In the end, the obtained results enable us to know to what degree geographic zones are represented by $\mathcal{E}$.

In the definition of the spatio-temporal representativeness, we say that an EBS system represents well a specific geographic zone for a given time interval, if it finds at least one event in $\mathcal{E_R}$. For this reason, its calculation requires fixing the spatial and temporal scales of the events in $\mathcal{E}$ (resp. $\mathcal{E_R}$) with $l_Z$ and $l_T$, i.e. $\mathcal{E}^{l_Z}_{l_T}$ (resp. $\mathcal{E_R}^{l_Z}_{l_T}$). Since there can be some reporting delay between the events of $\mathcal{E}$ and $\mathcal{E_R}$, we also consider in this calculation the previous (resp. next) time interval in order not to penalize an EBS system. For a given geographic zone, we perform this calculation for all the time intervals, and then we take their average to obtain a final score. This score is in the range $[0,1]$, where the score of $0$ (resp. $1$) indicates that $\mathcal{E}^{l_Z}_{l_T}$ never (resp. always) finds an event in $\mathcal{E_R}^{l_Z}_{l_T}$ for a given geographic zone.

For space matters, we explain in the Appendix how we calculate the spatio-temporal representativeness score of an event database $\mathcal{E}$ with respect to a gold standard database $\mathcal{E^R}$ (Section~\ref{subsecapx:QuantitativeEvaluationForSpatialDimension}).

\subsection{Temporal Dimension}
\label{subsec:Method_TemporalDimension}
For the temporal dimension, we include two evaluation assessments. The first one is a quantitative assessment based on the concept timeliness (Section~\ref{subsubsec:TemporalDimension_Timeliness}). The second one is a qualitative assessment related to the consistent periodic behavior of the events (Section~\ref{subsubsec:TemporalDimension_FullPartialPeriodicity}).

\subsubsection{Timeliness}
\label{subsubsec:TemporalDimension_Timeliness}

We start with the first comparison, which is based on the concept \textit{timeliness}~\cite{Bahk2015, Arsevska2016, Oliveira2018, Schwind2014, Barboza2013}. Barboza \textit{et al.}~\cite{Barboza2013} define this concept as the ability of identifying disease events in a timeframe enabling utilization of the information by decision makers to mitigate potentially dangerous situations as soon as possible.

In the literature, timeliness is measured as the time difference between the publication date of an event in an EBS system and that of the same event in a gold-standard database. Nevertheless, we model it with an exponential decay function in order to obtain a normalized score, as proposed in \cite{Liu2020, Lu2015, Claes2014}. Its calculation for an event database $\mathcal{E}$ of an EBS system is performed with respect to a gold standard database $\mathcal{E}_R$. This requires to know the binding of the events between $\mathcal{E}$ and $\mathcal{E}_R$, which is unknown in advance. To estimate such a binding we rely on the method described in Section~\ref{subsec:EventMatching}. In this method, for a given event $e \in \mathcal{E}$, we define a bijective function $f(e, \mathcal{E})$, which returns the putatively associated event $e'$ in $\mathcal{E}_R$ with $e \neq e'$. Then, when we repeat it for each event in $\mathcal{E}$, and we obtain the set $\overline{\mathcal{E}}$ of events with $\overline{\mathcal{E}} \subseteq \mathcal{E}$, which represents a subset of events having the correspondence with the events in $\mathcal{E}_R$. Note that not all events in $\mathcal{E}$ has a binding in $\mathcal{E}_R$. 
In the end, the obtained score is in the range $[0,1]$, where the score of $0$ (resp. $1$) indicates that an EBS system is never (resp. always) timely in the detection of the events in $\mathcal{E}_R$.

For space matters, we explain in the Appendix how we calculate the timeliness score of an event database $\mathcal{E}$ with respect to a gold standard database $\mathcal{E^R}$ (Section~\ref{subsubsecapx:QuantitativeEvaluationForTimeliness}).

\subsubsection{Full and Partial Periodicity}
\label{subsubsec:TemporalDimension_FullPartialPeriodicity}
Another interesting temporal dimension analysis is to check if there are any periodically occurred events (e.g., at least once every $n$ weeks), which are geographically close to each other. For instance, we know from the literature that some Avian Influenza events can occur seasonally due to migratory birds, or it can become endemic due to its persistence in some regions. Therefore, it can be useful to characterize the cyclic behavior of the epidemiological events by taking into account the spatial information. We perform this task by identifying periodic-frequent spatial patterns from the field of spatio-temporal frequent pattern mining~\cite{Plantevit2010, Kiran2019}. Next, we first introduce the necessary definitions and concepts.

As in Section~\ref{subsec:Method_SpatialDimension}, in the following, we also fix the spatial and temporal scales of the events in $\mathcal{E}$ (resp. $\mathcal{E_R}$) with $l_Z$ and $l_T$. Therefore, we investigate on the temporal aspects discussed above through $\mathcal{E}^{l_Z}_{l_T}$ (resp. $\mathcal{E_R}^{l_Z}_{l_T}$). 
Let $Z$ (resp. $T$) represent all spatial (resp. temporal) entities with respect to a spatial (resp. temporal) scale $l_Z$ (resp. $l_T$) in $\mathcal{E}^{l_Z}_{l_T}$, i.e. $Z = Dom_{\mathcal{E}} (D_Z, l_Z)$ (resp. $T = Dom_{\mathcal{E}} (D_T, l_T)$). Each element in $\mathcal{E}^{l_Z}_{l_T}$ is called transaction. Moreover, in each transaction, we call \textit{pattern} a set $X$ of spatial entities, with $X \subseteq Z$. If $X$ contains $k$ spatial entities, then it is called a $k$-pattern. A pattern $X$ is called \textit{spatial}, if the maximum distance between any two of its spatial entities is no more than the user-specified distance $\alpha$. That is, $X$ is a spatial pattern if $max(Dist(z_{p}, z_{q})| \forall z_{p}, z_{q} \in X) \leq \alpha$.

Furthermore, the number of transactions containing a spatial pattern $X$ in $\mathcal{E}^{l_Z}_{l_T}$ is called the support of $X$, and denoted as $sup(X)$. If this support is large, then one can naturally ask how recurrent $X$ is in $\mathcal{E}^{l_Z}_{l_T}$. Let $t_{i}^{X}$ and $t_{j}^{X}$ be two consecutive time intervals at which $X$ appears in $\mathcal{E}^{l_Z}_{l_T}$. The time difference between $t_{i}^{X}$ and $t_{j}^{X}$ is defined as an inter-arrival time of $X$, and defined as $t_{j}^{X} - t_{i}^{X}$. Let $T^{X}_{\iota}$ be the set of all inter-arrival times of $X$ in $\mathcal{E}^{l_Z}_{l_T}$. The recurrence of a spatial pattern $X$ is considered \textit{full periodic} (\textit{periodic} for short), if any value in the set $T^{X}_{\iota}$ is never no more than the user-specified maximum inter-arrival time $\iota$. The cardinality of $T^{X}_{\iota}$ in $\mathcal{E}^{l_Z}_{l_T}$ constitutes the period-support of $X$, denoted as $psup(X)$. In other words, $X$ periodically appears $psup(X)$ times within the data, and at least once every $\iota$ time intervals. Our aim in this section is to find all spatial patterns that periodically appear in $\mathcal{E}^{l_Z}_{l_T}$. We call them periodic spatial patterns.

\begin{table}
    \centering
    \scriptsize
    \begin{tabular}{|l|p{9cm}|}
        \hline
        \textbf{Country} & \textbf{Spatially Close Countries} \\\hline
        France & Italy, Spain, Switzerland, Germany, Belgium, Luxembourg \\\hline
        Italy & France, Switzerland, Germany, Slovenia, Austria, Liechtenstein \\\hline
        Spain & France, Portugal \\\hline
        Portugal & Spain \\\hline
        China & Mongolia, Russia, Italy, Afghanistan, Tajikistan, Kyrgyzstan, Nepal, India, North Korea \\\hline
        India & China, Pakistan, Nepal, Bhutan, Bangladesh \\\hline
        Nepal & China, India \\\hline
        Pakistan & Iran, Afghanistan, China, India \\\hline
    \end{tabular}
    \captionsetup{width=.9\linewidth}
     \caption{Spatial closeness between the considered countries used in the example illustrated in Table~\ref{tab:STpatterns_example}. For the sake of clarity, we consider only neighbor countries sharing a border as spatially close.}
     \label{tab:STpatterns_country_neighbors}
\end{table}

We illustrate all these concepts with the same example illustrated in Table~\ref{tab:STpatterns_example}. On top of that, since we are interested in spatial patterns, Table~\ref{tab:STpatterns_country_neighbors} depicts the spatial neighborhood of the countries with respect to the parameter $\alpha$. Overall, some countries (e.g. India) face against a long infection period, whereas the others (e.g. France) succeed in stopping quickly the propagation of the viruses. Regarding the frequency of the spatial patterns from Table~\ref{tab:STpatterns_example}, we have $sup(India)=8$ (the most frequent), whereas we have $sup(Nepal)=2$ and $sup(China)=2$ (the least frequent). Moreover, \textit{India} is the only periodic spatial pattern (with $psup(India)=7$) when $\iota=2$. Nevertheless, when we set $\iota=4$, in this case, the periodic spatial patterns are \textit{India} (with $psup(India)=7$), \textit{Portugal} (with $psup(Portugal)=3$), \textit{Spain} (with $psup(Spain)=3$) and \textit{Portugal-Spain} (with $psup({Portugal{-}Spain})=3$). Note that the patterns \textit{India-Portugal} and \textit{India-Spain} are also periodic (with $psup(India{-}Portugal)=3$ and $psup(India{-}Spain)=3$), but they do not fulfill the requirement of spatial closeness (see Table~\ref{tab:STpatterns_country_neighbors}).

For some cases, the periodicity condition can be too strict. For instance, in Table~\ref{tab:STpatterns_example} France (resp. Pakistan) appears in the first (resp. last) 3 transactions, which is also valuable information. To weaken this strict definition, we also consider the partial periodicity condition. In this weaker condition, it is sufficient for a spatial pattern to periodically appear only in \textit{some} transactions of $\mathcal{E}^{l_Z}_{l_T}$. Concretely, a spatial pattern $X$ is said to be a \textit{partial} periodic spatial pattern if its period-support $psup(X)$ is no less than the user-specified minimum period-support $\varrho$. For instance, if we take the same example illustrated in Table~\ref{tab:STpatterns_example}, all partial periodic spatial patterns for $\iota=2$ and $\varrho=2$ are illustrated in Table~\ref{tab:STpatterns_partial_periodic_results}. It is worth noticing that the parameter $\iota$ has a positive effect on the generation of partial periodic spatial patterns, while $\varrho$ has a negative effect on the number of patterns being generated from the database. Moreover, note that the input parameters $\iota$ and $\varrho$ can be both expressed in percentage or in count, respectively. For instance, when we set $\varrho = 1.0$ (resp. $\varrho < 1.0$), this amounts to generate full (resp. partial) periodic spatial patterns.

\begin{table}
\scriptsize
    \centering
    \begin{tabular}{|l|l|}
        \hline
        \textbf{Pattern} & \textbf{Periodic Support} \\\hline
        India & 7\\\hline
        Portugal & 3\\\hline
        Spain & 3\\\hline
        Portugal-Spain & 2 \\\hline
        Pakistan & 2 \\\hline
        India-Pakistan & 2\\\hline
        France & 2\\\hline
    \end{tabular}
    \captionsetup{width=.9\linewidth}
     \caption{All partial periodic spatial patterns obtained with $\iota=2$ and $\varrho=2$ based on the example illustrated in Table~\ref{tab:STpatterns_example}. The spatial closeness between the considered countries is defined in Table~\ref{tab:STpatterns_country_neighbors}.}
     \label{tab:STpatterns_partial_periodic_results}
\end{table}

In this work, given spatial and temporal scales $l_Z$ and $l_T$, we discover all full and partial periodic spatial patterns in  $\mathcal{E}^{l_Z}_{l_T}$ with respect to the input parameters $\iota$, $\varrho$ and $\alpha$ through the method \textit{ST-ECLAT} (ST for short), 
proposed in \cite{Kiran2019}. Particularly, we are interested in two use cases for obtaining these patterns. First, we want to know what spatial entities (e.g. countries) have consistently epidemiological events throughout the year. We call the obtained results \textit{continuous} periodic patterns. Second, we want to know what spatial entities (e.g. countries) have a seasonal effect and are exposed to disease events only for some period of time every year. We call the obtained results \textit{seasonal} (or \textit{yearly}) periodic patterns.

For space matters, we explain in the Appendix how we quantitatively evaluate the performance of $\mathcal{E}$ in terms of its ability to detect these continuous and seasonal periodic patterns with respect to a gold standard database $\mathcal{E^R}$ (Section~\ref{subsubsecapx:QuantitativeEvaluationForPeriodicity}).

\subsection{Thematic Dimension}
\label{subsec:Method_ThematicDimension}
In this section, we aim to evaluate EBS systems in terms of thematic entities they extract from the events. In other words, we want to know whether the dimensions $D_D$ and $D_H$ in an event database $\mathcal{E}$ provides fine- or coarse-grained information. Ideally, we expect $\mathcal{E}$ to provide very detailed information, as in a gold standard database $\mathcal{E_R}$. Note that this aspect is related to one of the relevant characteristics of an EBS system in the CDC's guideline, so-called \textit{completeness}~\cite{Barboza2013}.

In our evaluation, we want to discover the rich data relations between spatial, temporal and thematic entities with two use cases. In the first use case, we totally omit the temporal aspect, and we propose to find out what thematic entities characterize most a spatial entity. For instance, if we take the same example illustrated in Table~\ref{tab:event-example}, we might want to know where the specific Avian Influenza serotype H7N9 is more prevalent. Our second use case is the temporal version of the first one~\cite{Li2018}, in which we are interested in the periodic aspects, as in Section~\ref{subsubsec:TemporalDimension_FullPartialPeriodicity}. For instance, when several Avian Influenza events with particular spatial and thematic characteristics repeat themselves at regular intervals in the data, this would indicate an ongoing spreading pattern with specific characteristics. In this work, we propose to perform these two use cases within a single evaluation scheme through the identification of frequent patterns, as in Section~\ref{subsubsec:TemporalDimension_FullPartialPeriodicity}.

As opposed to Section~\ref{subsubsec:TemporalDimension_FullPartialPeriodicity}, there are several key differences in this section, because we consider the fact that each event in $\mathcal{E}$ can be described with different hierarchical event attributes. First, we do not fix any temporal, spatial or thematic scale on $\mathcal{E}$, and we ensure that each transaction in $\mathcal{E}$ corresponds to a single event. Second, a transaction in $\mathcal{E}$ does not simply consist of atomic spatial entities, it is rather represented by a tuple $Y = (d_Z, d_D, d_H)$, with $d_Z \in Dom_Z(\mathcal{E})$, $d_T \in Dom_T(\mathcal{E})$ and $d_H \in Dom_H(\mathcal{E})$. In this context, we call this tuple $Y$ \textit{multidimensional pattern}~\cite{Pinto2001}. Therefore, with the multidimensionality of the patterns, $\mathcal{E}$ is defined as $\mathcal{E} = \{(t_{1}, Y_1), (t_{2}, Y_2), \dots , (t_{|N_\mathcal{E}|}, Y_{|N_\mathcal{E}|})\}$, where $|N_\mathcal{E}|$ represents the size of $\mathcal{E}$, $Y_j$ is a multidimensional pattern and $t_{j}$ represents the timestamp of $Y_j$. Third, we adapt $\mathcal{E}$ to include various hierarchical information of the event attributes. To do so, we modify $\mathcal{E}$ by adding all ancestors in the associated hierarchy of every multidimensional pattern. In the end, each transaction consists of the original multidimensional pattern and its variants with all ancestors in the associated hierarchy. We denote this modified event database as $\mathcal{E}^+$. We illustrate how we obtain $\mathcal{E}^+$ from $\mathcal{E}$ with an example in Table~\ref{tab:Modified_event_database_example_for_thematic_STpatterns}.

\begin{table*}[ht] 
    \scriptsize
    \centering
    \begin{tabular}{|l|l|}
        \hline
        \textbf{Time} & \textbf{Multidimensional pattern} \\\hline
        $t_1$ & \textbf{(Paris, AI, bird)}, (Île de France, AI, bird), (France, AI, bird) \\\hline
        $t_2$ & \textbf{(Italy, AI, wild bird)}, (Italy, AI, bird) \\\hline
        $t_3$ & \textbf{(Spain, H5N1, wild bird)}, (Spain, HPAI, wild bird), (Spain, AI, wild bird), (Spain, H5N1, bird), (Spain, HPAI, bird), (Spain, AI, bird) \\\hline
    \end{tabular}
     \caption{Illustrative example of the modified version $\mathcal{E}^+$ of a subset of an event database $\mathcal{E}$ with three events. The first column indicates the timestamp of the events, and the second column describes the multidimensional patterns. Note that the second column contains all ancestors in the associated hierarchy of every multidimensional pattern. For the sake of clarity, we show in bold the elements in $\mathcal{E}$, before obtaining its modified version $\mathcal{E}^+$.} \label{tab:Modified_event_database_example_for_thematic_STpatterns}
\end{table*}

In this work, we perform the two use cases discussed above by discovering frequent multidimensional patterns, accompanied by the partial periodicity condition. When the temporal aspect is omitted, we simply calculate the support $sup(Y)$ of each multidimensional pattern $Y$ in $\mathcal{E}^+$, as such pattern always corresponds to a single tuple $(d_Z, d_D, d_H)$ in this context. We call them \textit{static} multidimensional patterns. When we take the partial periodicity condition into account, this amounts to find partial periodic multidimensional patterns, as it ensures that two multidimensional patterns appear in the same time interval in $\mathcal{E}^+$. We call them \textit{temporal} multidimensional patterns. In practice, we use the ST algorithm described in Section~\ref{subsubsec:TemporalDimension_FullPartialPeriodicity} to generate these \textit{static} and \textit{temporal} multidimensional patterns. The flexibility of ST is that when we set a very large inter-arrival time value to $\iota$, this allows us to omit the partial periodicity condition. 

For space matters, we explain in the Appendix how we quantitatively evaluate the performance of an event database $\mathcal{E}$ of an EBS system in terms of its ability to detect these static and temporal multidimensional patterns with respect to a gold standard database $\mathcal{E^R}$ (Section~\ref{subsecapx:QuantitativeEvaluationThematicDimension}).

\subsection{Source Dimension}
\label{subsec:Method_SourceDimension}
Finally, the last part of our evaluation framework is regarding online news and press agencies, that we call short news outlets or news sources, involved in the propagation of epidemiological information on the web.

All EBS systems rely partially or completely on various online news outlets for ensuring their monitoring of emerging infectious diseases across the world. Nevertheless, there are not enough studies that characterize and assess the news sources involved in EBS systems. 
For instance, Schwind \textit{et al.}~\cite{Schwind2014} point out that local news outlets are more likely to report ongoing epidemiological events than international media sources do. For this reason, we aim to identify and characterize important news outlets, and we propose in this section an evaluation scheme for the news outlets involved in the propagation of epidemiological information on the web. Our evaluation scheme consists of two different objectives. Our first objective is that we want to identify important news outlets for information dissemination. In our second objective, we are interested in the ability of news reporting in timely manner. In other words, we want to rank news outlets, which publish epidemiological events as fast as possible. In the following, we propose to perform these tasks through network analysis. Note that some EBS systems are designed to collect epidemiological data from both official and unofficial data sources. In order to a have fairer evaluation across EBS systems, we do a preprocessing step by eliminating the official data sources and keeping only unofficial ones within the data.

\subsubsection{Identification of Important News Outlets}
\label{subsubsec:IdentificationImportantNewsOutlets}
We say that a news outlet needs to fulfill two conditions in order to be considered as important. First, it reports epidemiological information that are reported by both local and international news media. In other words, if someone follows the news reported by an important news outlet, it means she receives sufficiently necessary epidemiological information for her country and nearby. Second, it also reports the events that are reported by important news outlets. We perform the task of identification of important news outlets through network analysis. We design our approach in two steps. First, we extract the news outlet network $G_\mathcal{E}$ from an event database $\mathcal{E}$, where nodes represent news outlets and edges describe the relations for node pairs. We do this process on the whole or a subset of data for each considered EBS platform. Then, we apply a well-suited centrality measure to rank the news outlets by their importance score. Next, we describe how we process these steps.

\begin{figure}[ht!]
    \centering
    \includegraphics[width=0.8\textwidth]{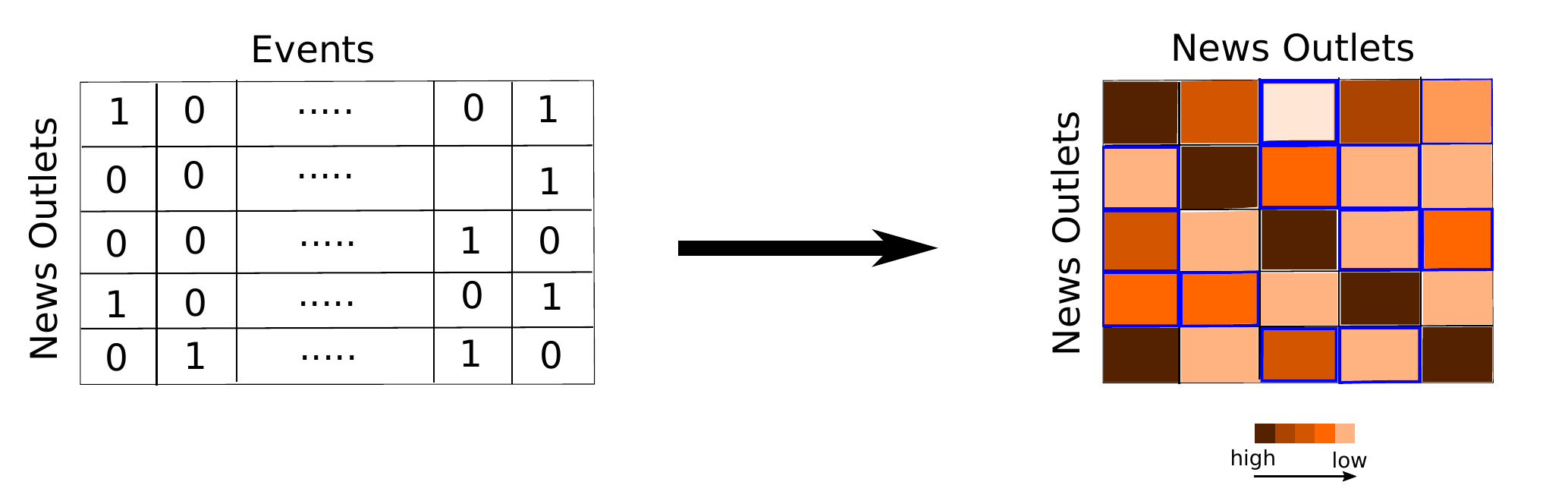}
    \captionsetup{width=.9\linewidth}
    \caption{Illustration of the network construction steps. In the first step, we build an initial matrix, in which we indicate for each event if the news outlets report it or not. In the second step, we convert the first matrix into a second one to encode to what degree the news outlets report the same events.}
    \label{fig:newslet-network-construction}
\end{figure}

Traditionally, most of the existing works in the literature extract a news outlet network, when citation information (i.e. what sources cite what other sources) between news outlets is available. 
Nevertheless, this information is hardly available in the data collected by EBS systems. For this reason, we propose to use a co-occurrence fraction counting method known from the field of scientometrics~\cite{Leydesdorff2017}, as also used in \cite{Choloniewski2019}. The construction of $G_\mathcal{E}$ is as follows and illustrated in Figure~\ref{fig:newslet-network-construction}. Let us say an EBS system monitor $|P|$ news outlets for $|E|$ distinct events. First, we construct a $|P| \times |E|$ matrix $A$, where the rows represent news outlets and the columns represents the distinct events detected by an EBS system. Each element of matrix A is defined as in Equation~\ref{eq:EventReporting_binary}.

\begin{equation}
    a_{ie} = 
    \begin{cases} 
    1, \mbox{if news outlet $i$ reports event $e$,} \\
    0, \mbox{otherwise}
    \end{cases}
    \label{eq:EventReporting_binary}
\end{equation}

Next, we transform the matrix $A$ into another $|P| \times |P|$ symmetric matrix $B$ to measure how frequent two news outlets report the same events. Each element $b_{ij}$ of matrix B is defined as in Equation~\ref{eq:newslet_network_edge_weights}.

\begin{equation}
b_{ij} = \sum\limits_{e=1}^{E} \frac{a_{ie} a_{je}}{a_{ie}^2}
\label{eq:newslet_network_edge_weights}
\end{equation}

In the end, we obtain a score of $1$ (resp. $0$), when two news outlets always (resp. never) report the same events, or a score in  $[0,1]$ otherwise. 

Finally, in the second step of our approach, we apply a centrality measure over $G_\mathcal{E}$ to identify important news outlets. A centrality measure aims to rank the vertices of a network by assigning them a score. The more central a vertex is, the larger score it has. In the literature, there is a large number of centrality measures, each having a particular objective. In this work, we propose to use the PageRank centrality for $G_\mathcal{E}$, as it is more suitable to our definition of important news outlets. In the end, we want to see how similar important news outlets are among multiple EBS systems. If it is very similar, this would indicate that they rely mostly on the same news sources. In the rest of the work, we denote the first $k$ most important news outlets from the PageRank centrality result by \text{PageRank}($G_\mathcal{E}$, k).

\subsubsection{Timely Detection}
\label{subsubsec:TimelyDetection}
One specific criteria that one may want to optimize in event detection is to minimize detection time (i.e. capturing an epidemiological event as soon as possible). Our second objective is related to this timeliness capability of the news outlets. We want to identify the news outlets, which are timely in event detection, and not those detecting as many events. To do so, we follow the work of Leskovec \textit{et al.}~\cite{Leskovec2007}. Their method first creates the news outlet network $G_\mathcal{E}$ from $\mathcal{E}$, then finds through their method CELF a set $A$ of news outlets, which minimizes detection time, while covering all the event set $E$. 

\begin{figure}[ht!]
    \centering  
    \includegraphics[width=0.9\textwidth]{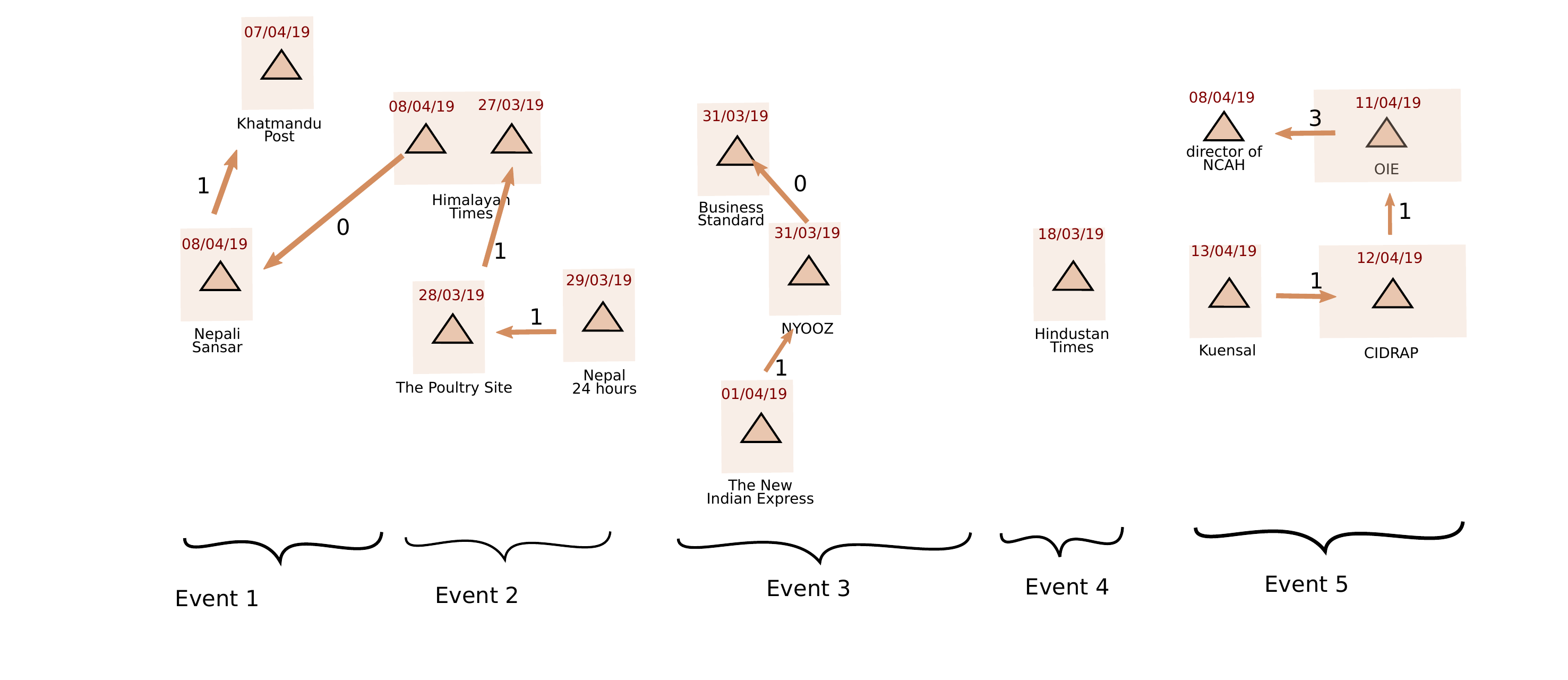}
    \captionsetup{width=.9\linewidth}
    \caption{Illustration of the event cascades for five events. The nodes represent news outlets publishing news documents, and the edges between them obey time order and the edge weights represent their time difference. For instance, the cascade for the first event starts at the news document published by \textit{Khatmandu Post}, and then the edges are sequentially created by adding other news documents in $S_e$ linking to it.}
    \label{fig:leskovec-network-construction}
\end{figure}

First, we extract our news outlet network $G_\mathcal{E}$ as follows. Let us suppose that the set $S_e$ of news outlets reports through their news documents the same event $e$ in an event database $\mathcal{E}$. For every event $e \in \mathcal{E}$, we create a path structure $P_e$, that we call \textit{cascade}, such that a news outlet in $S_e$ sequentially join the cascade $P_e$ by linking to other news outlets in $S_e$, whereby the edges obey time order and the weights of directed edges represent the time difference between two news documents. When we repeat this process for each event $e \in \mathcal{E}$, this gives us a network in the end. We illustrate this network creation for several events with an example in Figure~\ref{fig:leskovec-network-construction}.  

Then, we apply the CELF algorithm to $G_\mathcal{E}$ in order to identify a set $A$ of timely news outlets. This algorithm starts with the empty set $A_0 = \emptyset$ and iteratively adds in step $k$ the news outlet $s_k$ maximizing the marginal gain as in Equation~\ref{eq:marginal_gain_max_CELF}.

\begin{equation}
    s_k = \argmax_{s \in P \setminus A_{k-1}} R(A_{k-1} \cup \{s\}) - R(A_{k-1}).
    \label{eq:marginal_gain_max_CELF}
\end{equation}

The algorithm stops, once it has selected $k = |A|$ elements. The marginal gain is expressed for a subset $A$ of news outlets in terms of the function $R(A)$, which is used as a penalty reduction function. It is defined as in Equation~\ref{eq:penalty_reduction_function_CELF}.

\begin{equation}
    R(A) = \sum_{e \in E} P(e) \big( T_{max} - \min\limits_{s \in A} T(e,s) \big),
    \label{eq:penalty_reduction_function_CELF}
\end{equation}

where $T_{max}$ is time horizon, $P$ is a (given) probability distribution over the events and $T(e,s)$ represents the time delay in days, until news outlet $s$ participates in the event path $P_e$. Note that $T(e,s)$ equals $T_{max}$, if $s$ does not report event $e$. Moreover, in our context each event has uniform probability, therefore we omit $P(e)$ from the definition of $R(A)$. In the rest of the work, we denote the first $k$ most timely news outlets obtained from the CELF method for $G_\mathcal{E}$ by \text{CELF}($G_\mathcal{E}, k$).

For space matters, we explain in the Appendix how we quantitatively evaluate the performance of an event database $\mathcal{E}$ in terms of its ability to detect the important and timely news outlets (Section~\ref{subsecapx:QuantitativeEvaluationSourceDimension}).

\section{Experimental Setup}
\label{sec:ExperimentalSetup}
In this section, we define our experimental setup in order to illustrate how to use our framework and interpret
its results. We first present the selected EBS/IBS systems, to which we apply our framework (Section~\ref{subsec:Experimens_SelectedEBSSystems}). Then, we describe the input event data, as well as its processing (Section~\ref{subsec:Experimens_Data_EventProcessing}). 
The results are presented afterwards, in Section~\ref{sec:Results}.

\subsection{Selected EBS systems}
\label{subsec:Experimens_SelectedEBSSystems}
We show the relevancy of our framework on two well-known EBS systems PADI-web~\cite{Valentin2021a} and ProMED~\cite{Carrion2017}. Moreover, we use the reference gold standard database {Empres-i} from the World Organisation for Animal Health (WOAH) and the Food and Agriculture Organization of the United Nations (FAO)~\cite{empresi} to evaluate the performances of PADI-web and ProMED. Our choice of the {Empres-i} database is based on the fact that it is a well-populated official database for the main animal diseases, such as Avian Influenza and African Swine Fever~\cite{Farnsworth2010, Arsevska2018}. ProMED collects and organizes its disease events through 50 subject matter expert moderators from 34 countries~\footnote{\url{https://promedmail.org/team}}, who provide written commentary, giving the reader the necessary historical context and/or clinical background to understand the importance of the information being reported. ProMED also supply references to previous reported events and to the scientific literature for the sake of completeness. In principle, ProMED first searches for an official source (e.g. WOAH report) if it is available at the same time that an online news document is available. For this reason, ProMED relies on both official and unofficial sources for event detection. On the contrary, PADI-web is an automated surveillance system, which automatically collects online news documents with customized queries using the Google News aggregator, translates all non-English documents into English, classifies the documents, and extracts epidemiological information (diseases, dates, symptoms, hosts and locations) from the relevant news documents. PADI-web is currently integrated in the French Platform for Animal Health Surveillance (ESA Platform)~\cite{Arsevska2018}. 

We choose ProMED and PADI-web in our experiments for several reasons. First, each one belongs to a different EBS category: moderated vs. automated. Second, both EBS systems have a well-established surveillance system, since they are operational for a long time. They are currently collaborating with and used by national and supranational health authorities. Third, they are open-access tools. Finally, several works have assessed PADI-web and ProMED, separately~\cite{Schwind2014, Tsai2013} or together~\cite{Arsevska2016, Arsevska2018}. We base our discussion in Section~\ref{sec:Results} on these previous results, when possible.

\subsection{Event Data and Processing}
\label{subsec:Experimens_Data_EventProcessing}
The event datasets we use concern the Avian Influenza (AI) cases affecting bird species from 2019 to 2021. These AI cases can be high pathogenic AI (HPAI) or low pathogenic AI (LPAI). They are retrieved from PADI-web, ProMED and {Empres-i}, respectively. Regarding the PADI-web data, we rely only on those considered as relevant from PADI-web's automatic process. We chose a three-year study period (2019-2021) to sufficiently capture the space-time epidemiological characteristics of the AI events around the world. In order not to penalize ProMED we keep all its data provided by official and unofficial data sources, although PADI-web relies only on online news outlets. Nevertheless, for the sake of completeness, we compare both systems in Table~\ref{tab:results_spatiotemporal_representativeness_country_scale} and Figure~\ref{fig:radar_chart_all_quantitative_eval_results_promed_unofficial} by discarding the ProMED data provided by official data sources (i.e. WOAH reports). Furthermore, it is also worth noticing that automated EBS systems, such as PADI-web, might report false event information due to their automated location detection and extraction strategies. Evaluating the rate of reporting false events for such systems is not the scope of this work.

We process the collected raw event datasets by transforming them into normalized event databases, as explained in the Appendix (Section~\ref{secapx:CorpusEventExtraction}). During this processing, we deal with the different event definitions that PADI-web and ProMED have, which do not exactly match the one proposed in Section~\ref{subsec:EventRelatedDefinitionsNotations}, as the definition of an event can be different from one EBS system to another. Regarding ProMED, we only extract the information regarding news outlets from the raw news documents. Regarding PADI-web, the events are essentially disease-location pairs. Moreover, PADI-web extracts event-related thematic information for each collected news document, without relating them to any event. For this reason, we complete the minimally defined events with the extracted thematic entities, as detailed in Section~\ref{subsecapx:EventCompletion}. In the end, we obtain three normalized event databases for PADI-web, ProMED and Empres-i, denoted by $\mathcal{E}_{PW}$, $\mathcal{E}_{PM}$ and $\mathcal{E}_{EI}$, respectively. These normalized data are not publicly available due to third party restrictions, nevertheless, they are available on request~\footnote{\url{https://entrepot.recherche.data.gouv.fr/dataset.xhtml?persistentId=doi:10.57745/Y3XROX}}.

\section{Results}
\label{sec:Results}
We now assess, compare and discuss the performances of the considered EBS systems when applied to our framework on the Avian Influenza event databases $\mathcal{E}_{PW}$, $\mathcal{E}_{PM}$ and $\mathcal{E}_{EI}$.  In our experiments, the number of events by year for those event databases are shown in Table~\ref{tab:event-stats}. In total, there are $1515$, $786$ and $5229$ events for $\mathcal{E}_{PW}$, $\mathcal{E}_{PM}$ and $\mathcal{E}_{EI}$, respectively. We present the results in line with our four evaluation objectives: 1) spatial (Section~\ref{subsec:Results_SpatialDimension}), 2) temporal (Section~\ref{subsec:Results_TemporalDimension}), 3) thematic (Section~\ref{subsec:Results_ThematicDimension}) and 4) source (Section~\ref{subsec:Results_SourceDimension}) dimensions. Our source code is
publicly available\footnote{\url{github.com/arinik9/compebs}}.

\begin{table}
    \centering
    \scriptsize
    \begin{tabular}{|p{1.1cm}|p{1.5cm}|p{1.5cm}|p{1.5cm}|l|}
        \hline
        \textbf{EBS/IBS system} & \textbf{Number of events in 2019} & \textbf{Number of events in 2020} & \textbf{Number of events in 2021} & \textbf{Total} \\\hline
        PADI-web & 116 & 436 & 963 & 1515\\\hline
        ProMED & 28 & 245 & 513 & 786\\\hline
        {Empres-i} & 267 & 1539 & 3423 & 5229\\\hline
    \end{tabular}
    \captionsetup{width=.9\linewidth}
    \caption{Event statistics for PADI-web, ProMED and {Empres-i} in the Avian Influenza dataset.}
    \label{tab:event-stats}
\end{table}

\subsection{Spatial Dimension}
\label{subsec:Results_SpatialDimension}
\begin{figure*}
    \centering
    \begin{subfigure}[b]{0.78\textwidth}
    \centering
    \includegraphics[width=1.0\textwidth]{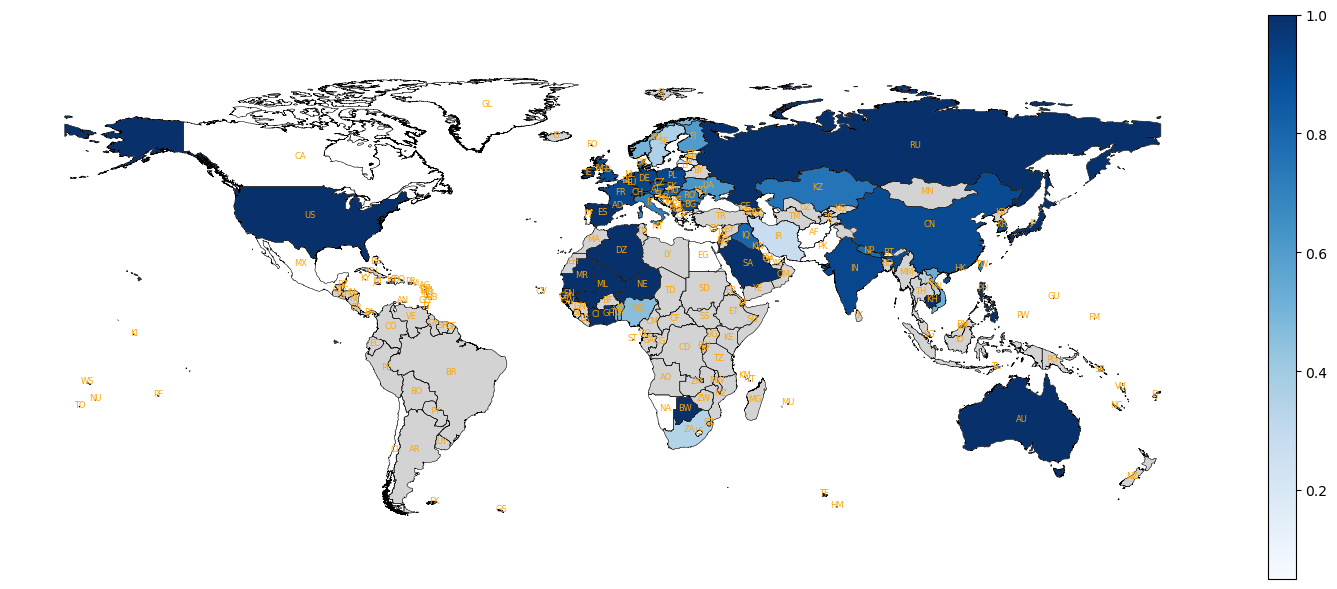}
    \caption{Spatio-temporal representativeness scores at country scale for PADI-web with respect to the results of Empres-i.}
    \label{subfig:spatio-temporal_representativeness_padiweb_country_scale}
    \end{subfigure}
    \begin{subfigure}[b]{0.78\textwidth}
    \centering
    \includegraphics[width=1.0\textwidth]{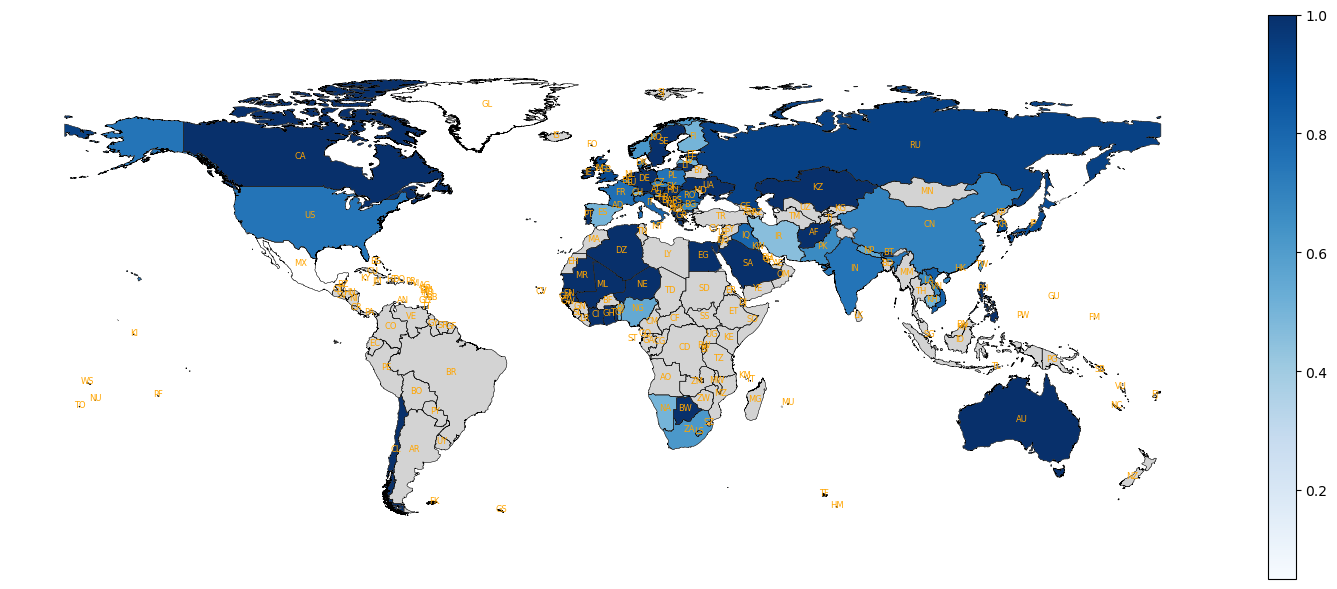}
    \caption{Spatio-temporal representativeness scores at country scale for ProMED with respect to the results of Empres-i.}
    \label{subfig:spatio-temporal_representativeness_promed_country_scale}
    \end{subfigure}
    \begin{subfigure}[b]{0.78\textwidth}
    \centering
    \includegraphics[width=1.0\textwidth]{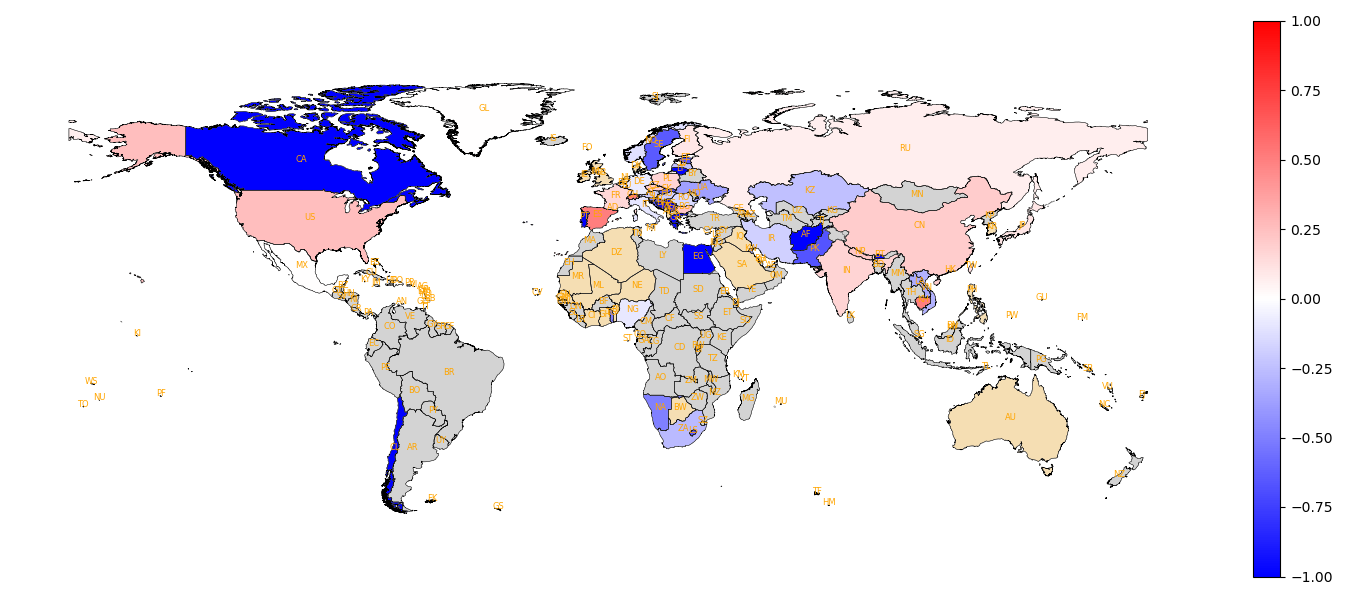}
    \caption{Spatio-temporal representativeness score differences between PADI-web (Figure~\ref{subfig:spatio-temporal_representativeness_padiweb_country_scale}) and ProMED (Figure~\ref{subfig:spatio-temporal_representativeness_promed_country_scale}) at country scale.}
    \label{subfig:spatio-temporal_representativeness_padiweb_vs_promed_country_scale}
    \end{subfigure}
    \captionsetup{width=.9\linewidth}
    \caption{Spatio-temporal Representativeness scores  at country scale for PADI-web and ProMED with respect to the results of Empres-i. In $(a)$ and $(b)$, the degree to which an EBS system covers the {Empres-i} events occurring in a country is shown with different blue scales, and it is shown in white when an EBS system never finds an event in $\mathcal{E}_{EI}$. In $(c)$, the score differences between $(a)$ and $(b)$ are shown. It is colored in blue (resp. red) when ProMED (resp. PADI-web) gives better spatio-temporal representativeness scores for a country and in yellow in case of non-zero equality. In all these plots, countries without an {Empres-i} event are indicated in gray.}
    \label{fig:spatio-temporal_representativeness_country_scale}
\end{figure*}

We evaluate through spatio-temporal representativeness how accurate EBS systems describe the distribution of events found in the gold standard {Empres-i} database in terms of place and time. We calculate the spatio-temporal representativeness scores for each region and country with monthly time intervals. In the following, we discuss only these scores at country scale and leave those at region scale in the Appendix, due to lack of space. We plot the scores at country scale in Subfigures~\ref{subfig:spatio-temporal_representativeness_padiweb_country_scale} and \ref{subfig:spatio-temporal_representativeness_promed_country_scale} (see Figures~\ref{figapx:spatio-temporal_representativeness_padiweb_region_scale} and \ref{figapx:spatio-temporal_representativeness_promed_region_scale} in the Appendix for region scale). In these plots, countries without an {Empres-i} event are indicated in gray, and the degree to which an EBS system covers the events occurring in a country is shown with different blue scales, where the large (resp. small) values are indicated in dark (resp. light) blue. When an EBS system never finds an event in $\mathcal{E}_{EI}$, it is shown in white.

We see from these plots that both PADI-web and ProMED report the events from a large number of countries, but they represent well only some of them. Moreover, although some countries (e.g. Ireland, Australia) are equally represented by PADI-web and ProMED, there are some discrepancies in the spatial focus of these EBS systems. For instance, PADI-web (resp. ProMED) better covers the USA, Spain, France, India and China (resp. South Africa, Vietnam, Kazakhstan, Ukraine and Sweden) than ProMED (resp. PADI-web). Furthermore, PADI-web never reports an event from Canada, Portugal and Afghanistan that ProMED covers well. All these similarities and differences at country scale are summarized in Table~\ref{tab:results_spatiotemporal_representativeness_country_scale}.

\begin{table}
    
    
        
    
    
    
    \centering
    \begin{tabular}{|p{1.5cm}|p{3.05cm}|p{3.05cm}|}
        \hline
         & \scriptsize \textbf{PADI-web covers} & \scriptsize \textbf{PADI-web never covers}  \\\hline
        \scriptsize \textbf{ProMED covers} & \scriptsize Ghana, Philippines, Saudi Arabia, Senegal, Mali, Hong Kong, Ivory Coast, Albania, Algeria, Netherlands, Niger, Mauritania, Botswana, Ireland, Australia (15 countries in total)$^\gamma$ & \scriptsize Chile, Bosnia and Herzegovina, Lithuania, Serbia, Bhutan, Lesotho, Namibia, Pakistan, Greece, Canada, Slovenia, Egypt, Afghanistan, Portugal (14 countries in total)\\\hline
        \scriptsize \textbf{ProMED never covers} & \scriptsize - (0 country in total) & \scriptsize Mexico, Greenland, Dominican Republic (3 countries in total)\\\hline
    \end{tabular}
    \captionsetup{width=.9\linewidth}
     \caption{Countries covered by PADI-web and ProMED according to the spatio-temporal representativeness scores.\\
     $\gamma$ Here, we focus only on the countries, where the spatio-temporal representativeness score is the maximum value of 1 for both PADI-web and ProMED. If we focus on the countries covered both PADI-web and ProMED instead, there are 58 countries in common.}     \label{tab:results_spatiotemporal_representativeness_country_scale}
\end{table}

Finally, we also plot the spatio-temporal representativeness score differences between PADI-web and ProMED in Subfigure~\ref{subfig:spatio-temporal_representativeness_padiweb_vs_promed_country_scale} (See Figure~\ref{figapx:spatio-temporal_representativeness_padiweb_vs_promed_region_scale} for region scale) to ease their comparison. In this figure, it is colored in blue (resp. red) when ProMED (resp. PADI-web) gives better spatio-temporal representativeness score for a country and in yellow in case of non-zero equality. We see from the figure that ProMED gives better scores than PADI-web does for the overwhelming majority of countries. The average spatio-temporal representativeness score over all these countries also confirms this superiority ($0.59$ vs. $0.80$ , see Section~\ref{subsecapx:QuantitativeEvaluationForSpatialDimension} in the Appendix for the calculation details). Note that when we consider the country and region scales together in the calculation of this score, 
it still confirms the superiority ($0.40$ vs. $0.55$), although the score is lower than that at country scale. This decrease is mostly because of the difficulty of geocoding task in event normalization, i.e. accurately assigning geographic coordinates to spatial entities, due to the ambiguity among place names~\cite{Ng2020}. 



In summary, both PADI-web and ProMED report the Avian Influenza events for a large number of countries. This is mostly due to the fact that Avian Influenza (with African Swine Fever) is one of the animal disease cases reported well by both systems~\cite{Arsevska2016}, and that the number of detected Avian Influenza events increases each year (see Table~\ref{tab:event-stats}). Nevertheless, there are some discrepancies in the spatial focus of these EBS systems. These discrepancies are also consistent with the previous works. For instance, in~\cite{Arsevska2016}, the official Avian Influenza events lie mostly in Central America, Africa (mostly Egypt, Nigeria and South Africa), Middle East and Asia. Although both PADI-web and ProMED cover these areas, the degree to which they report the events are different in~\cite{Arsevska2016}. Indeed, PADI-web covers more countries in Asia (particularly in China and India) than ProMED does, a point also highlighted in a ProMED's publication~\cite{Carrion2017}. Similarly, ProMED better covers Africa, Eastern Europe and Middle East than PADI-web does. 

\begin{table}
    
    
        
    
    
    \centering
    \begin{tabular}{|p{1.5cm}|p{3.05cm}|p{3.05cm}|}
        \hline
         & \scriptsize \textbf{PADI-web covers} & \scriptsize \textbf{PADI-web never covers}  \\\hline
        \scriptsize \textbf{ProMED covers} & \scriptsize Ghana, Senegal, Mali, Ivory Coast, Albania, Algeria, Mauritania, Botswana, Australia (9 countries in total)$^\gamma$ & \scriptsize Chile, Lithuania, Serbia, Namibia, Greece, Canada, Egypt, Afghanistan, Portugal (9 countries in total)\\\hline
       \scriptsize  \textbf{ProMED never covers} & \scriptsize Iran, Croatia, Switzerland, Laos, Philippines, Saudi Arabia, Hong Kong, Austria, Italy, Luxembourg, Spain (11 countries in total) & \scriptsize Mexico, Greenland, Bosnia and Herzegovina, Bhutan, Dominican Republic, Lesotho, Pakistan, Slovenia (8 countries in total)\\\hline
    \end{tabular}
    \captionsetup{width=.9\linewidth}
     \caption{Countries covered by PADI-web and ProMED according to the spatio-temporal representativeness scores, when we discard the ProMED data provided by official data sources (i.e. WOAH reports).\\
     $\gamma$ Here, we focus only on the countries, where the spatio-temporal representativeness score is the maximum value of 1 for both PADI-web and ProMED.}     \label{tab:results_spatiotemporal_representativeness_country_scale_promed_unofficial}
\end{table}

The most important factor that determines the events the EBS systems find is inevitably related to the online news outlets~\cite{Lyon2011}. We also investigate on this aspect in Section~\ref{subsec:Results_SourceDimension}. PADI-web relies only on the news aggregator Google News, whereas ProMED cooperates with 50 human moderators and curators from all around the world. Although these moderators rely on both the WOAH reports and online news outlets, the latter plays a substantial role for ProMED (338 out of 786 Avian Influenza events in our experiments). For the sake of completeness, we also compare in Table~\ref{tab:results_spatiotemporal_representativeness_country_scale_promed_unofficial} these systems in terms of spatio-temporal representativeness scores by discarding the ProMED data provided by official data sources (i.e. WOAH reports).  


\subsection{Temporal Dimension}
\label{subsec:Results_TemporalDimension}

We present the results of the temporal dimension in two parts: Timeliness (Section~\ref{subsubsec:Results_TimelinessAnalysis}) and periodicity (Section~\ref{subsubsec:Results_PeriodicityAnalysis}) analyses.

\subsubsection{Timeliness analysis}
\label{subsubsec:Results_TimelinessAnalysis}
We study how timely the EBS systems PADI-web and ProMED are compared to the {Empres-i} events, as well as the assessment of timeliness between them. Note that we perform this assessment based on the identification of the putatively associated events between a pair of event databases, as explained in Section~\ref{subsec:EventMatching}. The obtained statistics regarding these putatively associated events are shown in Table~\ref{tab:event_matching_statistics}.

\begin{table}
    \centering
    \scriptsize
    \begin{tabular}{|l|l|l|l|}
        \hline
        & \textbf{Number of putatively associated events} \\\hline
        PADI-web and {Empres-i} & 422 \\\hline
        ProMED and {Empres-i} & 469 \\\hline
        PADI-web and ProMED & 450 \\\hline
    \end{tabular}
    \caption{Event matching statistics in the Avian Influenza dataset.}
    \label{tab:event_matching_statistics}
\end{table}

\begin{figure*}
    \centering
    \begin{subfigure}[b]{0.31\textwidth}
    \centering
    \includegraphics[width=1.0\textwidth]{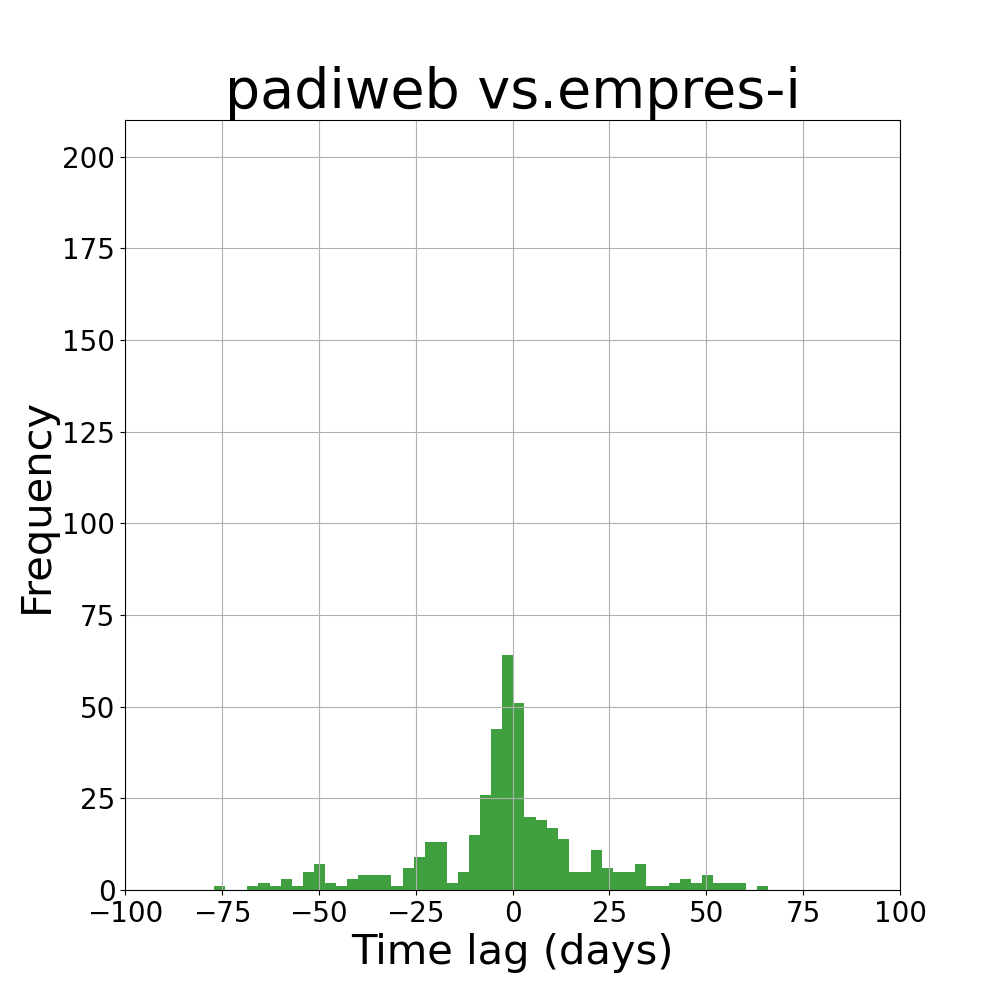}
    \caption{Time lags between PADI-web and Empres-i.}
    \end{subfigure}
    \hfill
    \begin{subfigure}[b]{0.31\textwidth}
    \centering
    \includegraphics[width=1.0\textwidth]{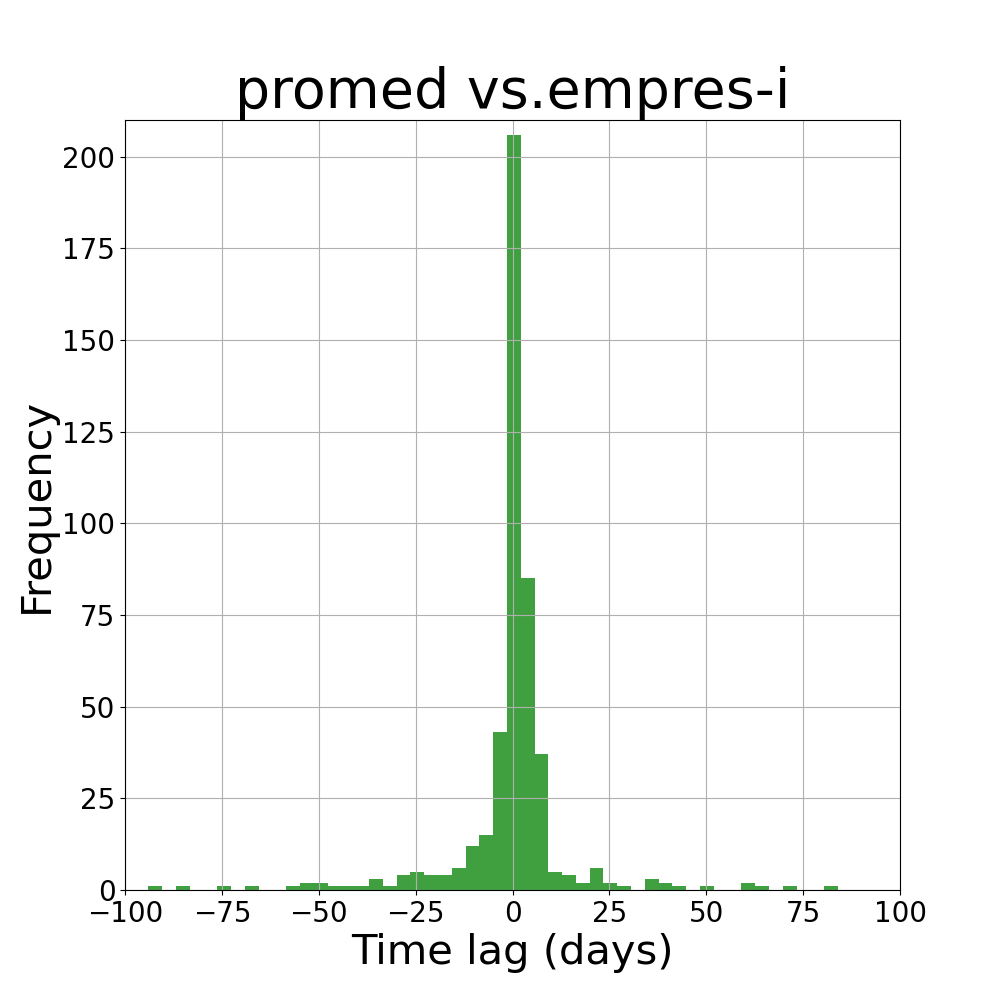}
    \caption{Time lags between ProMED and Empres-i.}
    \end{subfigure}
    \begin{subfigure}[b]{0.31\textwidth}
    \centering
    \includegraphics[width=1.0\textwidth]{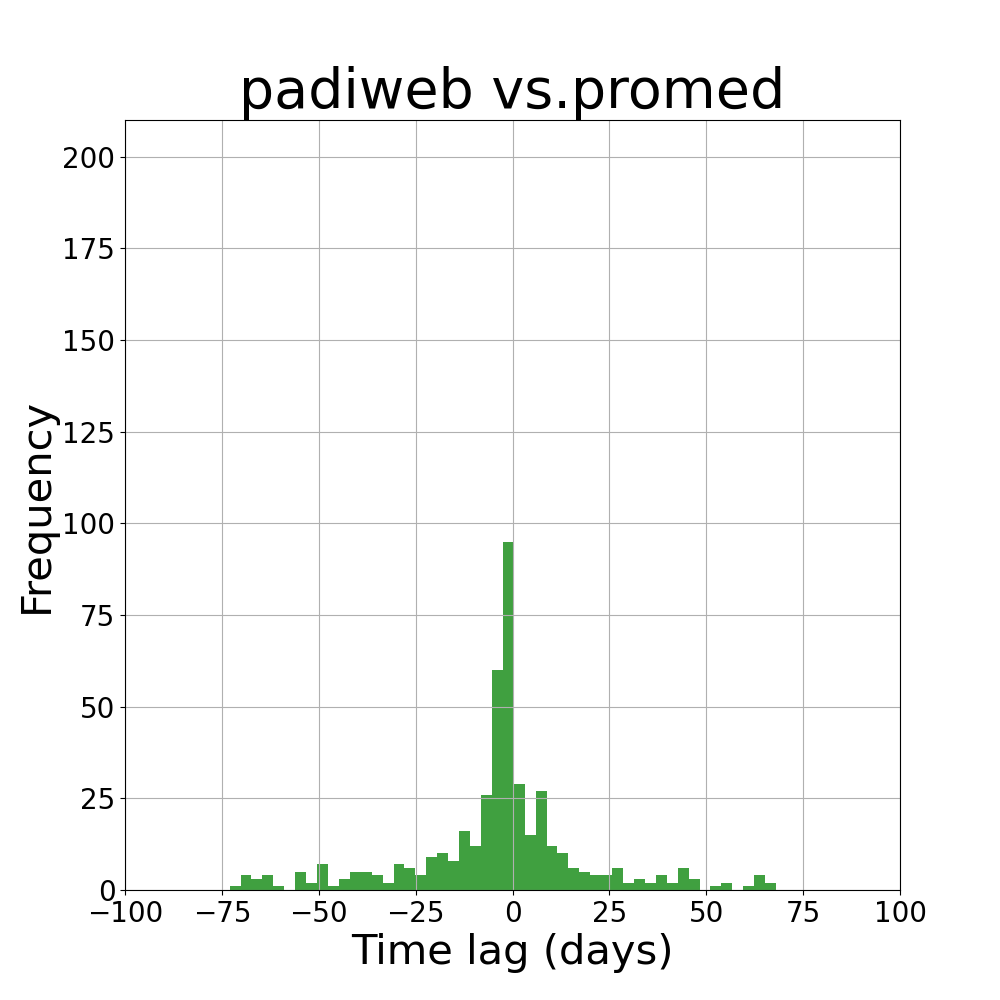}
    \caption{Time lags between PADI-web and ProMED.}
    \end{subfigure}
    \caption{Time lags of the putatively associated events between PADI-web and {Empres-i} (a), between ProMED and Empres-i, between PADI-web and ProMED. In these plots, when the first (resp. second) EBS system reports an event earlier than the second (resp. first) one, then this results in a negative (resp. positive) value.}
    \label{fig:TimeLags_AllPlatforms}
\end{figure*}

In Figure~\ref{fig:TimeLags_AllPlatforms}, we plot the time lag values for each pair of systems. In these plots, when the first (resp. second) system reports an event earlier than the second (resp. first) one, then this results in a negative (resp. positive) value. Moreover, we summarize the statistics in terms of timeliness in Table~\ref{tab:results_timeliness}. We can see from Figure~\ref{fig:TimeLags_AllPlatforms} and Table~\ref{tab:results_timeliness} that although both PADI-web and ProMED can be timely depending on events, PADI-web is more timely with respect to the {Empres-i} events compared to ProMED (49\% vs. 28\%). Indeed, the timeliness scores also confirm this superiority ($0.18$ vs. $0.12$, see Section~\ref{subsubsecapx:QuantitativeEvaluationForTimeliness} in the Appendix for the calculation details). Furthermore, when we look at the same events detected by PADI-web and ProMED (see the third row in Table~\ref{tab:results_timeliness}), PADI-web is also more timely (60\% vs. 33\%). The delay for ProMED is mostly related to its events collected from the official data sources (i.e. the WOAH reports). For instance, when we discard the ProMED data provided by official data sources (i.e. WOAH reports), the timely detection rate of ProMED increases from 28\% to 44\%. Nevertheless, when we focus only on the events, where PADI-web and ProMED are late, i.e. those reported after the {Empres-i} events, the average delay in days for ProMED is better compared to PADI-web (6.00 vs 14.8, here the less is better). This fact is also due to the official WOAH reports, as the publication dates of the WOAH-based ProMED events and those of the associated {Empres-i} events are very close, which reduces the average value.

\begin{table*}
    \centering
    \tiny
    \begin{tabular}{|p{2.5cm}|p{1.4cm}|p{1.4cm}|p{1.4cm}|p{1.4cm}||p{1.4cm}|p{1.4cm}|p{1.4cm}|p{1.4cm}|}
        \hline
         & \textbf{Number of events if the first system is in advance} & \textbf{Number of events if the first system is better and 30 days in advance} & \textbf{Number of events if the second system is in advance} & \textbf{Number of events if the second system is better and 30 days in advance} & \textbf{Average delay in days for the first system} & \textbf{Average delay in days for the second system} & \textbf{Timeliness score for the first system} & \textbf{Timeliness score for the second system}\\\hline
        \textbf{PADI-web vs. EMPRES-i} (422 events) & \cellcolor{gray!20} 209 (49\%) & \cellcolor{gray!20} 39 & 185 & 29 & \cellcolor{gray!20} 14.8 & 16.37 & 0.18 & \cellcolor{gray!20} 0.21\\\hline
        \textbf{ProMED vs. EMPRES-i} (469 events) & 134 (28\%) & \cellcolor{gray!20} 16 & \cellcolor{gray!20} 293 & 12 & \cellcolor{gray!20} 6.00 & 12.76 & 0.12 & \cellcolor{gray!20} 0.09\\\hline
        \textbf{PADI-web vs. ProMED} (450 events) & \cellcolor{gray!20} 273 (60\%) & \cellcolor{gray!20} 47 & 150 (33\%) & 30  & \cellcolor{gray!20} 17.08 & 15.04 & 0.15 & \cellcolor{gray!20} 0.23\\\hline
    \end{tabular}
    \caption{Timeliness summary statistics for the comparisons between PADI-web and Empres-i, between ProMED and Empres-i, between PADI-web and ProMED.}
     \label{tab:results_timeliness}
\end{table*}

Overall, PADI-web detects the 209 putatively associated events (49\%) before their publication in {Empres-i}, and those 39 of them are 30 days in advance. Likewise, ProMED detects the 134  putatively associated events (28\%) before {Empres-i}, and those 16 of them are 30 days in advance. Moreover, PADI-web (resp. ProMED) detects the 273 (resp. 150) putatively associated events before ProMED (resp. PADI-web), and those 47 (resp. 30) of them are 30 days in advance. Finally, our results for time lags and timeliness are also consistent with the previous works~\cite{Arsevska2016, Tsai2013}. Indeed, the performances of PADI-web and ProMED are also comparable in these works, and PADI-web is slightly more timely than ProMED.

\subsubsection{Periodicity Analysis}
\label{subsubsec:Results_PeriodicityAnalysis}
We now study how accurate EBS systems detect full or partial periodic continuous and seasonal patterns with different temporal scales based on the evolution of the epidemiological events.
We compare the obtained results based on the {Empres-i} dataset to see to what extent PADI-web and ProMED can capture similar patterns. 

\begin{figure*}
    \centering
    \includegraphics[width=0.9\textwidth]{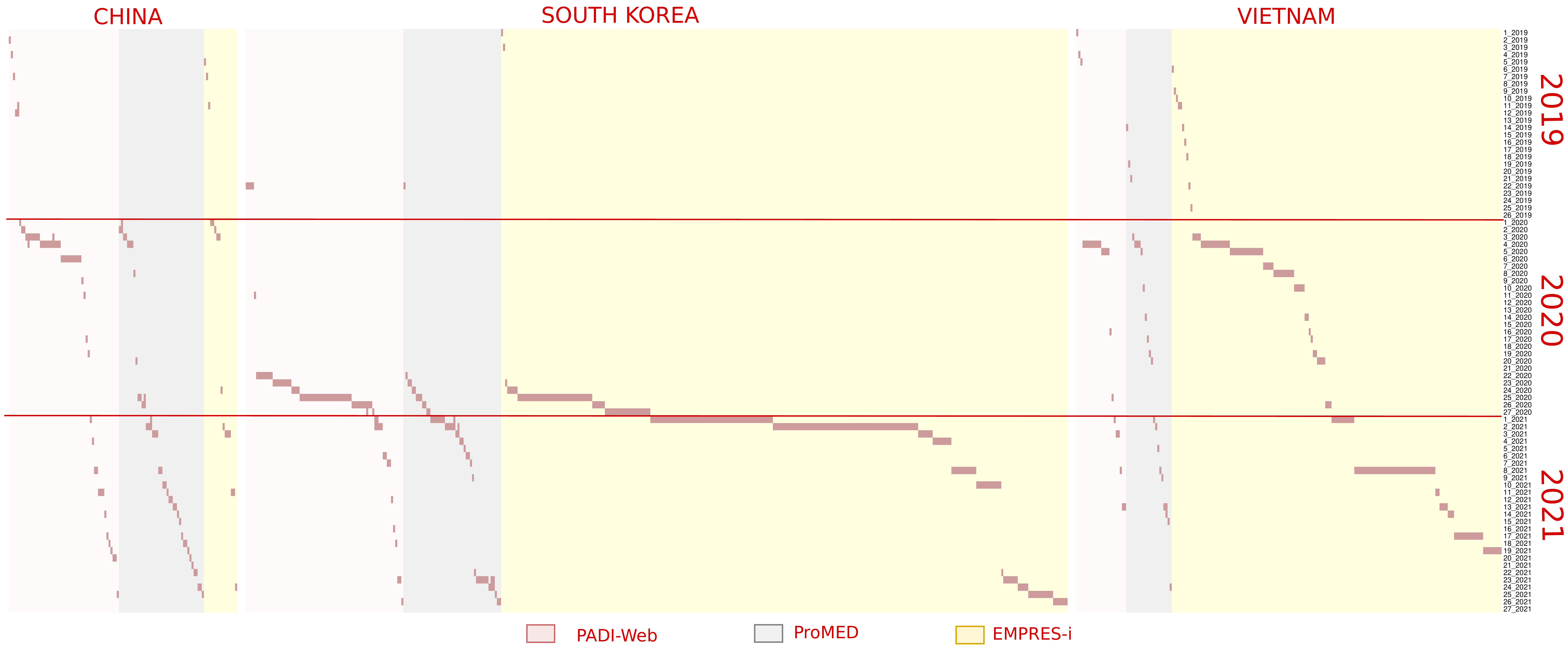}
    \captionsetup{width=.9\linewidth}
    \caption{Evolution of epidemiological events for China, South Korea and Vietnam from 2019 to 2021 with fine-grained temporal scale. The columns represent distinct epidemiological events provided by PADI-web (in pink), ProMED (in gray) and {Empres-i} (in yellow). The rows correspond to the bi-weeks of 2019, 2020 and 2021. Each cell of the matrix indicates the absence or presence of at least one event for a given time period and country. Only the cells in red indicate the presence of events. Finally, the columns (i.e. the events) are regrouped by country, as indicated on the top part of the plot.}
\label{fig:empresi_padiweb_promed_heatmap_biweekly_2019_2021}
\end{figure*}

To ease our discussion, we visualize the evolution of the events occurring in some countries of interest with fine-grained temporal scale from 2019 to 2021 in Figure~\ref{fig:empresi_padiweb_promed_heatmap_biweekly_2019_2021}. We describe it generically here, for matters of convenience. This figure is in a form of heatmap matrix. The columns represent distinct events provided by PADI-web (in pink), ProMED (in gray) and {Empres-i} (in yellow). The rows correspond to the bi-weeks of 2019, 2020 and 2021. Each cell of the matrix indicates the absence or presence of at least one event for a given time period and country. Only the cells in brown indicate the presence of events. Finally, the columns (i.e. the events) are regrouped by country, as indicated on the top part of the plot. 

We start with the full and partial periodic seasonal patterns. We obtain the results from PADI-web, ProMED and {Empres-i} by discretizing the time (resp. spatial) dimension into monthly intervals  (resp. country zones) and by applying the ST method with the parameters $\iota=12$, $\varrho=\{0.5, 1.0\}$ and $\alpha=1000km$. Recall that in our dataset the full periodicity, with $\varrho=1.0$, amounts to be the events occurring every year for the same time period from 2019 to 2021. Regarding the partial periodicity, with $\varrho=0.5$, a seasonal pattern is valid in our dataset, when the events occur during two consequent years between 2019 and 2021 for the same time period. In the comparison, we need to take into account the fact that the detection time for the same events can differ up to 30 days in average for PADI-web and ProMED with respect to the {Empres-i} events (see Figure~\ref{fig:TimeLags_AllPlatforms}). Therefore, it is reasonable to observe some time delay in the results. 

Table~\ref{tab:results_seasonal_periodic_patterns} shows the full and partial seasonal periodic frequent patterns of six countries for each EBS/IBS system to see when and where epidemiological events repeatedly occur every year from 2019 to 2021. These countries are China, South Korea, Vietnam, India, United Kingdom and France. In Table~\ref{tab:results_seasonal_periodic_patterns}, the {Empres-i} patterns detected by both PADI-web and ProMED are in orange, and it is colored in red (resp. blue) when only PADI-web (resp. ProMED) finds them. We see from the table that the results of PADI-web and ProMED are not very inline with the {Empres-i} seasonal patterns, and we summarize the comparison in two points.

\begin{table*}
    \centering
    \scriptsize
    \setlength{\tabcolsep}{3pt}
    \renewcommand{\arraystretch}{0.99}
    \begin{tabular}{|l|l|p{0.25cm}|p{0.25cm}|p{0.25cm}|p{0.25cm}|p{0.25cm}|p{0.25cm}|p{0.25cm}|p{0.25cm}|p{0.25cm}|p{0.25cm}|p{0.25cm}|p{0.25cm}||p{0.25cm}|p{0.25cm}|p{0.25cm}|p{0.25cm}|p{0.25cm}|p{0.25cm}|p{0.25cm}|p{0.25cm}|p{0.25cm}|p{0.25cm}|p{0.25cm}|p{0.25cm}|}
        \hline
           \textbf{EBS system} & \textbf{country} & \multicolumn{12}{c|}{\textbf{full periodicity (2019-2021)}} & \multicolumn{12}{c|}{\textbf{partial periodicity (2019-2021)}} \\
           & & \scalebox{.75}{Jan} & \scalebox{.75}{Feb} & \scalebox{.75}{Mar} & \scalebox{.75}{Apr} & \scalebox{.75}{May} & \scalebox{.75}{Jun} & \scalebox{.75}{Jul} & \scalebox{.75}{Aug} & \scalebox{.75}{Sep} & \scalebox{.75}{Oct} & \scalebox{.75}{Nov} & \scalebox{.75}{Dec} & \scalebox{.75}{Jan} & \scalebox{.75}{Feb} & \scalebox{.75}{Mar} & \scalebox{.75}{Apr} & \scalebox{.75}{May} & \scalebox{.75}{Jun} & \scalebox{.75}{Jul} & \scalebox{.75}{Aug} & \scalebox{.75}{Sep} & \scalebox{.75}{Oct} & \scalebox{.75}{Nov} & \scalebox{.75}{Dec}\\ \hline
        PADI-web & China & \cellcolor{gray!40} & \cellcolor{red!40} & & \cellcolor{gray!40} & & & &  & & &  & & \cellcolor{orange!40} & \cellcolor{orange!40} & & \cellcolor{gray!40} & & & & \cellcolor{gray!40} & \cellcolor{gray!40} & & & \\
        PADI-web & South Korea & & & & & & & & & & & & & & & & & & & & & & \cellcolor{gray!40} & \cellcolor{orange!40} & \cellcolor{orange!40}\\
        PADI-web & Vietnam & & \cellcolor{gray!40} & & & & & & & & & & & & \cellcolor{gray!40} & & & & & \cellcolor{orange!40} & & & & & \\
        PADI-web & India & \cellcolor{gray!40} & \cellcolor{red!40} & \cellcolor{gray!40} & & & & & & & & & \cellcolor{gray!40} & \cellcolor{orange!40} & \cellcolor{orange!40} & \cellcolor{gray!40} & & & & & & & & & \cellcolor{gray!40} \\
        PADI-web & United Kingdom & & & & & & & & & & & & \cellcolor{orange!40} & \cellcolor{gray!40} & \cellcolor{gray!40} & \cellcolor{gray!40} & & \cellcolor{gray!40} & & & & & & \cellcolor{orange!40} & \cellcolor{orange!40} \\ 
        PADI-web & France & & & & & & & & & & & &  & & & & & & & & & & & \cellcolor{orange!40} & \cellcolor{orange!40} \\\hline
        ProMED & China & & & & & & & & & & & &  & \cellcolor{orange!40} & \cellcolor{orange!40} & & \cellcolor{gray!40} & & & & & & \cellcolor{gray!40} & & \cellcolor{gray!40} \\
        ProMED & South Korea & & & & & & & & & & & & & & & & & & & & & & \cellcolor{gray!40} & \cellcolor{orange!40} & \cellcolor{orange!40} \\
        ProMED & Vietnam & & & & & & & \cellcolor{blue!40} & & & & & & & & & & & & \cellcolor{orange!40} & & \cellcolor{blue!40} & \cellcolor{blue!40} & & \\
        ProMED & India & & & & & & & & & & & & & \cellcolor{orange!40} & \cellcolor{orange!40} & \cellcolor{gray!40} & \cellcolor{orange!40} & & & & & & & & \\
        ProMED & United Kingdom & & & & & & & & & & & & \cellcolor{orange!40} & & & & & & & & & & & \cellcolor{orange!40} & \cellcolor{orange!40}\\
        ProMED & France & & & & & & & & & & & & \cellcolor{gray!40} & & & & & & & & & & & \cellcolor{orange!40} & \cellcolor{orange!40}\\ \hline
        {Empres-i} & China & & \cellcolor{red!40} & & & & & & & & & & & \cellcolor{orange!40} & \cellcolor{orange!40} & & & & & & & & & \cellcolor{gray!40} & \\
        {Empres-i} &  South Korea & & & & & & & & & & & & & & & & & & & & & & & \cellcolor{orange!40} & \cellcolor{orange!40} \\
        {Empres-i} & Vietnam & & & & & \cellcolor{gray!40} & & \cellcolor{blue!40} & \cellcolor{gray!40} & \cellcolor{gray!40} & & & & & & \cellcolor{gray!40} & \cellcolor{gray!40} & \cellcolor{gray!40}  & & \cellcolor{orange!40} & \cellcolor{gray!40} & \cellcolor{blue!40} & \cellcolor{blue!40} & & \cellcolor{gray!40} \\
        {Empres-i} & India & & \cellcolor{red!40} & & & & & & & & & & & \cellcolor{orange!40} & \cellcolor{orange!40} & & \cellcolor{orange!40} & & & & & & & & \\
        {Empres-i} & United Kingdom & & & & & & & & & & & & \cellcolor{orange!40} & & & & & & & & & & & \cellcolor{orange!40} & \cellcolor{orange!40} \\
        {Empres-i} & France & & & & & & & & & & & &  & & & & & & & & & & & \cellcolor{orange!40} & \cellcolor{orange!40} \\\hline
    \end{tabular}
    \captionsetup{width=.9\linewidth}
     \caption{Full and partial seasonal (or yearly) periodic patterns for PADI-web, ProMED and Empres-i. For the sake of comparison, we only show them only for five countries: China, South Korea, Vietnam, India, United Kingdom and France. The {Empres-i} patterns detected by both PADI-web and ProMED are colored in orange, and it is colored in blue (resp. red) when only ProMED (resp. PADI-web) finds them.}
     \label{tab:results_seasonal_periodic_patterns}
\end{table*}


First, we observe the full periodic seasonal patterns in the Empres-i data for some countries, and we expect PADI-web and ProMED to detect them. These countries are China, Vietnam, India, United Kingdom, Taiwan, South Africa, Bulgaria, Japan and Denmark (see Table~\ref{tab:results_seasonal_periodic_patterns} and Figure~\ref{fig:empresi_padiweb_promed_heatmap_biweekly_2019_2021} for some of them). PADI-web captures its full periodic seasonal patterns for the considered four countries of Table~\ref{tab:results_seasonal_periodic_patterns}, plus for Taiwan and Japan. In these patterns, PADI-web accurately detects only the pattern for the United Kingdom. Nonetheless, there are some discrepancies for the other countries. Indeed, it underrepresents (resp. overrepresents) Vietnam (resp. China and India). For instance, PADI-web overrepresents the events occurring in China, because particularly since 2020, with the rise of Covid-19 cases, media sources make much news about China and coronavirus. For instance, one of the news titles is "\textit{Chinese authorities say viral pneumonia outbreak is not SARS, MERS or bird flu}"\footnote{\url{www.reuters.com/article/us-china-pneumonia-idUSKBN1Z40G3}}. Overall, the average evaluation scores for PADI-web and ProMED based on the seasonal full periodic frequent patterns are $0.49$ and $0.17$, respectively (see Section~\ref{subsubsecapx:QuantitativeEvaluationForPeriodicity} in the Appendix for the calculation details).

Second, we see many more seasonal patterns for partial periodicity. This indicates that a country witnesses an event during two consecutive years from 2019 to 2021. On the one hand, ProMED is interestingly able to capture almost accurately the patterns for China, South Korea, India, United Kingdom and France. But, it largely underrepresents the patterns for Vietnam. On the other hand, PADI-web captures two seasonal patterns for Vietnam, but it still underrepresents it. Moreover, it also overrepresents the seasonal patterns for the United Kingdom, China and India. Overall, the average evaluation scores for PADI-web and ProMED based on the partial periodic seasonal frequent patterns are $0.51$ and $0.85$, respectively. When we consider the partial and full periodic frequent seasonal patterns together, we obtain the average scores of $0.50$ and $0.51$ for PADI-web and ProMED, respectively.


Next, we pass to continuous periodic patterns, i.e. the epidemiological events occurring consistently throughout the year, by applying the ST method with $\iota=2$, $\varrho=0.1$ and $\alpha=1000km$. Some of our results are shown in Table~\ref{tab:results_wekkly_monthly_periodic_patterns} and Figure~\ref{fig:empresi_padiweb_promed_heatmap_biweekly_2019_2021}. 
Table~\ref{tab:results_wekkly_monthly_periodic_patterns} shows the most 10 frequent partial weekly and monthly periodic continuous patterns at country scale for each EBS/IBS system from 2019 to 2021. We expect an EBS system to provide a similar ranking as in {Empres-i}. In Table~\ref{tab:results_wekkly_monthly_periodic_patterns}, the {Empres-i} partial continuous patterns detected by both PADI-web and ProMED are in orange, and it is colored in blue (resp. red) when only PADI-web (resp. ProMED) finds them. 

Overall, we observe some differences across the results, and we interpret them in two parts. We first analyze the monthly partial periodic continuous patterns. On the one hand, we see that the ranking of ProMED at monthly scale is much more in line with {Empres-i} compared to PADI-web, as it captures seven monthly {Empres-i} patterns (out of 10). On the other hand, PADI-web detects only few partial continuous patterns. Furthermore, both PADI-web and ProMED are able to detect the same three monthly {Empres-i} patterns, related to the recurrent events occurring in Germany, Netherlands and France. Finally, only PADI-web (resp. ProMED) is able to detect the patterns for Taiwan (resp. Russia, Vietnam, Sweden and Denmark). Overall, when we consider the whole ranking results, we obtain the evaluation scores of $0.86$ and $0.91$ for PADI-web and ProMED, respectively.

\begin{table*}
    \centering
    \tiny
    \begin{tabular}{|l|p{3cm}|p{2cm}||p{2cm}|p{2cm}||p{2cm}|p{3cm}|}
        \hline
        \textbf{Rank} & \textbf{PADI-web (month)} & \textbf{PADI-web (week)} & \textbf{ProMED (month)} & \textbf{ProMED (week)} & \textbf{EMPRES-i (month)} & \textbf{EMPRES-i (week)} \\\hline
        1 & China (16) & India (41) & \cellcolor{blue!40} Russia (15) & India (23) & \cellcolor{red!40} Taiwan (30) & Taiwan (66)\\\hline
        2 & India (15) & \cellcolor{red!40} France (30) & China (14) & \cellcolor{blue!40} Russia (23) & South Africa (23) & \cellcolor{orange!40} Germany (36)\\\hline
        3 & United Kingdom (14) & \cellcolor{orange!40} United Kingdom (25) & India (14) & \cellcolor{orange!40} South Korea (23) & \cellcolor{blue!40} Vietnam (21) & South Africa (29)\\\hline
        4 & \cellcolor{orange!40} France (13) & \cellcolor{orange!40} Germany (15) & \cellcolor{orange!40} Germany (12) & \cellcolor{orange!40} Germany (22) & \cellcolor{blue!40} Russia (14) & \cellcolor{blue!40} Russia (29)\\\hline
        5 & \cellcolor{orange!40} Germany (12) & Japan (13) & \cellcolor{orange!40} Netherlands (8) & China (21) & \cellcolor{blue!40} Sweden (13) & Sweden (23)\\\hline
        6 & Japan (11) & \cellcolor{orange!40} South Korea (13) & \cellcolor{blue!40} Vietnam (8) & Japan (17) & \cellcolor{orange!40} Germany (12) & Poland (23)\\\hline
        7 & Germany-United Kingdom (10) & N/A & \cellcolor{orange!40} France (7) & \cellcolor{orange!40} United Kingdom (11) & \cellcolor{orange!40} Netherlands (11) & \cellcolor{orange!40} United Kingdom (22)\\\hline
        8 & \cellcolor{orange!40} Netherlands (9) & N/A & South Korea (7) & N/A & Belgium (10) & \cellcolor{red!40} France (21)\\\hline
        9 & France-Germany (9) & N/A & \cellcolor{blue!40} Sweden (7) & N/A & \cellcolor{orange!40} France (10) & Germany-United Kingdom (20)\\\hline
        10 & \cellcolor{red!40} Taiwan (8) & N/A & \cellcolor{blue!40} Denmark (5) & N/A & \cellcolor{blue!40} Denmark (9) & \cellcolor{orange!40} South Korea (20)\\\hline
    \end{tabular}
     \caption{Most 10 frequent partial weekly and monthly periodic patterns for PADI-web, ProMED and Empres-i. The periodic support values for these patterns are indicated in parenthesis. These results are produced with the parameter values $\iota=2$, $\varrho=0.1$ and $\alpha=1000km$. $N/A$ indicates that there is no available entry. The {Empres-i} patterns detected by both PADI-web and ProMED are colored in orange, and it is colored in blue (resp. red) when only ProMED (resp. PADI-web) finds them.
}
     \label{tab:results_wekkly_monthly_periodic_patterns}
\end{table*}

Now, we pass to the weekly partial periodic continuous patterns. The identification of these patterns is harder compared to the monthly patterns, because this amounts to seek the events occurring at least once every two weeks. Indeed, PADI-web (resp. ProMED) can only find six (resp. seven) partial weekly periodic patterns in total. Moreover, we see that the obtained patterns are slightly different compared to the monthly patterns, hence they give another temporal vision of the EBS systems. For instance, South Korea is not that frequent at monthly scale in the {Empres-i} data, but it is one of the most 10 frequent weekly partial periodic patterns (see also Figure~\ref{fig:empresi_padiweb_promed_heatmap_biweekly_2019_2021}). 
Of the detected patterns by PADI-web and ProMED, four of them are also found in Empres-i's result. This shows that both systems have a comparable performance at weekly scale. Overall, when we consider the weekly and monthly partial patterns together, we obtain the average evaluation scores of $0.69$ and $0.75$ for PADI-web and ProMED, respectively (see Section~\ref{subsubsecapx:QuantitativeEvaluationForPeriodicity} in the Appendix for the calculation details).

To conclude this part, identifying the full and partial weekly, monthly and seasonal periodic patterns gives a different analysis perspective to assess the performances of the EBS systems. Overall, both PADI-web and ProMED have comparable results. Nevertheless, there are some substantial differences between them. When it comes to the seasonal patterns presented in Table~\ref{tab:results_seasonal_periodic_patterns}, on the one hand, ProMED finds less partial periodic seasonal patterns compared to PADI-web, but most of them are found in the {Empres-i}'s result. On the other hand, PADI-web finds more seasonal patterns, bu they are not as accurate as the patterns provided by ProMED. This is probably because several of them might be either false alerts, i.e. suspected cases being not confirmed by the national authorities, or directly erroneous due to the automatic processing framework of PADI-web. Consequently, it overrepresents some countries. When it comes to the weekly and monthly continuous patterns presented in Table~\ref{tab:results_wekkly_monthly_periodic_patterns}, the obtained results are barely in line with the Empres-i's results. Nonetheless, ProMED performs slightly better, since it correctly finds more monthly partial patterns. Overall, we obtain the final periodicity scores of $0.59$ and $0.63$ by combining both continuous and seasonal periodicity aspects for PADI-web and ProMED, respectively. Finally, our results are also partially confirmed by the previous works~\cite{Bhatia2021}. In \cite{Bhatia2021}, the authors measure the correlation between the weekly event time series derived from ProMED, HealthMap\footnote{similar to PADI-web, as they both are automated systems.} and an official source WHO data using Pearson’s correlation coefficient. They find out that the results derived from ProMED and HealthMap are moderately correlated with the ones reported by World Health Organization (WHO) on West African Ebola, and that there exist some substantial differences between them, particularly at the peak of the epidemics.

\subsection{Thematic Dimension}
\label{subsec:Results_ThematicDimension}

In this section, we study how detailed EBS systems provide the thematic information encoded in their data. We want to know how similar the frequent multidimensional patterns across EBS systems are. We compare the results based on the {Empres-i} dataset to see to what extent PADI-web and ProMED can capture similar patterns. To ease our discussion, we also visualize the relations between spatial and thematic entities with a \textit{chord diagram} in Figure~\ref{fig:Thematic_entity_chord_diagram_for_all_platforms}, which is found in the Appendix for space matters.

Table~\ref{tab:results_multidimensional_items} shows the most 13 frequent static and temporal multidimensional patterns for PADI-web, ProMED and Empres-i. We obtain these patterns with the parameters $\iota \in \{10, 30, \infty\}$ and $\varrho=1$ (in count). Note that the use of $\infty$ represents a very large value for eliminating the periodicity aspect from the method ST. This amounts to obtain the static version of frequent multidimensional patterns, without any temporal aspect. We describe Table~\ref{tab:results_multidimensional_items} generically here, for matters of convenience. Given a specific spatial scale, each row corresponds to a spatial entity and these entities are regrouped by a specific system. The columns represent the existing host entities in a specific hierarchical level, and they are regrouped by the existing disease entities in a specific hierarchical level. For instance, in Table~\ref{tab:results_multidimensional_items} we stick to the second level of hierarchy for spatial, disease and host entities (see Table~\ref{tab:ThematicDimension_TaxonomyClasses} in the Appendix for thematic taxonomy). Each cell can encode four different information. First, we display a dash character, when a specific multidimensional pattern, be static or temporal, is not frequent. Second, the statistics of a given multidimensional pattern is expressed in the format $x | y$. The first value $x$ corresponds to the static condition, and represents its frequency number (i.e. support) without considering the partial periodicity constraint. The second value corresponds to the temporal condition, and represents its periodic support with respect to the parameters $\iota$ and $\varrho$. Third, we also highlight with different gray scales to what degree a multidimensional pattern is partially periodic in the data at hand. For instance, in Table~\ref{tab:results_multidimensional_items}, we consider two different $\iota$ values, which are $10$ and $30$ days. The results for the former (resp. latter) are indicated in dark (resp. light) gray. Finally, we show the {Empres-i} multidimensional patterns detected by both PADI-web and ProMED in orange, and they are colored in blue (resp. red) when only PADI-web (resp. ProMED) finds them.

We can summarize Table~\ref{tab:results_multidimensional_items} in five points. First, as expected, {Empres-i} provides only fine-grained disease information. Interestingly, the data collected by ProMED is also fine-grained, whereas PADI-web provides mostly coarse-grained disease information (see also Figure~\ref{fig:Thematic_entity_chord_diagram_for_all_platforms} in the Appendix). Second, the overwhelming majority of the frequent multidimensional patterns provided by PADI-web, ProMED and Empres-i concern the HPAI cases. This fact highlights how national and international authorities prioritize the surveillance of HPAI cases, since it is highly contagious among birds, and can be deadly, especially for domestic poultry. Third, apart the unknown bird category, most of the frequent multidimensional patterns of {Empres-i} (resp. PADI-web and ProMED) concerns wild (resp. domestic) birds. On this point, we can say that the distribution of host categories are not very balanced, with a slight dominance for unknown bird category. Fourth, the overwhelming majority of the frequent static multidimensional patterns for all systems are strongly partial periodic, with $\iota=10$ days. Finally, when we compare the frequent multidimensional patterns across EBS systems, we observe that both PADI-web and ProMED detect few {Empres-i} patterns (four and two patterns for ProMED and PADI-web, respectively). This fact shows how the thematic data collected by PADI-web and ProMED can be different with respect to {Empres-i}. Interestingly, although PADI-web is currently collaborating with the French Platform for Animal Health Surveillance (see Section~\ref{subsec:Experimens_SelectedEBSSystems}), France is not in the first 13 frequent patterns for PADI-web. Overall, when we consider all the static and temporal frequent multidimensional patterns together, we obtain the ranking scores of $0.64$ and $0.63$ for PADI-web and ProMED, respectively (see Section~\ref{subsecapx:QuantitativeEvaluationThematicDimension} in the Appendix for the calculation details).



\begin{table*}
    \centering
    \tiny
    \renewcommand{\arraystretch}{1.5}
    \begin{tabular}{| p{0.5cm} | p{1.75cm} | p{0.9cm} p{0.9cm} p{0.95cm}| p{0.9cm} p{0.9cm} p{0.95cm}| p{0.9cm} p{0.9cm} p{0.95cm}|}
    \hline
      & \textbf{country} & \multicolumn{3}{c|}{\textbf{HPAI}} & \multicolumn{3}{c|}{\textbf{LPAI}} & \multicolumn{3}{c|}{\textbf{AI unknown}} \\
       & & \textbf{Domestic} & \textbf{Wild} & \textbf{Bird unknown} & \textbf{Domestic} & \textbf{Wild} & \textbf{Bird unknown} & \textbf{Domestic} & \textbf{Wild} & \textbf{Bird unknown} \\
      \hline
      \multirow{5}{*}{\rotatebox[origin=c]{90}{ \parbox[c]{1.5cm}{\centering \textbf{PADI-web}}}}
        & Japan & - & - & - & - & - & - & \cellcolor{gray!40} 32 | 31 & - & \cellcolor{gray!40} 12 | 10\\
        & United Kingdom & \cellcolor{gray!40} 8 | 5 & \cellcolor{gray!40} \textcolor{orange}{9 | 8} & - & - & - & - & - & \cellcolor{gray!40} 13 | 10 &\cellcolor{gray!40} 24 | 19\\
        & South Korea & - & - & \cellcolor{gray!40} \textcolor{red}{8 | 7} & - & - & - & \cellcolor{gray!40} 10 | 7 & \cellcolor{gray!40} 9 | 7 & \cellcolor{gray!40} 20 | 18\\
        & China & \cellcolor{gray!40} 11 | 10 & - & \cellcolor{gray!40} 9 | 7 & - & - & - & - & - & - \\
        & India & - & - & - & - & - & - & \cellcolor{gray!40} 9 | 6  & - & -\\
      \hline
      \multirow{10}{*}{\rotatebox[origin=c]{90}{ \parbox[c]{2.4cm}{\centering \textbf{ProMED}}}}
        & Germany & \cellcolor{gray!40} 6 | 2 & \cellcolor{gray!40} \textcolor{blue}{12 | 9} & -  & - & - & - & - & - & - \\
        & Vietnam & - & - & \cellcolor{gray!40} \textcolor{blue}{8 | 3} & - & - & - & - & - & -\\
        & Japan & \cellcolor{gray!40} 7 | 6 & - & - & - & - & - & - & - & -\\
        & Ireland & \cellcolor{gray!40} 5 | 3 & - & - & - & - & - & - & - & -\\
        & India & \cellcolor{gray!10} 4 | 2 & - & \cellcolor{gray!40} 3 | 1 & - & - & - & - & - & -\\
        & China & \cellcolor{gray!40} 3 | 2 & - & \cellcolor{gray!40} 4 | 1 & - & - & - & - & - & -\\
        & United Kingdom & - & \cellcolor{gray!40} \textcolor{orange}{3 | 1} & - & - & - & - & - & - & -\\
        & South Korea & - & \cellcolor{gray!40} \textcolor{blue}{3 | 2} & - & - & - & - & - & - & -\\
        & United States & - & - & - & - & - & - & \cellcolor{gray!40} 3 | 2 & - & -\\
        & Hungary & \cellcolor{gray!40} 3 | 1 & - &  - & - & - & - & - & - & -\\
      \hline
      \multirow{11}{*}{\rotatebox[origin=c]{90}{ \parbox[c]{2.4cm}{\centering \textbf{EMPRES-i}}}}
        & Hungary & - & - & \cellcolor{gray!40} 268 | 267 & - & - & - & - & - & -\\
        & Germany & - & \cellcolor{gray!40} \textcolor{blue}{267 | 265} & \cellcolor{gray!40} 41 | 38 & - & - & - & - & - & -\\
        & United Kingdom & - & \cellcolor{gray!40}  \textcolor{orange}{102 | 101} & - & - & - & - & - & - & -\\
        & Denmark & - & \cellcolor{gray!40} 96 | 95 & - & - & - & - & - & - & -\\
        & Vietnam & - & - & \cellcolor{gray!40} \textcolor{blue}{65 | 56} & - & - & - & - & - & -\\
        & Netherlands & - & \cellcolor{gray!40} 57 | 55 & - & - & - & - & - & - & -\\
        & Taiwan & - & - & - & \cellcolor{gray!40} 45 | 37 & - & 35 | - & - & - & -\\
        & Japan & - & - & \cellcolor{gray!40} 44 | 43 & - & - & - & - & - & -\\
        & Russia & - & - & \cellcolor{gray!40} 41 | 36 & - & - & - & - & - & -\\
        & Poland & - & - & \cellcolor{gray!40} 40 | 33 & - & - & - & - & - & -\\
        & South Korea & - & \cellcolor{gray!40} \textcolor{blue}{37 | 35} & \cellcolor{gray!10} \textcolor{red}{- | 33} & - & - & - & - & - & -\\
        & France & - & - & \cellcolor{gray!40} - | 32 & - & - & - & - & - & -\\
    \hline
    \end{tabular}\\
    \vspace{-0.2cm}
    \captionsetup{width=.9\linewidth}
    \caption{
    Most 13 frequent static and temporal multidimensional patterns. To ease our discussion, these patterns are only at country and solely concern the events occurring in 2020. When a multidimensional pattern is static, it is shown in white. When it is partial periodic with 10 days (resp. 30 days), it is shown in dark (resp. light) gray. The Empres-i patterns detected by both PADI-web and ProMED are colored in orange, and it is colored in blue (resp. red) when only ProMED (resp. PADI-web) finds them.}
    \label{tab:results_multidimensional_items}
\end{table*}

\subsection{Source dimension}
\label{subsec:Results_SourceDimension}
Finally, we assess how important and timely the news outlets involved in EBS systems for information dissemination. Next, we rank them in terms of these two objectives and see if PADI-web and ProMED obtain similar results. Overall, it is worth noticing that PADI-web (resp. ProMED) includes 480 (resp. 189) distinct news outlets. Only 63 of them are in common between PADI-web and ProMED. On top of the news outlets, ProMED also mostly relies on official reports from WOAH (472 events out of 786). For this reason, we also include WOAH in our analysis for ProMED. Nevertheless, we solely discuss the performances of the news outlets for a fair comparison.

    \begin{table*}
     \centering
     \tiny
        \renewcommand{\arraystretch}{1.5}
        \begin{tabular}{| p{0.5cm} | p{2cm} p{2cm} | p{2cm}  p{2cm} | p{2cm} p{2cm} |}
        \hline
          \textbf{rank} & \multicolumn{2}{c|}{\textbf{Asia}} & \multicolumn{2}{c|}{\textbf{Europe}} & \multicolumn{2}{c|}{\textbf{World}} \\
           & \textbf{PADI-web} & \textbf{ProMED} & \textbf{PADI-web} & \textbf{ProMED} & \textbf{PADI-web} & \textbf{ProMED} \\
          \hline
          1 & hindustantimes (IND) & WOAH & reuters & WOAH & \textbf{reuters} & WOAH \\
          2 & indiatimes (IND) & yna.co.kr (KOR) & heraldscotland (GBR) & rossaprimavera.ru (RUS) & hindustantimes (IND) & rossaprimavera.ru (RUS) \\
          3 & businessworld.in (IND) & outbreaknewstoday & thepoultrysite & tatar-inform.ru (RUS) & thepoultrysite & yna.co.kr (KOR) \\
          4 & reuters & \textbf{nippon (JPN)} & francetvinfo (FRA) & bbc (GBR) & indiatimes (IND) & \textbf{reuters}\\
          5 & middleeastmonitor & \textbf{newindianexpress (IND)}  & outbreaknewstoday & reuters & businessworld.in (IND) & \textbf{nippon (JPN)}\\
          6 & thepoultrysite & tuoitrenews.vn (VNM) & phys & foodingredientsfirst & heraldscotland (GBR) & niknews.mk.ua (UKR) \\
          7 & \textbf{nippon (JPN)} & niknews.mk.ua (UKR) & agriculture.com & cheshire-live.co.uk (GBR) & \textbf{outbreaknewstoday} & \textbf{outbreaknewstoday} \\
          8 & oneindia (IND) & outlookindia (IND) & dgwgo (GBR) & kazakh-zerno (KAZ) & \textbf{nippon (JPN)} & bbc\\
          9 & \textbf{newindianexpress (IND)} & thebeijinger (CHN) & republicain-lorrain.fr (FRA) & nltimes.nl (NLD) & middleeastmonitor & tass.ru (RUS) \\
          10 & indianexpress (IND) & russian.news.cn (CHN) & farminguk (GBR) & nv.ua (UKR) & agriculture.com & tatar-inform (RUS) \\
        \hline
    	\end{tabular}\\
    	\vspace{-0.2cm}
        \captionsetup{width=.9\linewidth}
    	\caption{PageRank centrality results for PADI-web, ProMED and Empres-i. The first (resp. second) part of the table corresponds to the results based on the events occurring only in Asia (resp. Europe).  In the last part, the results are produced from the whole dataset.}
        \label{tab:news_outlet_analysis_pagerank}
    \end{table*}


First, we compare PADI-web and ProMED in terms of their important news outlets for information dissemination, obtained with the PageRank algorithm. 
Table~\ref{tab:news_outlet_analysis_pagerank} shows the first 10 (resp. 9) news outlets having the largest PageRank scores for PADI-web (resp. ProMED) based on the events occurring in Asia, Europe and in the whole world, respectively. We can summarize the results in two points. First, overall, the results show that the most important news outlets for both sources are almost completely different. 
PADI-web relies mostly on the Indian news outlets for the events in Asia and French and British/Scottish ones for Europe, whereas spatially more diverse news outlets are in ProMED's results, with a slight prevalence for the Russian (and nearby countries such as Ukraine and Kazakhstan) and British/Scottish news outlets. Consequently, PADI-web and ProMED have only three news outlets in common. Finally, on top of national news outlets, several international ones, such as \textit{Reuters} and \textit{Outbreak News Today}, can also take an important role for information dissemination for PADI-web and ProMED. Nevertheless, their rankings can be very different. For instance, Reuters is the first news outlets for the world-wide events, whereas it is at 10\textsuperscript{th} place for ProMED. This also confirms us how different the news collection
strategies of PADI-web and ProMED are.

\begin{table*}
 \centering
    \tiny
    \renewcommand{\arraystretch}{1.5}
    \begin{tabular}{| p{0.5cm} | p{2cm} p{2cm} | p{2cm}  p{2cm} | p{2cm} p{2cm} |}
    \hline
      \textbf{rank} & \multicolumn{2}{c|}{\textbf{Asia}} & \multicolumn{2}{c|}{\textbf{Europe}} & \multicolumn{2}{c|}{\textbf{World}} \\
       & \textbf{PADI-web} & \textbf{ProMED} & \textbf{PADI-web} & \textbf{ProMED} & \textbf{PADI-web} & \textbf{ProMED} \\
      \hline
      1 & \textbf{indiatimes (IND)} & WOAH & thepoultrysite & WOAH & \textbf{indiatimes (IND)} & WOAH \\
      2 & hindustantimes (IND) & outbreaknewstoday & francebleu (FRA) & vetandlife.ru (RUS) & hindustantimes (IND) & \textbf{outbreaknewstoday} \\
      3 & \textbf{yna.co.kr KOR)} & \textbf{yna.co.kr (KOR)} & fwi.co.uk (GBR) & regnum.ru (RU) & thepoultrysite & vetandlife.ru (RUS) \\
      4 & thepoultrysite & koreaherald & \textbf{farminguk (GBR)} & tass.ru (RUS) & \textbf{yna.co.kr KOR)} & \textbf{yna.co.kr (KOR)} \\
      5 & \textbf{newindianexpress (IND)} & \textbf{indiatimes (IND)} & reuters & rossaprimavera.ru (RUS) & \textbf{reuters} & \textbf{reuters} \\
      6 & thehindu (IND) & \textbf{reuters} & albaniandailynews (ALB) & life.ru (RUS) & \textbf{outbreaknewstoday} & koreaherald (KOR) \\
      7 & indianexpress (IND) & \textbf{nippon (JPN)} & heraldscotland (SCO) & bbc (GBR) & wattagnet & regnum.ru (RUS) \\
      8 & \textbf{nippon (JPN)} & vetandlife.ru (RUS) & wattagnet & khaleejtimes (ARE) & francebleu (FRA) & tass.ru (RUS) \\
      9 & \textbf{reuters} & poultrymed & 20minutes.fr (FRA) & \textbf{farminguk (GBR)} & newindianexpress (IND) & \textbf{indiatimes (IND)} \\
      10 & kashmirobserver (IND) & \textbf{newindianexpress (IND)} & phys & aphascience.blog.gov.uk (GBR) & thehindu (IND) & rossaprimavera.ru (RUS) \\
    \hline
    \end{tabular}\\
    \vspace{-0.2cm}
    \captionsetup{width=.9\linewidth}
    \caption{Timely detection results for PADI-web, ProMED and Empres-i. The first (resp. second) part of the table corresponds to the results based on the events occurring only in Asia (resp. Europe). In the last part, the results are produced from the whole dataset.}
    \label{tab:news_outlet_analysis_timely_detection}
\end{table*}

Next, we pass to the results of timely news outlets involved in PADI-web and ProMED, obtained with the method CELF 
by limiting the output size to the first 30 news outlets. Similar to the previous results, Table~\ref{tab:news_outlet_analysis_timely_detection} shows only the first 10 (resp. 9) timely news outlets for PADI-web (resp. ProMED) based on the events occurring in Asia, Europe and in the whole world, respectively. Compared to the previous results in Table~\ref{tab:news_outlet_analysis_pagerank}, we see here that PADI-web and ProMED share more common news outlets in terms of timely detection. Furthermore, we observe that the most important news outlets in terms of PageRank score are not necessarily timely in event detection. In other words, we observe some inconsistency issues in the results obtained by the methods PageRank and CELF for PADI-web and ProMED. Recall that this consistency assessment allows us verifying whether news outlets playing a key role in epidemiological information dissemination are also timely in event detection. For instance, regarding the events occurring in Europe for PADI-web, the French news outlets \textit{France Bleu} and \textit{20 Minutes} appear only in Table~\ref{tab:news_outlet_analysis_timely_detection}, whereas we observe two other French news outlets in Table~\ref{tab:news_outlet_analysis_pagerank}. Nevertheless, this kind of inconsistencies seems not to appear much in the whole data, i.e. world scale. Indeed, the rank evaluation scores also confirm this last point, as we obtain the ranking scores of $0.96$ and $0.89$ for PADI-web and ProMED, respectively (see Section~\ref{subsecapx:QuantitativeEvaluationSourceDimension} in the Appendix for the calculation details).

\begin{table*}
 \centering
    \fontsize{6.7}{6.7}\selectfont
    \renewcommand{\arraystretch}{1.5}
    \begin{tabular}{| p{0.9cm} | p{1cm} | p{2.95cm} p{2.7cm} | p{3.2cm} p{3cm} | p{0.6cm}  p{0.6cm} |}
    \hline
      \textbf{Dimension} & \textbf{Description} & \multicolumn{2}{c|}{\textbf{Qualitative Evaluation}} & \multicolumn{2}{c|}{\textbf{Qualitative Evaluation}} & \multicolumn{2}{c|}{\textbf{Quantitative Eval.}} \\
       & & \centering \textbf{PADI-web (+)} & \centering \textbf{PADI-web (-)} & \centering \textbf{ProMED (+)} & \centering \textbf{ProMED (-)} & \textbf{PADI-web} & \textbf{ProMED} \\
      \hline
      Spatial & Spatio-temporal representativeness & \hspace{-0.4cm} \fuda{
                    \vspace{-0.2cm}
                    \item Event reporting for 56 countries (Table~\ref{tab:results_spatiotemporal_representativeness_country_scale}).
                    \item Well covering 21 countries.
                    } & \hspace{-0.4cm} \fuda{
                    \vspace{-0.2cm}
                    \item  Absence of events from 14 countries that ProMED covers.
                    \item Less reporting capacity for Eastern Europe and Middle East.
                    } & \hspace{-0.4cm} \fuda{
                    \vspace{-0.2cm}
                    \item Event reporting for 70 countries (Table~\ref{tab:results_spatiotemporal_representativeness_country_scale}).
                    \item Well covering 36 countries.
                    \item Thanks to the WOAH reports, covering the countries better than PADI-web.
                    }
                    & \hspace{-0.4cm} \fuda{
                    \vspace{-0.2cm}
                    \item Absence of events from 3 countries that {Empres-i} reports.
                    \item Less reporting capacity for Western Europe, Asia and the USA.
                    }
                    & 0.40
                    & \textbf{0.55}
                    \\\hline
      \multirow{2}{*}{Temporal} & Timeliness & \hspace{-0.4cm} \fuda{
                    \vspace{-0.2cm}
                    \item Timely detection performance: $49\%$ (Table~\ref{tab:results_timeliness}).
                    } & \hspace{-0.4cm} \fuda{
                    \vspace{-0.2cm}
                    \item 14.8 days delay in average (by focusing solely on late events).
                    } & \hspace{-0.4cm} \fuda{
                    \vspace{-0.2cm}
                    \item Timely detection performance: $28\%$ (Table~\ref{tab:results_timeliness}).
                    } 
                    & \hspace{-0.4cm} \fuda{
                    \vspace{-0.2cm}
                    \item 6 days delay in average (by focusing solely on late events).
                    }
                    & \textbf{0.18}
                    & 0.12
                    \\\cline{2-6}
       & Periodicity & \hspace{-0.4cm} \fuda{
                    \vspace{-0.2cm}
                    \item Detecting more accurate full periodic seasonal patterns than ProMED does (Table~\ref{tab:results_seasonal_periodic_patterns}).
                    \item Detecting weekly and montyly partial periodic patterns for 2 counttries that ProMED does not (Table~\ref{tab:results_wekkly_monthly_periodic_patterns}).
                    } & \hspace{-0.4cm} \fuda{
                    \vspace{-0.2cm}
                    \item Detecting few weekly partial continuous patterns.
                    \item Detecting less accurate monthly partial periodic continuous patterns than ProMED does.
                    } & \hspace{-0.4cm} \fuda{
                    \vspace{-0.2cm}
                    \item  Detecting more accurate partial periodic seasonal and monthly partial periodic continuous patterns than PADI-web does (Table~\ref{tab:results_seasonal_periodic_patterns}).
                    \item Detecting weekly and montyly partial periodic patterns for 4 counttries that ProMED does not (Table~\ref{tab:results_wekkly_monthly_periodic_patterns}).
                    }
                    & \hspace{-0.4cm} \fuda{
                    \vspace{-0.2cm}
                    \item Detecting less accurate full periodic seasonal patterns than PADI-web does.
                    \item Detecting few weekly partial continuous patterns.
                    }
                    & 0.59
                    & \textbf{0.63}
                    \\\hline
      Thematic & Static and temporal multidimensional patterns & \hspace{-0.4cm} \fuda{
                    \vspace{-0.2cm}
                    \item Detecting 1 frequent multidimensional pattern that ProMED does not (Table~\ref{tab:results_multidimensional_items}).
                    \item Balanced distribution of domestic and wild bird cases.
                    \item Strongly partial periodicity for its frequent multidimensional patterns.
                    } & \hspace{-0.4cm} \fuda{
                    \vspace{-0.2cm}
                    \item Providing less detailed disease and host information than ProMED does.
                    \item Absence of LPAI cases.
                    } & \hspace{-0.4cm} \fuda{
                    \vspace{-0.2cm}
                    \item Providing more detailed disease and host information than PADI-web does.
                    \item Detecting 3 frequent multidimensional patterns that PADI-web does not (Table~\ref{tab:results_multidimensional_items}).
                    \item Strongly partial periodicity for its frequent multidimensional patterns.
                    }
                    & \hspace{-0.4cm} \fuda{
                    \vspace{-0.2cm}
                    \item Providing events concerning mostly domestic birds, i.e. less balanced distribution.
                    \item Absence of LPAI cases.
                    }
                    & \textbf{0.64}
                    & 0.63
                \\\hline
      Source & Important and timely news outlets & \hspace{-0.4cm} \fuda{
                    \vspace{-0.2cm}
                    \item Relying on more timely important news outlets than ProMED Tables~\ref{tab:news_outlet_analysis_pagerank} and \ref{tab:news_outlet_analysis_timely_detection}.
                    \item Using different national and international news outlets than ProMED.
                    } & \hspace{-0.4cm} \fuda{
                    \vspace{-0.2cm}
                    \item Relying mostly on the Indian, French and British/Scottish news outlets.
                    \item Relying less timely important news outlets than PADI-web.
                    } & \hspace{-0.4cm} \fuda{
                    \vspace{-0.2cm}
                    \item Using different national and international news outlets than PADI-web.
                    \item Relying on WOAH, which is important and timely Tables~\ref{tab:news_outlet_analysis_pagerank} and \ref{tab:news_outlet_analysis_timely_detection}.
                    }
                    & \hspace{-0.4cm} \fuda{
                    \vspace{-0.2cm}
                    \item Relying mostly on the
                    Russian (also, nearby countries such as Kazakhstan) and British news outlets.
                    \item Relying on less timely important news outlets than PADI-web.
                    }
                    & \textbf{0.96}
                    & 0.89
                    \\
    \hline
    \end{tabular}\\
    \vspace{-0.2cm}
    \caption{Summary of findings regarding all evaluation results between PADI-web and ProMED. }\label{tab:summary_table_all_quantitative_eval_results}
\end{table*}

\begin{figure}[ht!]
    \centering
    \includegraphics[width=0.4\textwidth]{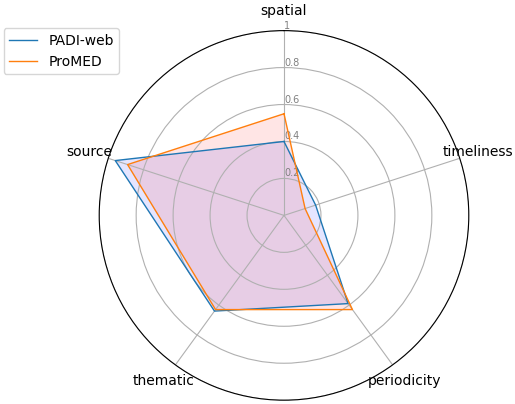}
    \captionsetup{width=.9\linewidth}
    \caption{Radar chart summarizing all quantitative evaluation results between PADI-web and ProMED.}
\label{fig:radar_chart_all_quantitative_eval_results}
\end{figure}

\begin{figure}[ht!]
    \centering
    \includegraphics[width=0.4\textwidth]{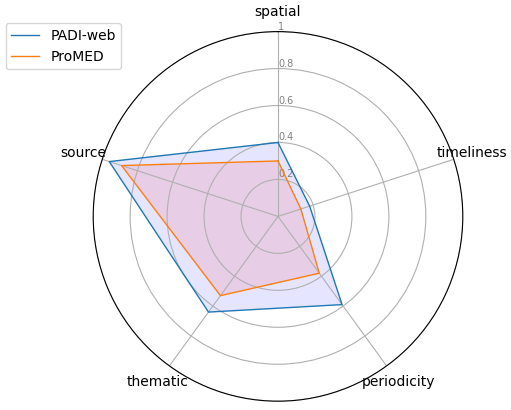}
    \captionsetup{width=.9\linewidth}
    \caption{Radar chart summarizing all quantitative evaluation results between PADI-web and ProMED, when we discard the ProMED data provided by official data
sources (i.e. WOAH reports).}
\label{fig:radar_chart_all_quantitative_eval_results_promed_unofficial}
\end{figure}

To conclude this part, we show that PADI-web and ProMED rely mostly on different important and timely news outlets. This is mostly because the Indian, French and British/Scottish news outlets take an important role for PADI-web, whereas these are mostly the Russian (and nearby countries such as Ukraine and Kazakhstan) and British/Scottish news outlets for ProMED. Moreover, both EBS systems also rely on the same international news outlets (e.g. Reuters), nevertheless these news outlets do not contribute to both EBS systems in the same manner. All these results suggest for these EBS systems to include more spatially more diverse news outlets for a greater geographic coverage (e.g. Baidu for Chinese news). Finally, we present a summary of findings in Table~\ref{tab:summary_table_all_quantitative_eval_results} and Figure~\ref{fig:radar_chart_all_quantitative_eval_results} based on all obtained evaluation results from this section and the previous ones. We see that PADI-web and ProMED seem to be complementary. PADI-web (resp. ProMED) performs slightly better for the timeliness and source (resp. spatial and periodicity) dimensions, and they have a comparable performance for the thematic dimension. For the sake of completeness, we also compare in Figure~\ref{fig:radar_chart_all_quantitative_eval_results_promed_unofficial} these EBS systems in terms of the presented five dimensions by discarding the ProMED data
provided by official data sources (i.e. WOAH reports).

\section{Conclusion}
\label{sec:Conclusion}


In this article, we have presented a new evaluation framework to identify the strengths and drawbacks of EBS systems in terms of epidemic surveillance. This evaluation is very valuable from the epidemiological standpoint, since it allows end-users to select the most appropriate EBS system(s) for an effective surveillance of a particular situation. We want not only to compare EBS systems, but also to produce results that the end user can easily interpret. For this purpose, we proposed a two-step framework based on our review of the literature. It first transforms the raw input event data into a set of normalized distinct events, then conducts a descriptive retrospective analysis of these events with four objectives: spatial, temporal, thematic and source analysis. We illustrated its relevance by applying it to an Avian Influenza dataset collected by PADI-web, ProMED and {Empres-i}. We showed that our framework allows identifying the strengths and drawbacks of the considered EBS systems. For some of our evaluation aspects, our results confirm the findings already published in the literature. For others, the systematic nature of our approach uncovers new findings for the considered EBS systems.

Our work could be extended in several ways. First, our method can be applied systematically to other EBS systems and other animal diseases, for the sake of completeness. This would give a better overview of the capabilities of the existing systems. Second, the source dimension can be better evaluated in Section~\ref{subsec:Method_SourceDimension}, if we can obtain an appropriate gold standard dataset. This would imply to conduct an extensive work like in~\cite{Ye2019}, but tailored to Epidemic Intelligence. Third, our evaluation framework focuses only on a descriptive retrospective analysis. Nevertheless, it would be also valuable to extend this work with a predictive analysis to see to what extent the existing EBS data can give an insight on the short or long term future using past event information, accompanied by domain-specific data (e.g. animal mobility models, environmental data). Some examples are risk mapping~\cite{Jones2008, Stevens2013, Paul2016, Durand2017} and epidemic forecasting with sparse data~\cite{Kandula2018}. Finally, due to the generic nature of our evaluation framework, it can be also applied to other spatio-temporal systems with similar properties (e.g. natural disaster surveillance systems), so this could constitute another perspective. 

\paragraph{Acknowledgments.} This study was partially funded by EU grant 874850 MOOD and is catalogued as MOOD063. The contents of this publication are the sole responsibility of the authors and do not necessarily reflect the views of the European Commission. We thank the ProMED-mail staff for sharing with us their dataset, as well as for their constructive comments to improve the content of this paper. ProMED is a program of the International Society for Infectious Diseases (ISID).

\phantomsection\addcontentsline{toc}{section}{References}
\printbibliography

\newpage
\begin{appendices}

\section{Event similarity}
\label{secapx:EventSimilarity}
Throughout this work, we need to assess the similarity of two events. In this section, we explain how we perform this similarity calculation in the presence of hierarchical data. The main idea is that we consider two events to be similar, if 1) their event attributes, with $D_i \in \{D_Z, D_D, D_H\}$, are identical or hierarchically linked in $H_i$, and 2) their event dates are very close. In the similarity assessment, we first calculate the similarity for each event attribute, and then we sum up the obtained values in order to get the final score. Note that we do not take the dimension $D_S$ into account in this calculation.

Before introducing this calculation, let us define some additional notations and definitions. First, for convenience, we denote the depth (or level) of a node $v$ in the hierarchy $H_i$ by $l_i(v)$. Likewise, the depth of the closest common ancestor of $v_1$ and $v_2$ is denoted by $\bar{l}_i(v_1, v_2)$. In $H_i$, a path connecting two nodes from the root (i.e. the most general level) to the leaves (i.e. the most specific level) (resp. from the leaves to the root) gives the specialization relation (resp. the generalization relation). Given an element $x \in Dom(D_i)$ and the associated hierarchy $H_i$, we denote by $x^\uparrow$ (respectively $x^\downarrow$) the set containing $x$ along with all generalizations (respectively specializations) of $x$ with respect to $H_i$ that belong to $Dom(D_i)$. Based on this, we also define the specificity relation, which allows comparing the attributes of two events being at different hierarchical levels. Namely, for a pair of event attributes $d_i$ and $d'_i$, with $i \in \{Z, T, D, H\}$, and associated with $e = (d_Z, d_T, d_D, d_H, d_S)$ and $e' = (d'_Z, d'_T, d'_D, d'_H, d'_S)$, event attribute $d_i$ is said to be more specific than $d'_i$, denoted by $d_i \preceq d'_i$, if $d'_i \in d_{i}^\uparrow$. For instance, in Table~\ref{tab:hierarchical-event-example}  \textit{H7N9} is more specific than \textit{avian influenza} in $D_D$, i.e. \textit{H7N9} $\preceq$ \textit{avian influenza}.

Then, let $e$ and $e'$ be two events, for which we calculate the similarity. For the dimension $D_i \in \{D_Z, D_D, D_H\}$, we use an ontology-based semantic similarity measure proposed in \cite{Rodriguez2003}, but tailored for our purposes. In this measure, we consider that two literal values $x$ and $y$ of a dimension $D_i \in \{D_Z, D_D, D_H\}$, with $x, y \in Dom(D_i)$, are similar, if they are identical or hierarchically linked in $H_i$, i.e. $y \preceq x \vee x \preceq y$. Otherwise, we assign a large negative value for a penalization. Concretely, we calculate the similarity between $x$ and $y$ in Equation~\ref{eq:SemanticSimilarityMeasure} as

\begin{equation}
	sim_i(x, y) = \lambda^i_{x,y} \frac{2 \bar{l}_i(x, y)}{l_i(x) + l_i(y)} - (1-\lambda^i_{x,y}) \sigma_i, 
	\label{eq:SemanticSimilarityMeasure}
\end{equation}

where the variable $\lambda^i_{x,y}$ takes a binary value, where $1$ means $x$ and $y$ are identical or hierarchically linked in $H_i$, and $0$ otherwise. Finally, the variable $\sigma_i$ is a large penalization factor, used when $\lambda = 0$. Note that the obtained score without the penalization factor is in the range $[0,1]$. But, in the end, the score is not normalized from one side, i.e. $[-\sigma_i,1]$.

Furthermore, we handle the temporal dimension $D_T$ in a different manner. We calculate the similarity between two dates in Equation~\ref{eq:DateSim} as

\begin{equation}
	sim_T(t_1, t_2) = 1 - \frac{|t_2 - t_1|}{L},
	\label{eq:DateSim}
\end{equation}
where $|t_1 - t_2|$ represents temporal distance in days and $L$ corresponds to a time delay which controls how the temporal distance between two dates are small enough. In this work, we set $L$ to $21$ (i.e. 3 weeks). Finally, we compute the final similarity score by summing the individual similarity scores.

\begin{equation}
	sim(e, e') = \sum\limits_{\substack{i \in \{Z, T, D, H\}\\d_i \in e, d'_i \in e'}} sim_i(d_i, d'_i),
	\label{eq:FinalSim}
\end{equation}

\section{Extraction of Event Database}
\label{secapx:CorpusEventExtraction}
In this section, we detail how we produce an event database $\mathcal{E}$ from the events collected by an EBS/IBS system. First, we need to clarify our terminology, as we use the term \textit{event} in two different contexts. We call the events being usually extracted at news document-level \textit{document events}. Since information about events can be scattered over different news documents in the corpus, this implies that several copies of a specific event can co-exist in the end. The fusion of these copies can produce fully-fledged event description and it is simply called \textit{corpus event}.

The input of our process is a set of document events, accompanied by their associated news documents. As mentioned in the main manuscript that the definition of an event can be different from one system to another, the minimalist one being a disease-location pair. For this reason, when the event definition of a system does not match the one proposed in Section~\ref{subsec:EventRelatedDefinitionsNotations}, we process the associated news documents in order to complete the missing information (Section~\ref{subsecapx:EventCompletion}). Then, we normalize the document events that can be used for evaluation and comparison purposes. In this context, the normalization of event attributes is an important step, since it allows transforming a raw text into one of well-defined taxonomy classes (Section~\ref{subsecapx:EventNormalization}). Finally, since the information about events can be scattered over different news documents, the last step consists in aggregating/fusing all the event information in order to produce fully-fledged event descriptions, i.e. cross-document information fusion (Section~\ref{subsecapx:CorpusEventConstruction}). In the end of our process, we produce a set of normalized corpus events. To simplify our processing, we suppose that the input data are collected for a particular disease and that each news document contains at least one event.

\subsection{Document Event Completion}
\label{subsecapx:EventCompletion}
In this section, when the event definition of an EBS system does not match the one proposed in Section~\ref{subsec:EventRelatedDefinitionsNotations}, we process the associated news documents in order to complete the missing information, if possible, assuming that the position of each extracted event is known in its associated news document. In the following, we explain this process in two parts: 1) Thematic and source entity extraction (Section~\ref{subsubsecapx:ThematicAndSourceEntityExtraction}) and 2) document event completion (Section~\ref{subsubsecapx:EventCompletion}). Note that these tasks are different from the event extraction methods~\cite{Xiang2019, Valentin2021}, which are not the scope of this work.

\subsubsection{Thematic and Source Entity Extraction}
\label{subsubsecapx:ThematicAndSourceEntityExtraction}

What we propose is a classical preprocessing for information extraction in EBS systems through the application of natural language processing (NLP)~\cite{Ng2020}. In the first step, we extract thematic and source-related information from the header, title and raw content of each news document. We first start with the header part. We retrieve the publication date and the news outlet publishing the news document at hand.

Then, we extract disease and host entity information from the title and the raw content of a news document. We consider a title is a special sentence of the document at hand, which summarizes the content with a few keywords. For this reason, we detail the extraction of event-related keywords together for the title and the raw content. First, the raw textual content is preprocessed by sentence splitting, tokenization and lemmatization using SpaCy~\cite{Honnibal2017}. We perform our entity extraction for each sentence, as well as the title. In each sentence, we extract by building the lexicon of host and disease names. We extract thematic entities with them by keyword search~\cite{Goel2020, Valentin2021}.

\subsubsection{Event Completion}
\label{subsubsecapx:EventCompletion}
In the event completion task, we can find multiple events in the same news document. Indeed, it is possible that a sentence can contain multiple events, or events can be located in the different part of the document. In the following, we follow the work of \cite{Yan2010} for the event completion task. To simplify our processing, we assume that the position of each extracted event is known in its associated news document.

As in~\cite{Yan2010}, we treat each sentence containing all essential event attributes as head sentence. To identify the head sentences we need to locate (host, time, host, disease) entities in the same sentence. 
The event completion starts from a head sentence and continues up to $k$ next sentences. Although a head sentence needs to contain all essential information, its subsequent sentences can contain more detailed thematic information about an event. At the end of the process, for each news document, we have a set of document events defined as in Section~\ref{subsec:EventRelatedDefinitionsNotations}.

\subsection{Event Normalization}
\label{subsecapx:EventNormalization}
This task consists in normalizing the attributes of each event in an event database, assuming that the events are defined as in Section~\ref{subsec:EventRelatedDefinitionsNotations}. As mentioned before, a normalization task aims at transforming a raw text into one of well-defined taxonomy classes, which results in hierarchical information.

We start with the normalization of spatial entities, also called \textit{geocoding}. This task consists in assigning geographic coordinates to spatial entities. This can be a challenging task due to the ambiguity among place names~\cite{Ng2020}. For instance, Avignon is a regional county municipality in Quebec/Canada, but also a city in France. In this work, we perform the geocoding task with the gazetteer GeoNames\footnote{\url{www.geonames.org/}}. For a given query of spatial entity, GeoNames outputs a ranked list of most appropriate geographic coordinates associated with the input text. Although the first result is usually the desired/correct geographic coordinates, we still need to apply a disambiguation technic to resolve the place ambiguity issue. This disambiguation is easily solved, if we know the associated country. Indeed, EBS systems often provide the country information of a spatial entity in an event. Otherwise, we employ the following heuristic strategy.

First, we find the corresponding news document, and we extract the country information of the spatial entities mentioned in the title, because we consider that a title is a special sentence of the document at hand, which summarizes the content with a few main keywords. If there is not any spatial entity in the title, we look for a nationality mention (e.g. French farm). In the worst case, we use the country information associated with the news outlet (e.g. France for Le Monde\footnote{\url{www.lemonde.fr}}), thanks to a publicly available dataset~\cite{Roberts2021}. The rationale behind this processing is according to pragmatic rules when one reads a news document of a given country~\cite{Lejeune2010}. Often, a city name is explicitly mentioned without citing the associated country name. Nevertheless, if the news document is about another country, it is more likely to be mentioned so that the reader has fewer chances of misunderstanding. We extract spatial entities from the title using SpaCy~\cite{Honnibal2017} and a lookup table. This lookup table is used to locate the spatial entities which are not detected by SpaCy. It can be constructed with a gazetteer and extracts spatial entities with keyword search. We also extract nationalities using a second lookup table. Then, we finalize the geocoding task from the ranked list provided by GeoNames by selecting the first result whose associated country is the same as the one extracted from the title.

\begin{table}
	\scriptsize
    \centering
	\renewcommand{\arraystretch}{1.2}
	\begin{tabular}{| p{1.25cm} | p{1.5cm} p{1.5cm} p{3cm} |}
    	\hline
        Entity type (Level 0) & Level 1 & Level 2 & Level 3 \\\hline
        \multirow{10}{*}{\textbf{Host}}
            & \multirow{2}{*}{\textbf{human}} & male & -- \\
            & & female & -- \\\cline{2-4}
            & \multirow{4}{*}{\textbf{mammal}} & arctoidea & seal, etc. \\
            & & canidae & fox, dog, etc.\\
            & & equidae & horse, donkey, mule, etc. \\
            & & camelidae & lama, alpaca, etc.\\\cline{2-4}
            & \multirow{2}{*}{\textbf{avian}} & wild & eagle, falcon, gull, hawk, black swan, crow, etc.\\
            & & domestic & chicken, turkey, fowl, ostrich, house crow, mute swan, etc.\\
           & \multirow{2}{*}{\textbf{mosquito}} & culex & culex pipiens, culex tarsalis, etc.\\
            & & aedes & aedes aegypti, aedes vexans \\\hline 
        \multirow{3}{*}{\textbf{Disease}}
            & \multirow{2}{*}{\textbf{Avian Infl.}} & low pathogenic & other serotypes \\
            & & high pathogenic & H5N1, H7N9, H5N6, H5N8 \\\cline{2-4}
            & \multirow{1}{*}{\textbf{West Nile}} & -- & -- \\
        \hline
	\end{tabular}
    \captionsetup{width=.9\linewidth}
    \caption{Adjusted taxonomy classes associated with the disease and host dimensions for Avian Influenza and West Nile Virus.}
\label{tab:ThematicDimension_TaxonomyClasses}
\end{table}

In terms of the temporal dimension $D_T$, the temporal expressions are usually in a well-structured format. The only task we need to do is to correctly estimate this format~\cite{Strotgen2010}. For instance, a date can be written in \textit{MM-DD-YYYY} or \textit{DD-MM-YYYY}. We do this estimation by checking the whole data. Finally, we normalize the temporal expressions according to the TIMEX3 annotation standard.


On what regards the thematic dimensions $D_D$ and $D_H$, we normalize these disease and host names against the NCBI Taxonomy database~\cite{Federhen2011}, using a manually composed table of species name synonyms. Moreover, we tailor some of these NCBI taxonomies to diseases of interest. This is because the NCBI database is supposed to be generic enough in order to be used for any application domain. Nevertheless, for a particular area of expertise (e.g. Epidemic Intelligence), the taxonomy classes would be better constructed. For instance, categorizing the bird types as \textit{wild} and \textit{domestic}, or the Avian Influenza serotypes as \textit{high} and \textit{low pathogenic} based on their ability to cause mortality in birds, is a valuable information for epidemiological standpoint. We illustrate some of the adjusted taxonomy classes for thematic and disease entities in Table~\ref{tab:ThematicDimension_TaxonomyClasses}.

\subsection{Corpus Event Construction}
\label{subsecapx:CorpusEventConstruction}
The process of extracting corpus events requires identifying the news documents reporting the same document events. To do so, we apply an overlapping clustering method. The idea of our method is that when an event has less precise information, then it is more likely to belong to different clusters. In the beginning of our method, every document event constitutes its own cluster. We start the method iterations with the events with most detail spatial entity information, and continue with one granularity level lower in each iteration. In each iteration and for each event of this iteration, we find a set of similar events of the same or higher granularity level. We assess the similarity between events, as described in Section~\ref{secapx:EventSimilarity}. This score is the sum of four scores reflecting the similarities of geographical locations, infection dates, disease names and host species by taking their hierarchical levels into account. When a set of similar events are found for a given event, then we put them in the same cluster. In each cluster update, we perform an information fusion step based on a voting approach. Namely, for each event attribute except date, we take the most frequent value. For the temporal attribute, we take the oldest infection date.

\section{Quantitative Evaluation of EBS Systems}
\label{secapx:QuantitativeEvaluation}

In this section, we explain how we quantitatively evaluate the performance of an event database $\mathcal{E}$ of an EBS system with respect to a gold standard database $\mathcal{E_R}$. In this evaluation, the results usually take the form of a ranked list. For this reason, we first need to define a ranking evaluation measure to evaluate the quality of the ranked lists with respect to those produced by a gold standard database (Section~\ref{subsecapx:RankingEvaluation}). Then, we pass to the quantitative evaluations for the spatial (Section~\ref{subsecapx:QuantitativeEvaluationForSpatialDimension}), temporal (Section~\ref{subsecapx:QuantitativeEvaluationForTemporalDimension}), thematic (Section~\ref{subsecapx:QuantitativeEvaluationThematicDimension}) and source (Section~\ref{subsecapx:QuantitativeEvaluationSourceDimension}) dimensions. 

Before presenting the quantitative evaluations, we need to define the selection operator $\sigma$ for the dimensions $D_{Z}$ and $D_{T}$ as $\sssigma^{Dom(D_Z) \in Z}_{Dom(D_T) \in T} (\mathcal{E})$ ($\sssigma^{Z}_{T} (\mathcal{E})$ for short), for two sets $Z$ and $T$ of literal values defined for $D_{Z}$ and $D_{T}$. This operator is used to select a subset of the input data $\mathcal{E}$. For the sake of simplicity, when the selection operator is not applied on $D_Z$ or $D_T$, we simply omit it from the notation $\sigma$. For instance, $\sssigma^{\{France\}} (\mathcal{E}^{country}_{week})$ represents a set of weekly events at country level occurring only in France. Finally, the number of remaining events after the selection is calculated by taking the cardinality, i.e. $|\sssigma^{\{France\}} (\mathcal{E}^{country}_{week})|$.

\subsection{Evaluation with Ranking}
\label{subsecapx:RankingEvaluation}
Throughout this framework, we compare the results produced by the event database $\mathcal{E}$ of an EBS system with those from a gold standard database $\mathcal{E}_R$, assuming that $\mathcal{E}_R$ is available to use. As we will see in the following sections, these results usually take the form of a ranked list. For instance, such a list might represent the periodic patterns ordered by their frequency in an event database $\mathcal{E}$. Furthermore, since $\mathcal{E}$ has a limited data compared to a gold standard database $\mathcal{E}_R$, these two ranked lists are usually of different size, and might contain disjoint values. For this reason, we need a specific ranking evaluation measure to evaluate the quality of the ranked list produced by an EBS system, called \textit{candidate list} and denoted by $L_\mathcal{E}$, with respect to that produced by a gold standard database, called \textit{reference list} and denoted by $L_{\mathcal{E}_R}$. In this work, we use the normalized F-measure $F(L_\mathcal{E}, L_{\mathcal{E}_R})$~\cite{Kishida2005, Valentin2021, Valentin2022}, which is the combination of normalized precision $P(L_\mathcal{E}, L_{\mathcal{E}_R})$ and recall $R(L_\mathcal{E}, L_{\mathcal{E}_R})$. These last two are based on the difference between the sum of ranks of the candidate and reference lists.

The calculation of the normalized precision and recall between $L_\mathcal{E}$ and $L_{\mathcal{E}_R}$, of sizes $N_{L_\mathcal{E}}$ and $N_{L_{\mathcal{E}_R}}$, requires extracting a list of common values found in both lists, that we call \textit{relevant list} of size $N_\Lambda$. Concretely, we calculate $F (L_\mathcal{E}, L_{\mathcal{E}_R})$ between this relevant list and the initial reference list. Finally, we define a bijection $r : \{ 1, 2, \dots, N_\Lambda \} \rightarrow \{ 1, 2, \dots, N_{L_\mathcal{E}} \}$, such that it gives the rank of the $i$\textsuperscript{th} element of the relevant list in the reference list. Based on this, the \textit{normalized recall} is defined as in Equation~\ref{eq:Rnorm_definition}.

\begin{equation}
    R (L_\mathcal{E}, L_{\mathcal{E}_R}) = 1 - \frac{\sum\limits_{i=1}^{N_\Lambda} r(i) - \sum\limits_{i=1}^{N_\Lambda} i}{N_\Lambda N_{L_\mathcal{E}} + N_\Lambda^2}.
    \label{eq:Rnorm_definition}
\end{equation}

When we transform the index $i$ representing each position of elements into $\log (i)$, it is called \textit{normalized precision}, which is calculated as in Equation~\ref{eq:Pnorm_definition}.

\begin{equation}
    P (L_\mathcal{E}, L_{\mathcal{E}_R}) = 1 - \frac{\sum\limits_{i=1}^{N_\Lambda} \log (r(i)) - \sum\limits_{i=1}^{N_\Lambda} \log (i)}{\log (C(N_{L_\mathcal{E}}, N_\Lambda))},
    \label{eq:Pnorm_definition}
\end{equation}

where $C(N_{L_\mathcal{E}}, N_\Lambda) = \frac{N_{L_\mathcal{E}}!}{N_\Lambda! (N_{L_\mathcal{E}}-N_\Lambda)!}$. Finally, $F (L_\mathcal{E}, L_{\mathcal{E}_R})$ is the harmonic mean of $P (L_\mathcal{E}, L_{\mathcal{E}_R})$ and $R (L_\mathcal{E}, L_{\mathcal{E}_R})$, which is calculated as

\begin{equation}
    F (L_\mathcal{E}, L_{\mathcal{E}_R}) = \frac{2 R(L_\mathcal{E}, L_{\mathcal{E}_R}) P(L_\mathcal{E}, L_{\mathcal{E}_R}) }{ R(L_\mathcal{E}, L_{\mathcal{E}_R}) + P(L_\mathcal{E}, L_{\mathcal{E}_R}) }.
    \label{eq:Fmeasure_norm_definition}
\end{equation}

\subsection{Quantitative Evaluation for Spatial Dimension}
\label{subsecapx:QuantitativeEvaluationForSpatialDimension}
In this section, we describe how we compute the spatio-temporal representativeness score of an event database $\mathcal{E}$ with respect to a gold standard database $\mathcal{E^R}$. The definition of the spatio-temporal representativeness requires fixing the spatial and temporal scales, denoted by $l_Z$ and $l_T$, respectively. Depending on these scales, we first discretize the spatial and temporal dimensions over a set $Z$ of geographic zones and a set $T$ of time intervals. Then, for a particular disease, we say that an event database $\mathcal{E}$ of an EBS system represents well a specific geographic zone $z$ for a given time interval $t$, if it finds at least one event in $\mathcal{E^R}$, which is defined in Equation~\ref{eq:spatio-temporal_event_existence} as 

\begin{equation}
    \mathbb{1}_{l_T,t}^{l_Z,z} (\mathcal{E}) = \begin{cases}
                                      1 , & |\sssigma^{\{z\}}_{\{t\}} (\mathcal{E}^{l_Z}_{l_T})| > 0 \\
                                      0, & \text{otherwise}.
                                      \end{cases}
    \label{eq:spatio-temporal_event_existence}
\end{equation}

The calculation of the spatio-temporal representativeness scores is done with respect to a gold standard database $\mathcal{E}_R$. Let $Z$ be the set of spatial entities found in $\mathcal{E}_R$ with respect to the spatial scale $l_Z$, i.e. $Z = Dom_{\mathcal{E}_R} (D_Z, l_Z)$. Also, let $T$ a set of time intervals with respect to the temporal scale $l_T$, i.e. i.e. $T = Dom_{\mathcal{E}_R} (D_T, l_T)$. For a given geographic zone $z \in Z$, its spatio-temporal representativeness score for $\mathcal{E}$ with respect to $\mathcal{E}_R$ is calculated as
\begin{equation} 
    \footnotesize
    \begin{split}
        \Phi^{l_Z, z}_{l_T} (\mathcal{E}, \mathcal{E}_R) = & 1 - \frac{1}{|T|} \sum_{t \in T} \max \bigg( 0, \mathbb{1}_{l_T,t}^{l_Z,z} (\mathcal{E}_R) \\
        & - \max( \mathbb{1}_{l_T,t-1}^{l_Z,z} (\mathcal{E}), \mathbb{1}_{l_T,t}^{l_Z,z} (\mathcal{E}), \mathbb{1}_{l_T,t+1}^{l_Z,z} (\mathcal{E}) ) \bigg).
        \label{eq:spatio-temporal_representativeness_score_for_zone_z}
    \end{split}
\end{equation}

In Equation~\ref{eq:spatio-temporal_representativeness_score_for_zone_z}, the term after the subtraction operator calculates an error score, which is similar to the mean absolute error. The difference is that when an EBS system detects an event, which is not detected in $\mathcal{E}_R$, is not considered as an error. Furthermore, since there can be some reporting delay between the events of $\mathcal{E}$ and $\mathcal{E}_R$, we also consider in Equation~\ref{eq:spatio-temporal_representativeness_score_for_zone_z} the previous (resp. next) time interval $t-1$ (resp. $t+1$) in order not to penalize $\mathcal{E}$. The obtained score is in the range $[0,1]$, where the score of $0$ (resp. $1$) indicates that an EBS system never (resp. always) finds an event in $\mathcal{E^R}$ for a given geographic zone. Finally, we obtain a single spatio-temporal representativeness score for $\mathcal{E}$ by taking the average of the obtained scores over all geographic zones $Z$, as in Equation~\ref{eq:spatio-temporal_representativeness_score}. 

\begin{equation}
        \Phi^{l_Z}_{l_T} (\mathcal{E}, \mathcal{E}_R) = \frac{1}{|Z|} \sum_{z \in Z} \Phi^{l_Z, z}_{l_T} (\mathcal{E}, \mathcal{E}_R).
        \label{eq:spatio-temporal_representativeness_score}
\end{equation}

This score is tied to particular $l_Z$ and $l_T$ scales. If we consider several different scale values, then we obtain an average value as in Equation~\ref{eq:spatio-temporal_representativeness_score_for_several_scales}, i.e.

\begin{equation}
    \Phi^{L_Z}_{L_T} (\mathcal{E}, \mathcal{E}_R) = \frac{1}{|L_Z| |L_T|} \sum\limits_{l_Z \in L_Z} \sum\limits_{l_T \in L_T} \Phi^{l_Z}_{l_T} (\mathcal{E}, \mathcal{E}_R),
    \label{eq:spatio-temporal_representativeness_score_for_several_scales}
\end{equation}

where $L_Z$ and $L_T$ are two sets of values for spatial and temporal scales, respectively.

\subsection{Quantitative Evaluation for Temporal Dimension}
\label{subsecapx:QuantitativeEvaluationForTemporalDimension}
In this section, we describe how we compute the timeliness (Section~\ref{subsubsecapx:QuantitativeEvaluationForTimeliness}) and periodicity (Section~\ref{subsubsecapx:QuantitativeEvaluationForPeriodicity}) scores of an event database $\mathcal{E}$ with respect to a gold standard database $\mathcal{E^R}$.

\subsubsection{Quantitative Evaluation for Timeliness}
\label{subsubsecapx:QuantitativeEvaluationForTimeliness}
In this section, we describe how to compute the timeliness score between an event database $\mathcal{E}$ and a gold standard database $\mathcal{E^R}$. First, we calculate in Equation~\ref{eq:temporal_distance} the temporal distance between an event $e \in \mathcal{E}_R$ and its corresponding one $e' \in \mathcal{E}$ as

\begin{equation} %
\delta (e, e') = \begin{cases}
                       t(e') - t(e), & t(e') \geq t(e)\\
                       0,            & t(e') < t(e),
                 \end{cases}
\label{eq:temporal_distance}
\end{equation}
    
where the function $t(e)$ gives the date of a given event $e$. Note that in Equation~\ref{eq:temporal_distance} when an EBS system finds an event before the corresponding report date in $\mathcal{E}_R$, we do not reward or penalize it, and put a score of $0$. Then, we calculate the timeliness score for $\mathcal{E}$ with respect to $\mathcal{E}_R$ as

    \begin{equation}
            \Psi (\mathcal{E}, \mathcal{E}_R) = \frac{1}{|\overline{\mathcal{E}}|} \sum_{e \in \overline{\mathcal{E}}} 1 - \exp^{-\frac{\delta (f(e, \mathcal{E}_R), e)}{L}},
        \label{eq:timeliness}
    \end{equation}

where the variable $L$ in Equation~\ref{eq:timeliness} indicates the maximum expected reporting delay, which controls the temporal similarity of two events. In this work, we set $L$ to 21 days.

\subsubsection{Quantitative Evaluation for Periodicity}
\label{subsubsecapx:QuantitativeEvaluationForPeriodicity}
In this section, we explain how we quantitatively evaluate the performance of an event database $\mathcal{E}$ of an EBS system in terms of its ability to detect the continuous and seasonal periodic patterns with respect to a gold standard database $\mathcal{E^R}$. In the following, we first evaluate an event database $\mathcal{E}$  for each pattern type separately, then we take an average of the obtained scores.

We first start with the evaluation of $\mathcal{E}$ in terms of its ability to detect \textit{continuous} full and partial periodic patterns with respect to a gold standard database $\mathcal{E^R}$. Let $l_Z$ and $l_T$ be spatial and temporal scales for $\mathcal{E}$ and $\mathcal{E^R}$. Moreover, let $I$ and $P$ be also two sets of parameter values for $\iota$ and $\varrho$ in \text{ST}. On top of that, the parameter value $\alpha$ controls the spatial closeness between the spatial entities in \text{ST}. Then, we propose to calculate the evaluation score $\widetilde{F}^{l_Z}_{l_T} (\mathcal{E}, \mathcal{E}_R, I, P, \alpha)$, which corresponds to the average score of $F$, as defined in Section~\ref{subsecapx:RankingEvaluation}, over the combination of all the considered parameters. It is calculated in Equation~\ref{eq:Fnorm_periodicity} as 

\begin{equation} 
    \footnotesize
    \begin{split}
    \widetilde{F}^{l_Z}_{l_T} (\mathcal{E}, \mathcal{E}_R, I, P, \alpha) = & \frac{1}{|I| |P|} \sum\limits_{\iota \in I} \sum\limits_{\varrho \in P} \\
    & F(\text{ST}(\mathcal{E}^{l_Z}_{l_T}, \iota, \varrho, \alpha), \text{ST}(\mathcal{E_R}^{l_Z}_{l_T}, \iota, \varrho, \alpha))).
    \label{eq:Fnorm_periodicity}
    \end{split}
\end{equation}

As we see, Equation~\ref{eq:Fnorm_periodicity} is calculated for particular spatial and temporal scales. We want to make our evaluation more robust by considering several scale values for dimensions $D_Z$ and $D_T$. To consider this aspect, we define two sets $L_Z$ and $L_T$ of values for spatial and temporal scales, respectively. Then, we compute the final average score for continuous full and partial periodic patterns in Equation~\ref{eq:Fnorm_periodicity_for_several_spatial_temporal_scales} as

\begin{equation} 
    \footnotesize
    \begin{split}
    \widetilde{F}^{L_Z}_{L_T} (\mathcal{E}, \mathcal{E}_R, I, P, \alpha) = \frac{1}{|L_Z| |L_T|} \sum\limits_{l_Z \in L_Z} \sum\limits_{ l_T \in L_T} \widetilde{F}^{l_Z}_{l_T} (\mathcal{E}, \mathcal{E}_R, I, P, \alpha).\hspace{0.6cm} \label{eq:Fnorm_periodicity_for_several_spatial_temporal_scales}
    \end{split}
\end{equation}

Second, we propose to perform the evaluation of $\mathcal{E}$ in terms of its ability to detect \textit{seasonal} full and partial periodic patterns with respect to a gold standard database $\mathcal{E^R}$. We propose to do it, similar to the spatio-temporal representativeness evaluation defined in Equation~\ref{eq:spatio-temporal_representativeness_score}. To detect such a seasonal effect within the data, we need to apply the method ST to a subset of $\mathcal{E}$, where the events occur only within a particular time interval over several years, and we repeat this process for all the subsets. For instance, a subset of data can represent all the events occurring in every January, whatever the year is, and we repeat similar selections for the other months. To do so, in Equation~\ref{eq:temporal_pattern_existence} we first need to define if a spatial pattern $x$ is in the output of $\text{ST}(\sssigma_{t} (\mathcal{E}^{l_Z}_{l_T}), \iota, \varrho, \alpha)$ for a given time interval $t$ and the input parameters $\iota$, $\varrho$, $\alpha$, i.e.

\begin{equation}
    \mathbb{1}^{l_Z}_{l_T, t} (\mathcal{E}, x, \iota, \varrho, \alpha) = \begin{cases}
                                      1 , & x \in \text{ST}(\sssigma_{t} (\mathcal{E}^{l_Z}_{l_T}), \iota, \varrho, \alpha) \\
                                      0, & \text{otherwise}.
                                      \end{cases}
    \label{eq:temporal_pattern_existence}
\end{equation}

Then, let $X$ be the set of all spatial patterns found in the reference ranking result of $\text{ST}(\sssigma_{t}(\mathcal{E_R}^{l_Z}_{l_T}, \iota, \varrho, \alpha))$ with respect to $l_Z$, $l_T$, $\iota$, $\varrho$ and $\alpha$. Based on this, we compute in Equation~\ref{eq:spatio-temporal_representativeness_score_for_periodicity} to what degree the set $X$ is discovered in $\mathcal{E}$ through ST for a particular time interval $t$.

\begin{equation}
    \footnotesize
    \begin{split}
        \widetilde{\Phi}^{l_Z}_{l_T, t} (\mathcal{E}, \mathcal{E}_R, \iota, \varrho, \alpha) = & \frac{1}{|X|} \sum_{x \in X} \max \bigg( \mathbb{1}^{l_Z}_{l_T, t-1} (\mathcal{E}, x, \iota, \varrho, \alpha), \\
        & \mathbb{1}^{l_Z}_{l_T, t} (\mathcal{E}, x, \iota, \varrho, \alpha), \mathbb{1}^{l_Z}_{l_T, t+1} (\mathcal{E}, x, \iota, \varrho, \alpha) \bigg),
        \label{eq:spatio-temporal_representativeness_score_for_periodicity}
    \end{split}
\end{equation}

 Then, let $T$ be all the literal values of a given temporal scale $l_T$ in $\mathcal{E}_R$, i.e. $T = Dom_{\mathcal{E_R}} (D_T, l_T)$.  We apply Equation~\ref{eq:final_seasonal_periodicity_score} to obtain an average score with respect to $T$, $I$ and $P$ for seasonal full and partial periodic patterns as

\begin{equation}
    \widetilde{\Phi}^{l_Z}_{l_T} (\mathcal{E}, \mathcal{E}_R, I, P, \alpha) = \frac{1}{|T| |I| |P|} \sum\limits_{t \in T} \sum\limits_{\iota \in I} \sum\limits_{\varrho \in P} \widetilde{\Phi}^{l_Z}_{l_T, t} (\mathcal{E}, \mathcal{E}_R, \iota, \varrho, \alpha).
    \label{eq:final_seasonal_periodicity_score}
\end{equation}

Similar to Equation~\ref{eq:Fnorm_periodicity_for_several_spatial_temporal_scales}, we compute the final average score for seasonal full and partial periodic patterns in Equation~\ref{eq:final_seasonal_periodicity_score_for_several_spatial_temporal_scales} by considering several values for spatial and temporal scales to make our evaluation more robust.
\begin{equation}
    \widetilde{\Phi}^{L_Z}_{L_T} (\mathcal{E}, \mathcal{E}_R, I, P, \alpha) = \frac{1}{|L_Z| |L_T|} \widetilde{\Phi}^{l_Z}_{l_T} (\mathcal{E}, \mathcal{E}_R, I, P, \alpha).\label{eq:final_seasonal_periodicity_score_for_several_spatial_temporal_scales}
\end{equation}

Finally, we calculate in Equation~\ref{eq:final_periodicity_score} the final periodicity score $\widetilde{\Gamma} (\mathcal{E}, \mathcal{E}_R)$ for both continuous and seasonal periodic patterns as

\begin{equation}
\widetilde{\Gamma} (\mathcal{E}, \mathcal{E}_R) = \frac{\widetilde{F}^{L_Z}_{L_T} (\mathcal{E}, \mathcal{E^R}, I, P, \alpha) + \widetilde{\Phi}^{l_Z}_{l_T} (\mathcal{E}, \mathcal{E}_R, I, P, \alpha) }{2}.
\label{eq:final_periodicity_score}
\end{equation}

\subsection{Quantitative evaluation for thematic dimension}
\label{subsecapx:QuantitativeEvaluationThematicDimension}
In this section, we explain how we evaluate the modified version $\mathcal{E}^+$ of an event database $\mathcal{E}$ in terms of its ability to detect \textit{static} and \textit{temporal} multidimensional patterns with respect to the modified version $\mathcal{E_R}^+$ of a gold standard database. 

Let $I$ and $P$ be the parameter values for \text{ST}. Note that the set $I$ of values allows handling both the static and temporal cases, as a very large inter-arrival time value for $\iota$ amounts to omit the partial periodicity condition in \text{ST}. Then, we propose to calculate the evaluation score $\Omega(\mathcal{E}^+, \mathcal{E_R}^+, I , P, \alpha)$, which corresponds to the average score of $F$, as defined in Section~\ref{subsecapx:RankingEvaluation}, over the combination of all the considered parameters. It is calculated in Equation~\ref{eq:Fnorm_periodicity_thematic} as 

\begin{equation} 
    \footnotesize
    \begin{split}
    \Omega(\mathcal{E}^+, \mathcal{E_R}^+, I , P) = & \frac{1}{|I| |P|} \sum\limits_{\iota \in I} \sum\limits_{\varrho \in P} F(\text{ST}(\mathcal{E}^+, \iota, \varrho), \\
    & \text{ST}(\mathcal{E_R}^+, \iota, \varrho))).
    \label{eq:Fnorm_periodicity_thematic}
    \end{split}
\end{equation}

\subsection{Quantitative evaluation for source dimension}
\label{subsecapx:QuantitativeEvaluationSourceDimension}
Despite the existence of several prominent works aiming at ranking a large number of news outlets in terms of various objectives (e.g. news quality, popularity~\cite{Ye2019}), to the best of our knowledge, none of them is specialized in Epidemic Intelligence. This means that, unlike the previous studied dimensions, we are not able to compare the obtained results with a gold standard database.

For this reason, we propose to evaluate an EBS system for the source dimension from a different perspective. Instead of evaluating the quality of the results, we rather assess the consistency of the results obtained in Sections~\ref{subsubsec:IdentificationImportantNewsOutlets} and~\ref{subsubsec:TimelyDetection} one another. The rationale is that for an effective surveillance system, we expect news outlets playing a key role in epidemiological information dissemination (as detected in Section~\ref{subsubsec:IdentificationImportantNewsOutlets}) to be those which are timely in event detection (as in Section~\ref{subsubsec:TimelyDetection}). Otherwise, this inconsistency amounts to decrease the main strength of an EBS system, which is its timeliness. 

Concretely, given an event database $\mathcal{E}$ we consider the ranked list produced for a timely detection objective (Section~\ref{subsubsec:TimelyDetection}) as a reference list, and compare it with that obtained in Section~\ref{subsubsec:IdentificationImportantNewsOutlets} by applying Equation~\ref{eq:Fmeasure_norm_definition}, as shown in Equation~\ref{eq:Fnorm_sorce_dimension}.

\begin{equation}
    \widehat{F} (\mathcal{E}) = F (\text{PageRank}(G_\mathcal{E}, k), \text{CELF}(G_\mathcal{E}, k)))
    \label{eq:Fnorm_sorce_dimension}
\end{equation}

\section{Additional Results}
\label{sec-apx:AdditionalResults}

\subsection{Additional Results for Spatio-temporal Representativeness}
\begin{landscape}
\begin{figure}[ht!]
    \includegraphics[width=1\linewidth]{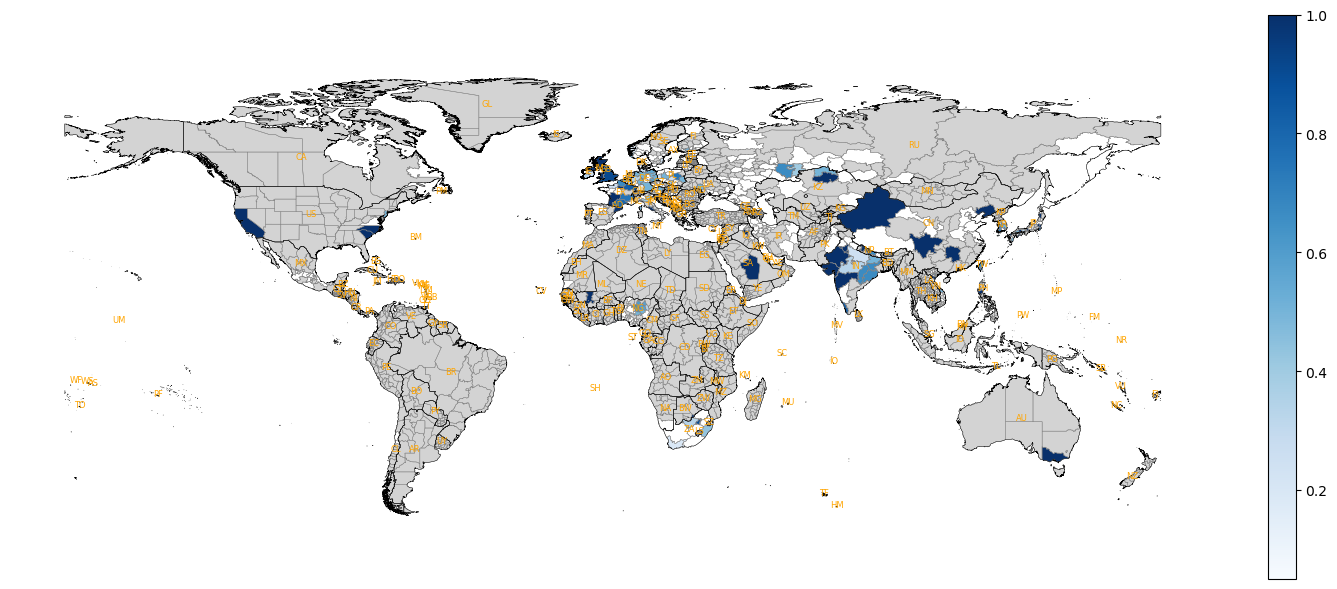}
    \caption{Spatio-temporal Representativeness scores at region scale for PADI-web with respect to the results of Empres-i. The degree to which an EBS system covers the {Empres-i} events occurring in a country is shown with different blue scales, and it is shown in white when an EBS system never finds an event in $\mathcal{E}_{EI}$. Moreover, countries without an {Empres-i} event are indicated in gray.}
    \label{figapx:spatio-temporal_representativeness_padiweb_region_scale}
\end{figure}
\end{landscape}

\begin{landscape}
\begin{figure}[ht!]
    \includegraphics[width=1\linewidth]{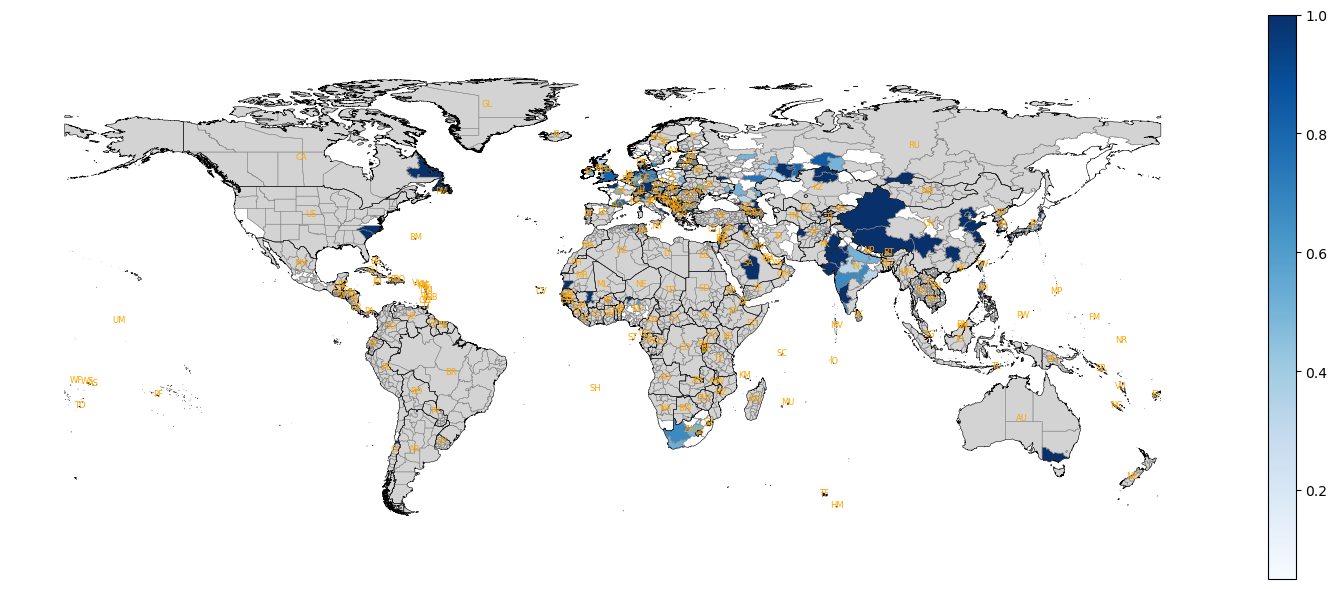}
    \caption{Spatio-temporal Representativeness scores at country scale for ProMED with respect to the results of Empres-i. The degree to which an EBS system covers the {Empres-i} events occurring in a country is shown with different blue scales, and it is shown in white when an EBS system never finds an event in $\mathcal{E}_{EI}$. Moreover, countries without an {Empres-i} event are indicated in gray.}
    \label{figapx:spatio-temporal_representativeness_promed_region_scale}
\end{figure}
\end{landscape}

In this section, we complete
our discussion in Section~\ref{subsec:Results_SpatialDimension} with the spatio-temporal representativeness scores obtained at region scale with monthly time intervals (Equation~\ref{eq:spatio-temporal_representativeness_score_for_zone_z}).  We plot the scores at region scale in Figures~\ref{figapx:spatio-temporal_representativeness_padiweb_region_scale} and \ref{figapx:spatio-temporal_representativeness_promed_region_scale}. In these plots, regions without an {Empres-i} event are indicated in gray, and the degree to which an EBS system covers the events occurring in a region is shown with different blue scales, where the largest values are in dark blue and the smallest ones in light blue. When an EBS system never finds an event in $\mathcal{E}_{EI}$, it is shown in white.

\begin{landscape}
\begin{figure}[ht!]
    \includegraphics[width=1\linewidth]{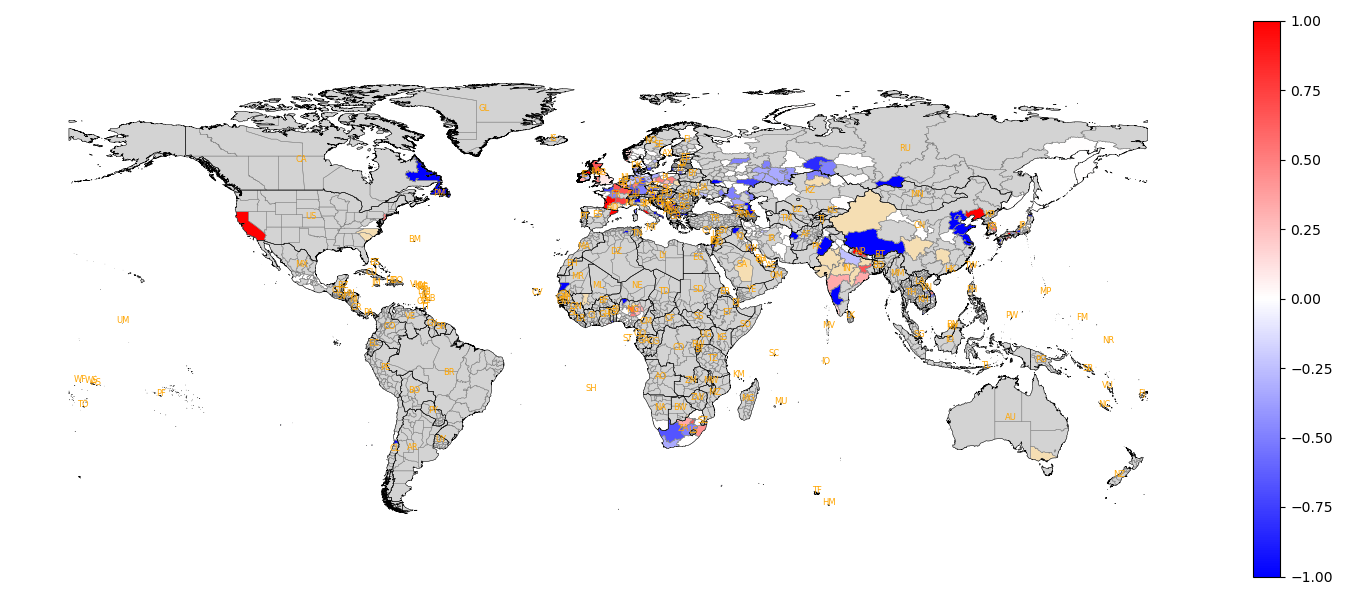}
    \caption{Spatio-temporal Representativeness score differences between PADI-web (Figure~\ref{figapx:spatio-temporal_representativeness_padiweb_region_scale}) and ProMED (Figure~\ref{figapx:spatio-temporal_representativeness_promed_region_scale}) at region scale. The regions are colored in blue (resp. red) when ProMED (resp. PADI-web) gives better spatio-temporal representativeness scores for a region and in yellow in case of non-zero equality. Moreover, countries without an {Empres-i} event are indicated in gray.}
    \label{figapx:spatio-temporal_representativeness_padiweb_vs_promed_region_scale}
\end{figure}
\end{landscape}

Moreover, we also plot the score differences in Figure~\ref{figapx:spatio-temporal_representativeness_padiweb_vs_promed_region_scale} to ease the comparison between PADI-web and ProMED. In this figure, it is colored in blue (resp. red) when ProMED (resp. PADI-web) gives better spatio-temporal representativeness score for a country and in yellow in case of non-zero equality.

\subsection{Additional Results for Thematic Dimension}
\begin{landscape}
\begin{figure}[ht!]
    \centering
    \begin{subfigure}[b]{0.43\linewidth}
    \centering
    \includegraphics[width=0.95\textwidth]{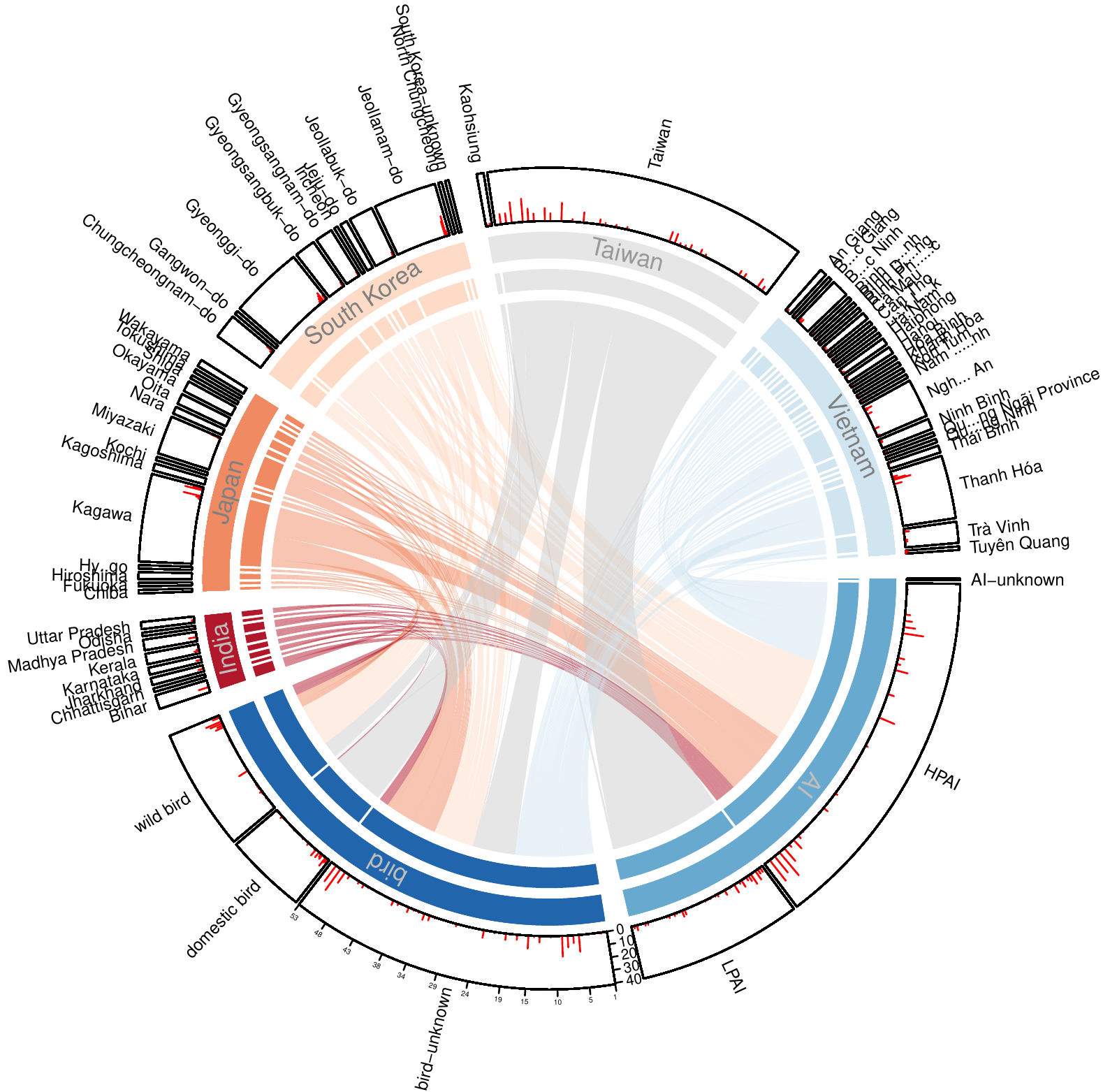}
    \caption{Empres-i}
    \label{subfig:chord_diagram_empres-i}
    \end{subfigure}
    \hfill
    \begin{subfigure}[b]{0.43\linewidth}
    \centering
    \includegraphics[width=0.95\textwidth]{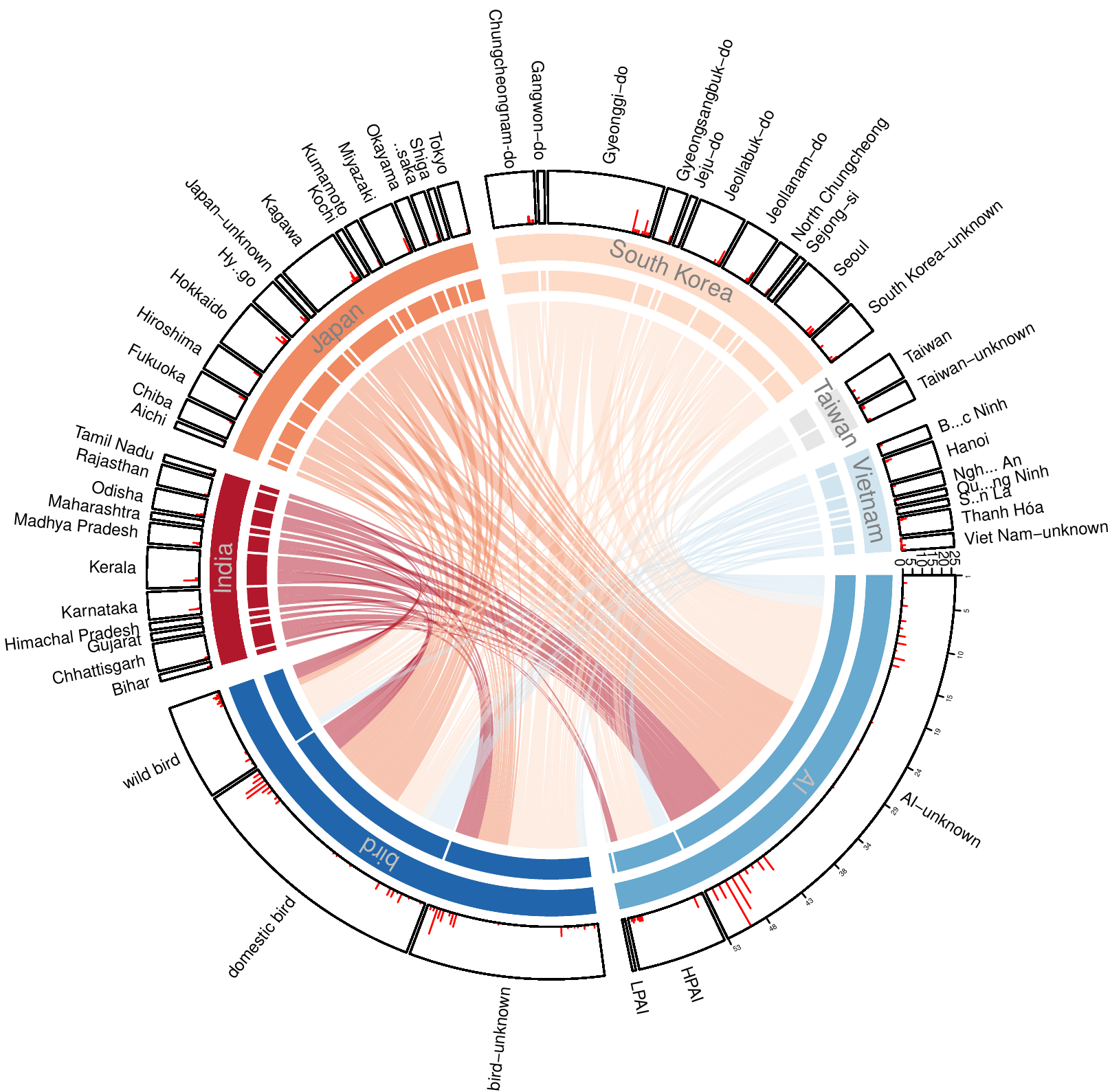}
    \caption{PADI-web}
    \label{subfig:chord_diagram_padiweb}
    \end{subfigure}
    \begin{subfigure}[b]{0.34\linewidth}
    \vspace{-3.25cm}
    \centering
    \includegraphics[width=0.95\textwidth]{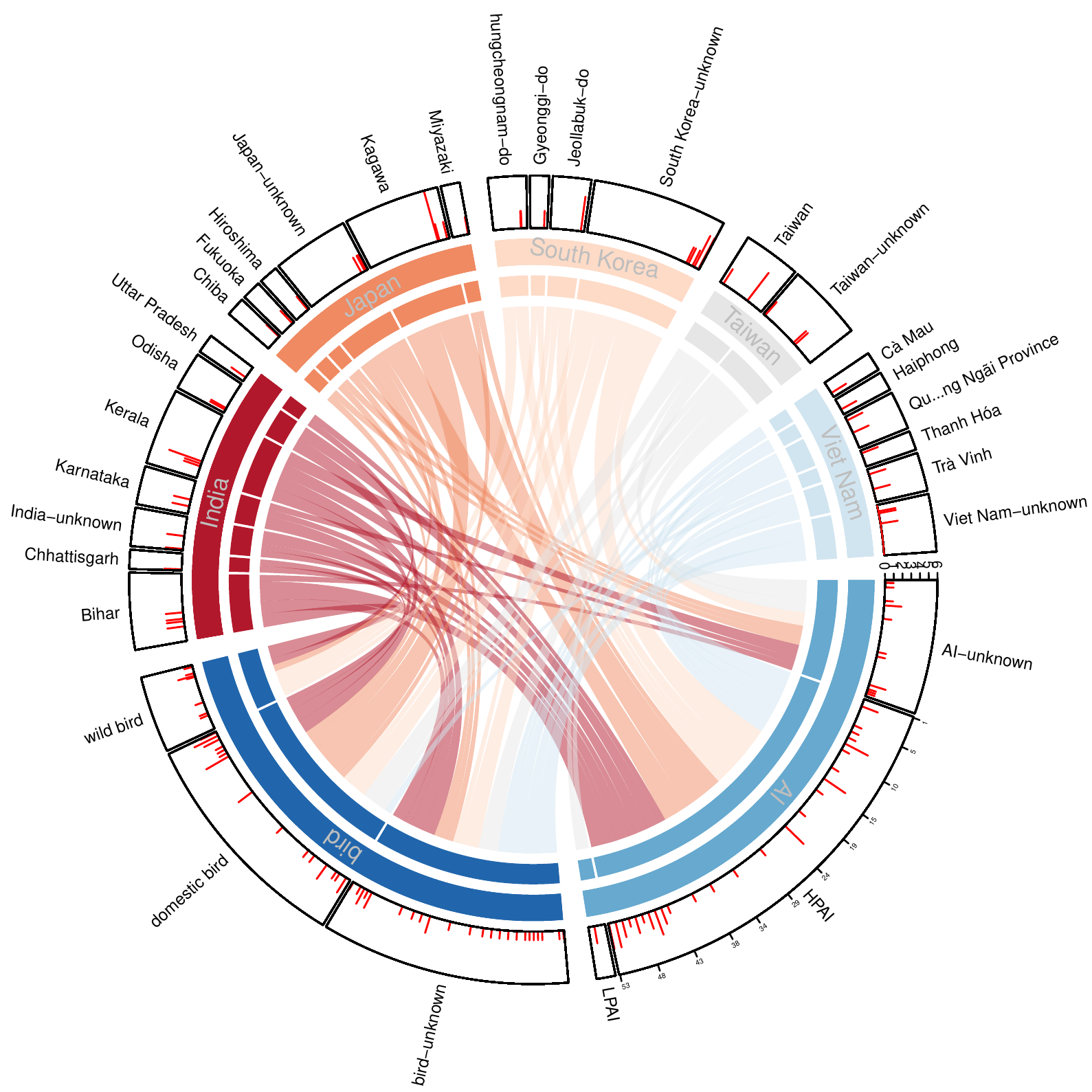}
    \caption{ProMED}
    \label{subfig:chord_diagram_promed}
    \end{subfigure}
    \caption{Visualization of the relations between spatial and thematic entities with a chord diagram for {Empres-i} (a), PADI-web (b) and ProMED (c). In these diagrams, we focus only on the Avian Influenza events occurring in Asian countries during 2020.
    }  \label{fig:Thematic_entity_chord_diagram_for_all_platforms}
\end{figure}
\end{landscape}

To ease our discussion in Section~\ref{subsec:Results_ThematicDimension}, in this section, we visualize in Figure~\ref{fig:Thematic_entity_chord_diagram_for_all_platforms} the relation between spatial and thematic entities with a \textit{chord diagram}. The advantage of this diagram is that we combine two aspects: hierarchy and temporal information. In each diagram, the connections between spatial and thematic entities are shown in the center of the circle. Each entity is represented by a fragment on the outer part of the circular layout. These entities are organized in a hierarchical way. In the most outer part of the circular layout, we show through a histogram how a given entity is involved in the events over time. In the histogram, the $x$-axis represents the weekly time intervals and the $y$-axis indicates the number of occurrences of a given entity in the events. Note that the same entities are colored in the same way across all the plots in order to ease the comparison.

We can compare the considered EBS systems in three aspects in these plots. The first aspect concerns the height of the segments, since it indicates the amount of information provided by a given EBS system. By looking at these segments, we can easily observe the frequent entities for each EBS system. For instance, the most frequent country for ProMED is India, whereas India is only the third most frequent country for PADI-web, after South Korea and Japan. Nevertheless, both EBS systems do not detect most of the events occurring in Taiwan. The second aspect concerns the links between entities. For instance, in ProMED we have more information regarding the details of Avian Influenza, compared to PADI-web. Nevertheless, both EBS systems do not detect most of the LPAI events. Finally, the third aspect is related to the temporal information, illustrated in the most outer layer of the circle. For instance, we see from Subfigure~\ref{subfig:chord_diagram_empres-i} that the Empres-i LPAI events occur throughout the year, whereas the HPAI ones occur only in the beginning and the end of the year.

\end{appendices}

\end{document}